\documentclass[12pt,english]{article}

\usepackage[letterpaper,margin=1in]{geometry}
\usepackage{mathptmx}
\usepackage{array}
\usepackage{float}
\usepackage{amsmath}
\usepackage{amsthm}
\usepackage{amssymb}
\usepackage{graphicx}
\usepackage{bm}
\usepackage{setspace}
\usepackage{natbib}

\usepackage[hidelinks]{hyperref}
\usepackage{tabularx}
\usepackage{placeins}
\usepackage{booktabs}
\usepackage{caption}
\usepackage{subfig}
\usepackage{makecell}
\usepackage{multirow}
\usepackage{longtable}
\usepackage{threeparttable}
\usepackage{pdflscape}

\setcitestyle{round}

\usepackage{babel}

\begin{document}
\begin{spacing}{1.5}
\title{The Quarter-Hour Effect: Periodic Algorithmic Trading and Return Predictability in Cryptocurrency Futures\thanks{We thank seminar participants at Duke University and Yonsei School of Business for valuable comments, and Daejin Kim for helpful comments and suggestions. We are grateful to Wade Kimbrough for undertaking an independent data validation and replication study. Hansen gratefully acknowledges funding provided by a grant from Ripple's University Blockchain Research Initiative (UBRI). The funder had no role in the research design, analysis, or publication decision.}}
\author{Chan Kim\thanks{Korea Information Society Development Institute, Department of AI Policy Research, 18, Jeongton-ro, Deoksan-eup, Jincheon-gun, Chungcheongbuk-do, 27872, Republic of Korea. E-mail: kchan3540@gmail.com.} \and Peter Reinhard Hansen\thanks{Corresponding author. University of North Carolina, Chapel Hill, Department of Economics, Gardner Hall, Chapel Hill, NC 27599, USA. E-mail: hansen@unc.edu.}}
\date{\today}
\maketitle
\vspace{-10mm}
\begin{abstract}
Cryptocurrency markets exhibit periodic bursts in volatility and volume at one-minute, five-minute, and quarter-hour marks. Using trade data for six Binance perpetual contracts, we link these bursts to algorithmic participation: trade-size roundness declines sharply during them. The Autocorrelation Map, a clock-phase-resolved display, reveals serial dependence in order flow and returns at quarter-hour openings that conventional measures obscure. Opening returns are predictable out of sample, while opening order imbalance predicts returns over four to twelve hours, with much weaker effects at finer clock-time frequencies. Together, these findings characterize periodic algorithmic trading and its cross-frequency variation.
\end{abstract}
\noindent \textit{\small{}Keywords:}{\small{} Cryptocurrency, High-Frequency Data, Market Microstructure, Algorithmic Trading, Return Predictability.}

\noindent \textit{\small{}JEL Classification:}{\small{} C58, G12, G14, G17}
\end{spacing}

\section{Introduction}\label{sec:introduction}

Algorithmic trading plays a prominent role in price formation and liquidity provision in modern electronic markets (\citealp{HendershottEtAl:2011}). It is widely known that some algorithmic strategies operate on standardized calendar-time bars, which is a convention shared by exchange APIs, charting platforms, technical-indicator defaults, and much of the academic literature on financial markets. When this temporal frame is shared across market participants, periodic bursts of trading activity at bar boundaries emerge as a structural feature of the market environment rather than an idiosyncratic choice of any single strategy.

\begin{figure}[!htbp]
    \centering
    \includegraphics[width=0.3\textwidth]{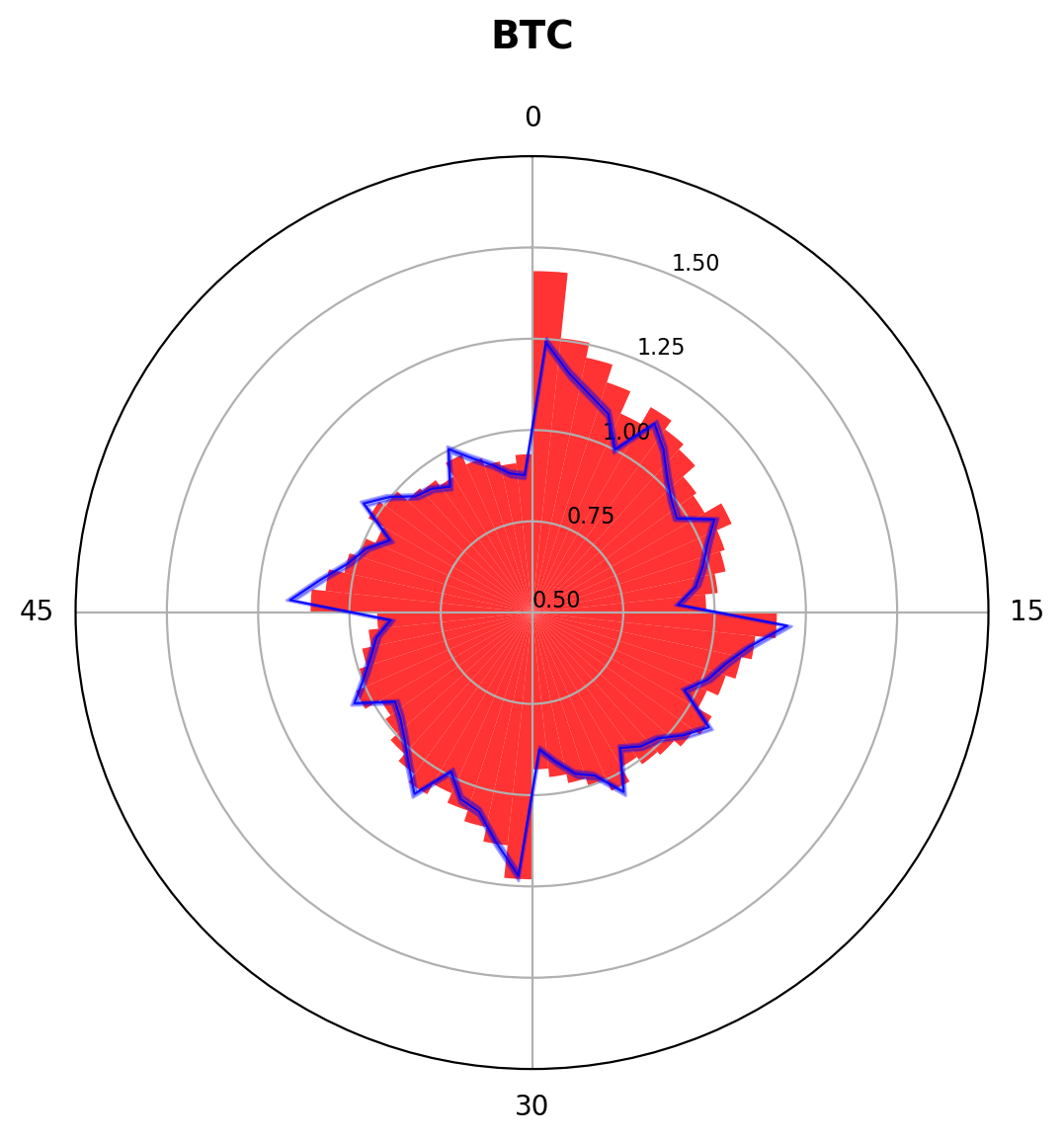}
    \includegraphics[width=0.3\textwidth]{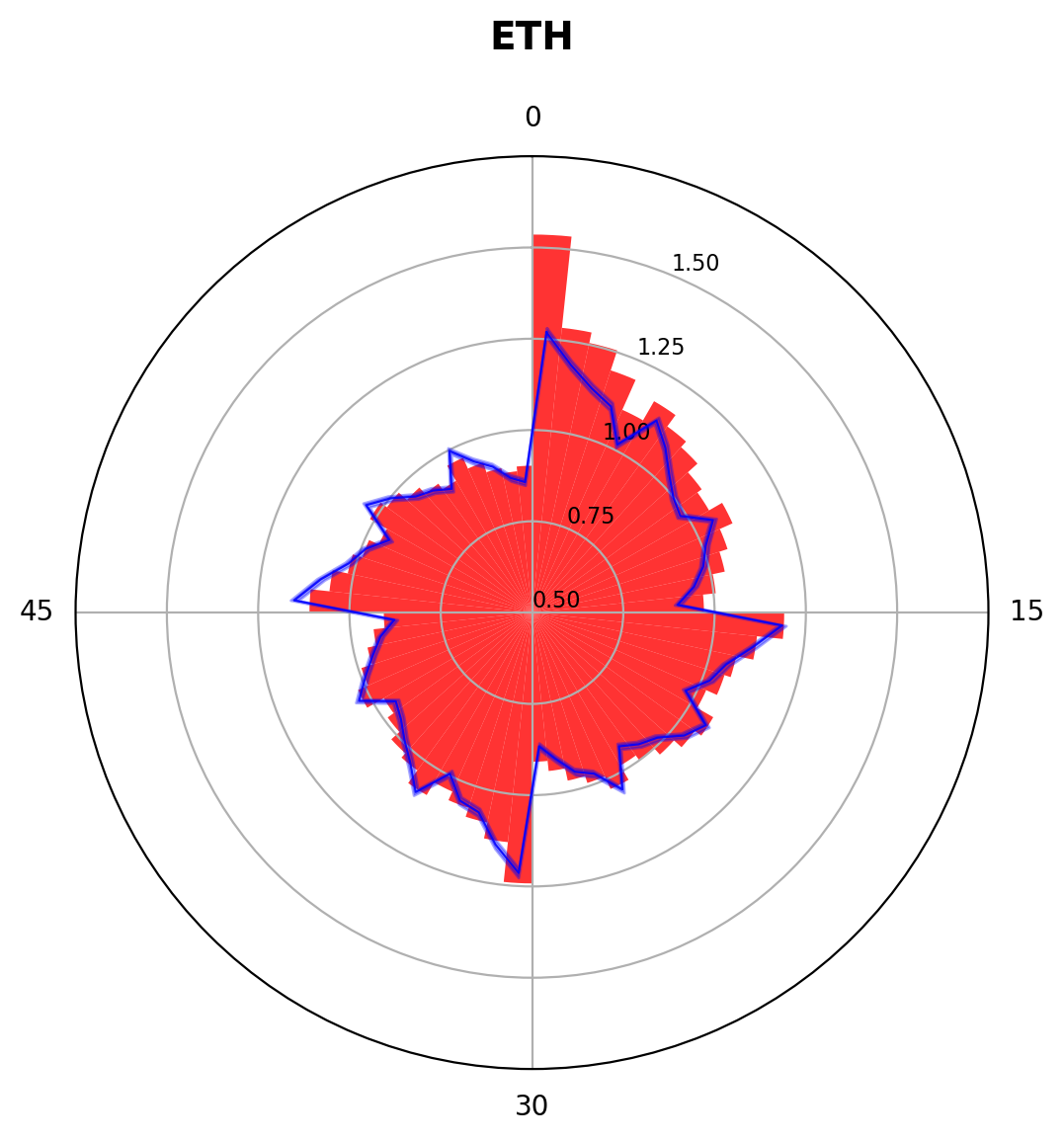}
    \includegraphics[width=0.3\textwidth]{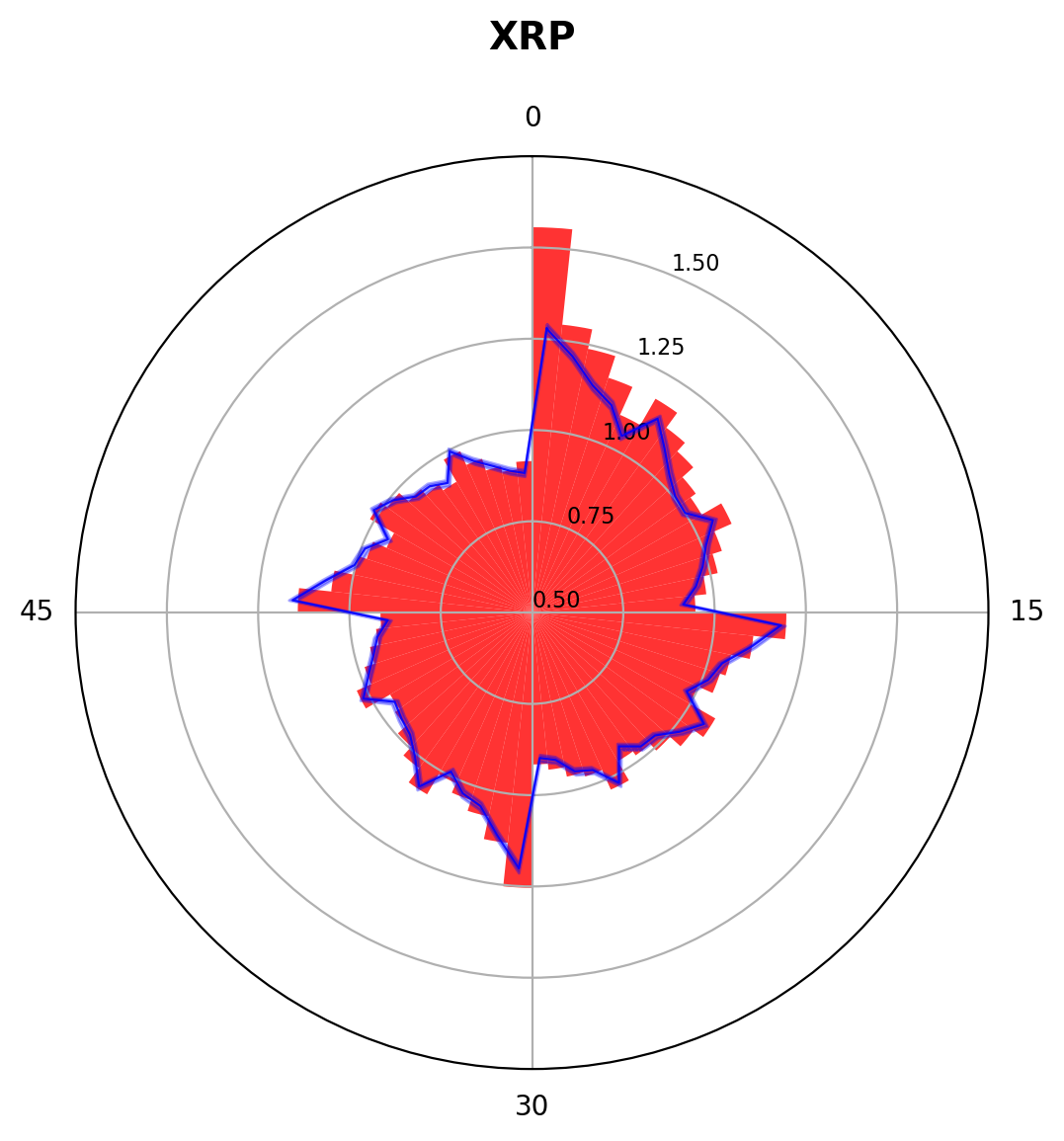}
    \includegraphics[width=0.3\textwidth]{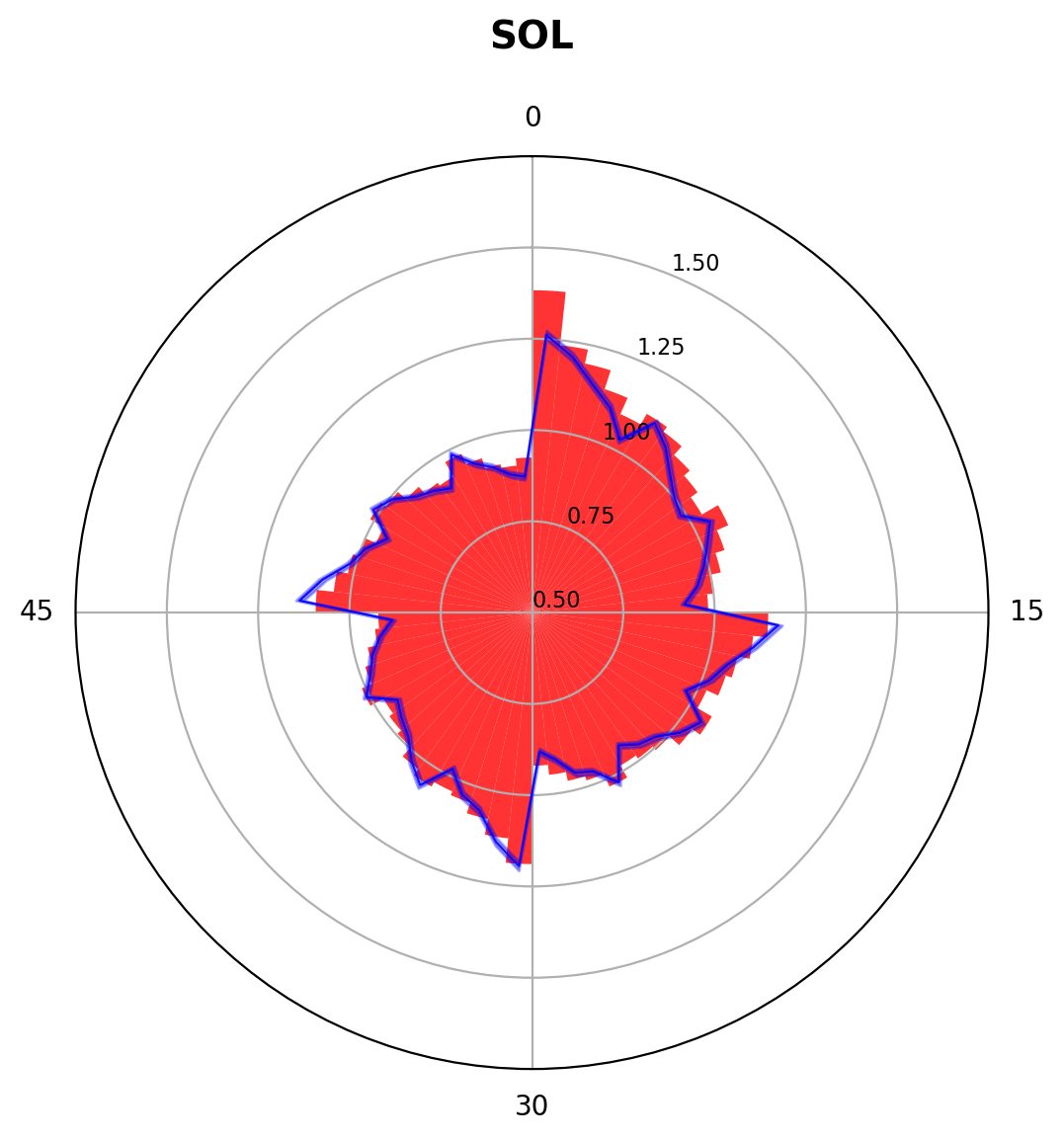}
    \includegraphics[width=0.3\textwidth]{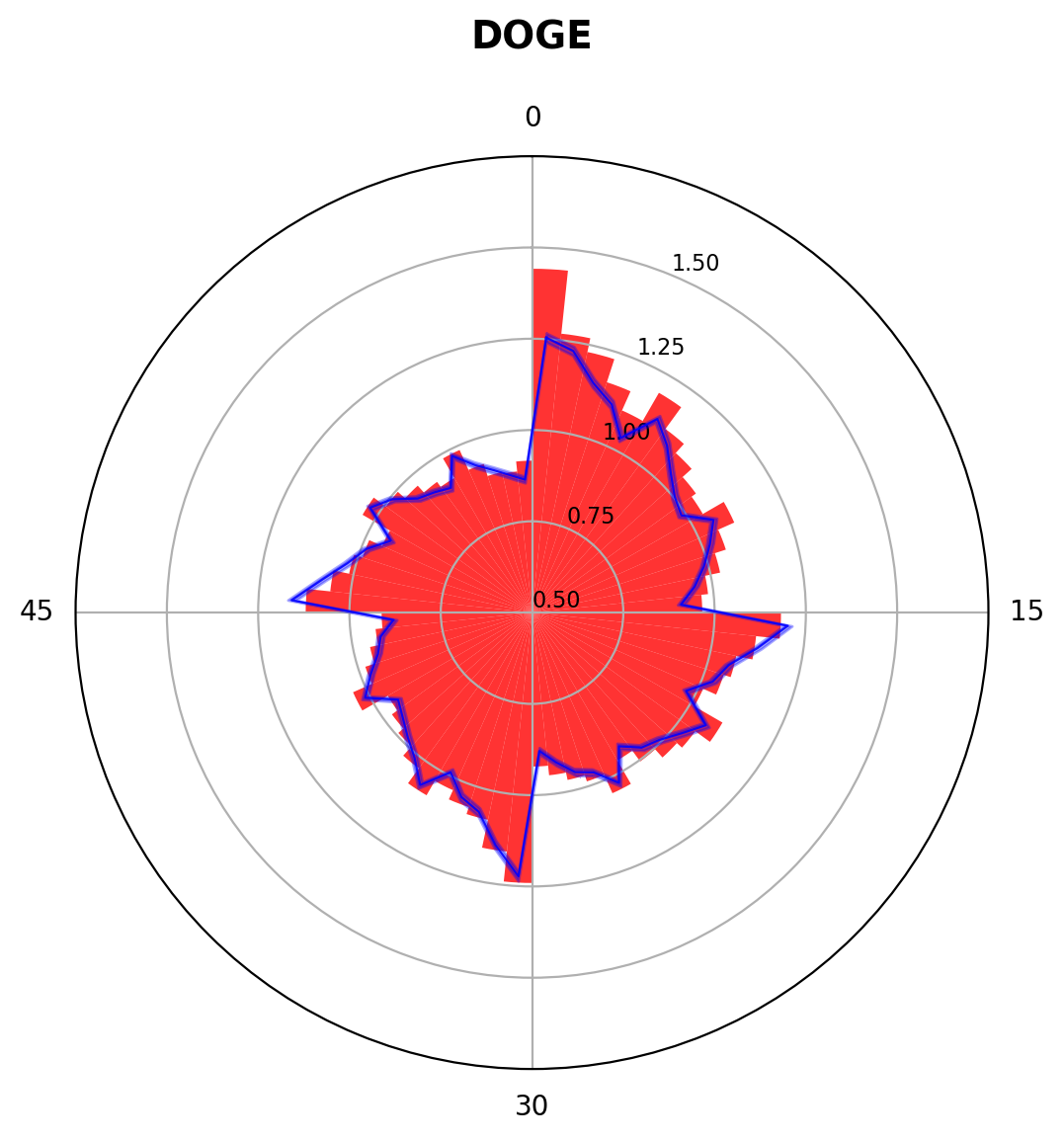}
    \includegraphics[width=0.3\textwidth]{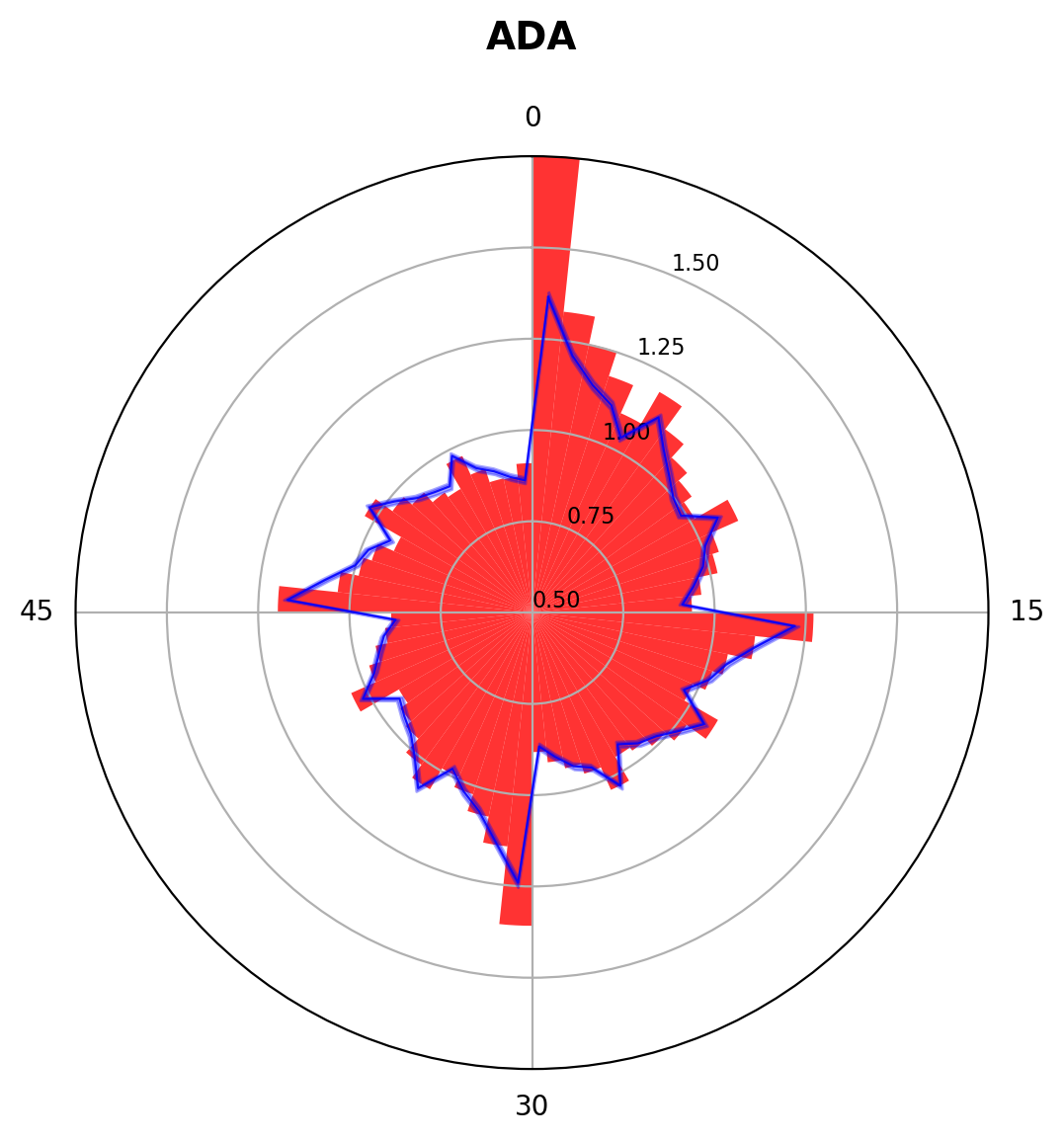}
    \caption{Minute-of-the-hour patterns in absolute returns (solid lines) and trading volume (bars) for six major Binance USDT perpetual futures contracts: BTC, ETH, XRP, SOL, DOGE, and ADA. Absolute returns are used as proxies for volatility. Both absolute returns and trading volume are normalized by their respective one-hour trailing averages. The pronounced spikes at 0, 15, 30, and 45 minutes reveal a strong recurring intrahour pattern across assets. Sample period: January 1, 2021 to October 31, 2024.}
    \label{fig:moh_nightingale}
\end{figure}

Cryptocurrency markets are no exception. As illustrated in the polar plots of Figure \ref{fig:moh_nightingale}, the most liquid cryptocurrency contracts exhibit pronounced star-shaped patterns: trading activity and short-horizon price variation concentrate into sharp bursts at every minute, every fifth minute, every quarter-hour, and most prominently at the top of the hour. \citet{HansenKimKimbrough:2024} document these nested periodicities for cryptocurrencies, and related patterns appear in various financial markets.\footnote{\citet{HasbrouckSaar:2013}, \citet{BroussardNikiforov:2014}, \citet{MuravyevPicard2022}, \citet{ChenChanChang2022}, \citet{Wuetal2025}, and \citet{Shynkevich2026} document related periodic patterns.} Their regularity is too systematic to be a benign timing artifact, and it raises broader questions about how algorithmic execution is reshaping market microstructure: what generates these bursts, how it propagates through order flow and returns, whether the resulting patterns are systematically forecastable, and whether boundary order flow contains information about subsequent price movement. We address these questions through four complementary analyses.

Relative to \citet{HansenKimKimbrough:2024}, who document that these periodicities exist, we take them as given and characterize their source, propagation, and predictability. Our analysis makes four contributions to the literature on periodic algorithmic trading and cryptocurrency market microstructure.

First, we provide high-frequency diagnostic evidence that periodic trading bursts are associated with algorithmic participation. Prior work uses daily round-size trade ratios as proxies for algorithmic intensity (\citealp{Congetal2023,Wuetal2025}). We take this diagnostic to high frequency, showing that round-size order shares decline sharply at the exact 10-second windows in which periodic bursts occur. This co-movement ties the bursts themselves to a behavioral signature of algorithmic order flow, providing sharper indirect evidence that they are associated with heightened algorithmic activity.

Second, we show that clock-time periodicities are not merely different resolutions of the same phenomenon. We introduce the Autocorrelation Map (ACM), a sign-based, time-domain display of clock-phase-resolved autocorrelations that admits an exact decomposition of the conventional robust autocorrelation function into phase-specific components. It reveals phase-specific dependence in order flow and returns concentrated around the quarter-hour grid, and placebo phase shifts confirm that this dependence is tied to the true boundary phase rather than to arbitrary 15-minute spacing. Distinct clock-time schedules therefore generate distinct dependence patterns.

Third, we document out-of-sample predictability of quarter-hour opening returns. Prior equity studies document periodic return predictability around economically salient trading schedules, including month-end liquidity needs, infrequent rebalancing, and market opening or closing flows \citep{HestonKorajczykSadka2010, Bogousslavsky2016, Etulaetal2020, Gaoetal2018}. Our setting differs in that the pattern recurs at generic quarter-hour marks that are ordinarily not tied to an asset-specific economic event, but instead coincide with the calendar-time bar grid embedded in trading infrastructure. The opening 10-second return is predictable out of sample from its boundary-aligned lags, and technical indicators add significant incremental information, so this predictability is not exhausted by same-phase recurrence. This evidence links the phase-specific dependence revealed by the ACM to economically meaningful short-horizon forecastability.

Fourth, we show that quarter-hour order imbalance contains medium-horizon predictive content for future returns. Much of the market microstructure literature infers informed trading and price discovery from contemporaneous price-impact measures, spread dynamics, or VAR-based decompositions of permanent and transitory price movements (\citealp{Stoll1989,Hasbrouck1991,BarclayWarner1993,HuangStoll1997,Madhavan2000,WangYang2015,MuravyevPicard2022,Brogaardetal2022}). We instead test whether order flow at specific clock-time boundaries predicts subsequent returns. Horizon-specific regressions indicate that this predictive content is concentrated at the quarter-hour frequency, and a two-stage decomposition shows that the longer-horizon component is mainly associated with the part of boundary order imbalance spanned by observable price-volume state variables.

Our findings also have implications for scheduled algorithmic traders and liquidity providers. Scheduled traders may condition execution timing on forecasts of quarter-hour opening returns or randomize execution around scheduled times to reduce synchronization and make their timing less predictable. Liquidity providers, in turn, may adjust quotes and inventory ahead of quarter-hour openings, where aggregate order flow and short-horizon returns become partly predictable.

The remainder of the paper proceeds as follows. Section 2 describes the dataset and the construction of high-frequency variables. Section 3 applies the trade-size roundness diagnostic. Section 4 introduces the Autocorrelation Map, which visualizes clock-phase autocorrelations, and documents phase-specific dependence in signed order flow and returns. Section 5 evaluates the out-of-sample forecastability of quarter-hour opening returns. Section 6 examines the informational content of quarter-hour order flow and decomposes its origin. Section 7 concludes.

\section{Data and Market Setting}\label{sec:data}

This study uses aggregate trade data from the Binance USDT-margined perpetual futures market, the largest crypto-derivatives venue by trading volume throughout our sample period.\footnote{Binance led all crypto-derivatives venues by trading volume in each year from 2021 to 2024; see CoinGecko, \emph{Market Share of Crypto Derivatives Exchanges by Trading Volume} (\url{https://www.coingecko.com/research/publications/crypto-derivatives-exchanges-market-share}).} Unlike traditional futures contracts, perpetual contracts have no predetermined expiration date; their prices are continuously anchored to a spot-market index. When the futures price diverges from the index, periodic funding payments are imposed on position holders to incentivize convergence; settlements occur every eight hours at fixed UTC times.

We use aggregate trade data on six large and liquid non-stablecoin cryptocurrencies, namely Bitcoin (BTC), Ethereum (ETH), XRP, Solana (SOL), Dogecoin (DOGE), and Cardano (ADA), covering January 1, 2021 through October 31, 2024.\footnote{Although Binance officially makes these data available for download through its website, we identified rare missing-data episodes in the official files; Binance had corrected these files by March 16, 2026. Their frequency was negligible and immaterial to our analysis.}  We select these assets for their long and liquid trading histories on Binance Futures. Each record includes the timestamp (in milliseconds), price, quantity, and the \texttt{isBuyerMaker} indicator. The \texttt{isBuyerMaker} variable is set to True when the buyer acts as the maker (executing at the best bid) and False when the buyer acts as the taker (executing at the best ask).

\subsection{Construction of High-Frequency Variables}\label{subsec:variables}

We aggregate millisecond-level trades into high-frequency calendar-time intervals to study intrahour periodicity. Let 
$t$ index these intervals. We define the price $P_t$ as the last traded price observed within interval $t$. We compute the corresponding log return as:
$$
r_t = \log(P_t) - \log(P_{t-1})
$$

We construct our measure of signed order flow using the \texttt{isBuyerMaker} indicator to infer trade direction. A trade is classified as buyer-initiated if the buyer is the taker (\texttt{isBuyerMaker} is False) and seller-initiated if the buyer is the maker (\texttt{isBuyerMaker} is True). Let $V_k$ denote the volume of trade $k$ within interval $t$, and let $D_k \in \{+1, -1\}$ indicate trade direction ($+1$ for buyer-initiated, $-1$ for seller-initiated). The net signed order flow for interval $t$ is
$$
\operatorname{OF}_t = \sum_{k \in t} V_k D_k
$$

We also construct the (volume-normalized) order imbalance:
$$
\operatorname{OI}_t = \frac{\operatorname{OF}_t}{\sum_{k \in t} V_k} \in [-1,1],
$$
which captures directional pressure independently of total trading volume, enabling comparison across intervals and assets with heterogeneous activity levels.

In addition to returns and order flow, we extract calendar-time coordinates from each observation. From the timestamp of interval $t$, we derive the date $d_t \in \{1, \ldots, 1400\}$ (one index per calendar day in the sample), the hour $h_t \in \{0, 1, \ldots, 23\}$, the minute-of-the-hour $m_t \in \{0, 1, \ldots, 59\}$, and the within-minute 10-second subperiod $b_t \in \{1, \ldots, 6\}$. The choice of 10-second windows is motivated in Section \ref{sec:PeriodicAlgo}.

\subsection{Descriptive Statistics}\label{subsec:desc_stats}

\begin{table}[!htbp]
\centering
\caption{Summary Statistics}
\label{tab:summary}
\small
\begin{tabularx}{\textwidth}{l *{7}{>{\raggedleft\arraybackslash}X}}
\toprule
Contract & \shortstack{Avg Daily\\Trades (M)} & \shortstack{Avg Daily\\DV (\$M)} & \shortstack{Avg Trade\\Size (\$)} & \shortstack{Zero 10s\\Bins (\%)} & \shortstack{$|r|_{10s}$\\(bps)} & \shortstack{$|r|_{1m}$\\(bps)} & \shortstack{Buyer\\Share} \\
\midrule
BTC & 1.54 & 14,579 & 9,441 & 0.01 & 1.879 & 5.436 & 0.499 \\
ETH  & 1.24 & 7,178 & 5,792 & 0.01 & 2.369 & 6.782 & 0.499 \\
XRP  & 0.29 & 957 & 3,308 & 0.09 & 2.939 & 8.128 & 0.503 \\
SOL  & 0.57 & 1,454 & 2,540 & 0.21 & 3.851 & 10.880 & 0.499 \\
DOGE  & 0.52 & 870 & 1,686 & 0.56 & 3.612 & 10.062 & 0.498 \\
ADA  & 0.29 & 544 & 1,881 & 0.45 & 3.159 & 8.686 & 0.497 \\
\bottomrule
\end{tabularx}
\medskip
\begin{minipage}[t]{\textwidth}
\setstretch{1}\footnotesize
\textit{Notes:} Sample period is January 1, 2021 to October 31, 2024 (1,400 calendar days). The unit of observation is a 10-second bar from Binance USDT-margined perpetual futures. Avg Daily Trades and Avg Daily DV are total divided by 1,400 calendar days. Avg Trade Size is total dollar volume divided by total trade count. $|r|_{10s}$ and $|r|_{1m}$ are mean absolute log returns. Buyer Share is the fraction of buyer-initiated trades.
\end{minipage}
\end{table}

Table \ref{tab:summary} reports summary statistics for our sample of six cryptocurrency perpetual futures contracts. The table presents the average daily number of trades, average daily dollar volume, average trade size, the frequency of zero-trade 10-second intervals, mean absolute interval returns (as proxies for volatility), and the fraction of buyer-initiated trades. 

Trading activity is highly concentrated in Bitcoin (BTC) and Ethereum (ETH), which average \$14.6 billion and \$7.2 billion in daily volume, respectively. These two dominant assets execute over a million trades per day, with average trade sizes substantially larger than those of the altcoins. Consequently, zero-trade 10-second intervals are almost absent for these contracts (0.01\% zero-trade bins). In contrast, the smaller-cap contracts (XRP, SOL, DOGE, ADA) exhibit lower aggregate volumes, smaller average trade sizes, and a higher frequency of zero-trade bins, reaching up to 0.56\% for DOGE. Furthermore, the smaller-cap assets exhibit substantially higher short-term volatility, with 1-minute absolute returns roughly double those of BTC. Finally, across all assets, the unconditional share of buyer-initiated trades is effectively symmetric, remaining close to 50\%. Daily minima, medians, and maxima of these aggregates are reported in Appendix Table~\ref{tab:daily_range}. The daily ranges are wide: the number of trades varies across days by a factor of roughly forty for BTC and by up to several hundred for the smaller contracts.

\section{Trade-Size Roundness as a Diagnostic of Algorithmic Participation}\label{sec:PeriodicAlgo}

Periodic patterns in trading intensity provide a window into the behavioral and technological regularities of modern financial markets. Prior research has documented such patterns across diverse asset classes: \citet{MuravyevPicard2022} report sub-minute periodicity in message arrivals in U.S. equity markets; \citet{ChenChanChang2022} identify a one-minute cycle in foreign-exchange trading activity; and \citet{Wuetal2025} show pronounced spikes in trade counts at 30-second, one-minute, and five-minute frequencies in both U.S. and Chinese equity markets. Recent cryptocurrency evidence also documents periodic trading activity and price discovery at subsecond frequencies, including within the first 100 milliseconds of the second \citep{Shynkevich2026EL,Shynkevich2026}.

Extending this evidence to the cryptocurrency market, we document comparable periodicity in trading activity, characterized by distinct surges at the 1-, 5-, and 15-minute marks, consistent with \citet{HansenKimKimbrough:2024}. While we attribute these broad patterns to algorithmic trading, the 15-minute cycle exhibits unique characteristics. The objective of this section is to disentangle these intrahour patterns and provide diagnostic evidence that the pronounced surges at quarter-hour boundaries are associated with a distinct form of periodic algorithmic activity.

\subsection{Intrahour Periodicity in Trading Activity}\label{subsec:intrahour_periodicity}

\begin{figure}[!htbp]
    \centering
    \includegraphics[width=0.32\textwidth]{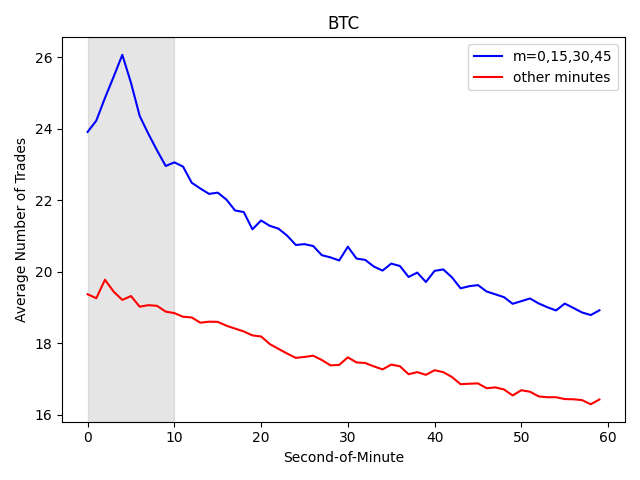}
    \includegraphics[width=0.32\textwidth]{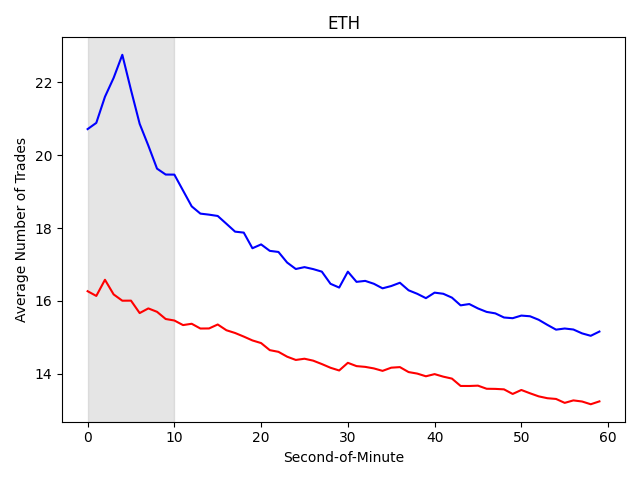}
    \includegraphics[width=0.32\textwidth]{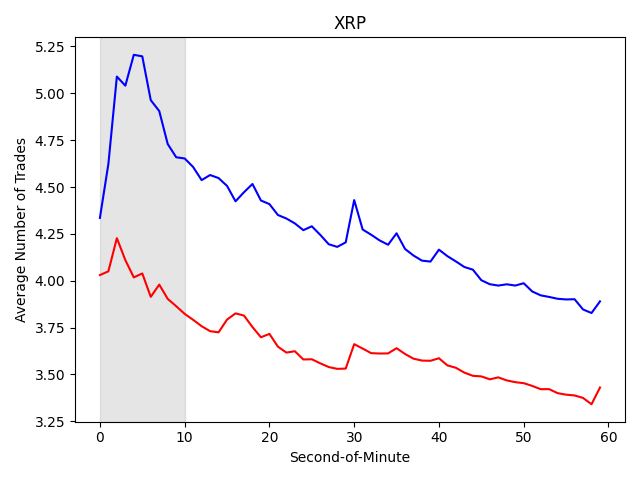}
    \includegraphics[width=0.32\textwidth]{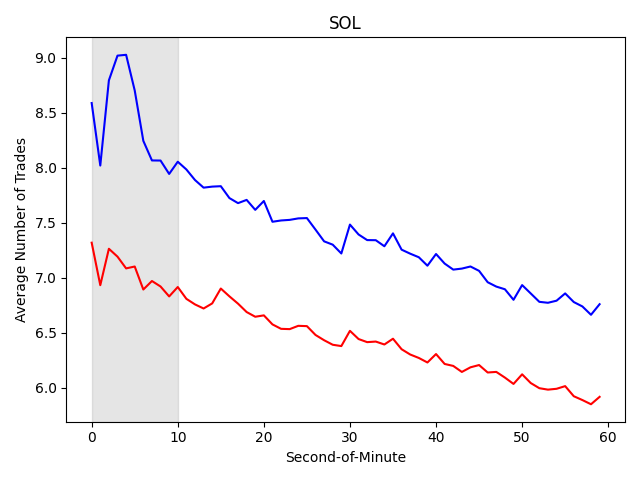}
    \includegraphics[width=0.32\textwidth]{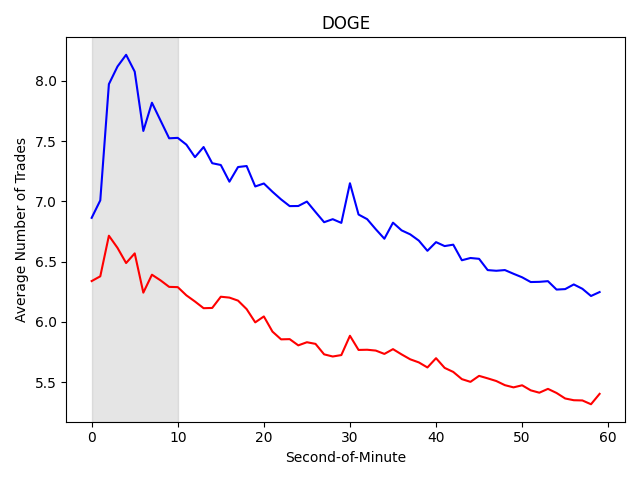}
    \includegraphics[width=0.32\textwidth]{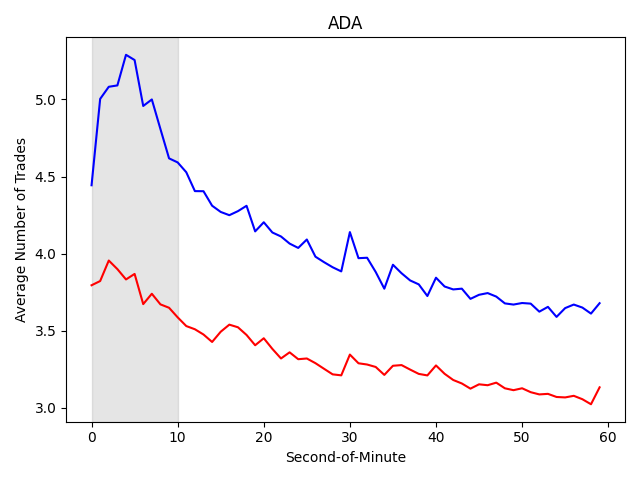}
    \caption{Each panel plots the average number of trades by second of the minute for BTC, ETH, XRP, SOL, DOGE, and ADA. The blue line averages activity during quarter-hour minutes (0, 15, 30, and 45), while the red line averages activity during all other minutes. Trading activity at quarter-hour marks rises sharply in the opening seconds and reaches its maximum within the first 10 seconds, indicated by the shaded region. This interval is defined as the peak window ($b=1$) in the analysis that follows. Sample period: January 1, 2021 to October 31, 2024.}
    \label{fig:TradingSecondOfMinute}
\end{figure}

Trading activity exhibits a nested intrahour structure, with bursts at one-minute, five-minute, quarter-hour, and hourly boundaries that become more pronounced at more salient clock times (Figure \ref{fig:moh_nightingale}; Appendix Figure~\ref{fig:TradingSecondOfHour}). To localize these bursts within each minute, Figure \ref{fig:TradingSecondOfMinute} plots average trade counts by second for quarter-hour and other minutes. Activity rises immediately after the minute boundary in both groups, but the increase is substantially larger at quarter-hour marks and peaks within the first 10 seconds. We therefore define this interval as the peak window.

Beyond trade counts, these bursts carry elevated turnover and volatility: averaged across the six contracts, the first 10 seconds of quarter-hour minutes exhibit roughly 26\% more trades, 32\% higher dollar volume, and 26\% larger absolute returns than the corresponding interval of ordinary minutes, all significant at the 1\% level (Appendix Table~\ref{tab:qh_burst}).

\subsection{Round-Number Bias as an Algorithmic Signature}\label{sec:rounding_bias}

We test whether the quarter-hour volume spikes originate from algorithmic agents by examining round-number bias in trade sizes. Human traders frequently submit orders at round numbers, consistent with cognitive ease and anchoring (\citealp{GarveyWu2014,VignaOsipovich2018}), whereas algorithmic traders are systematically less prone to such behavior. Building on this distinction, we measure the share of trades whose sizes end in trailing zeros (TZ) and test whether this share drops during the 10-second peak windows.

Let $\operatorname{TZshare}(z)_t$ denote the share of trades within interval $t$ whose sizes end with at least $z \in \{1, 2, 3\}$ trailing zeros, conditional on those trades being mechanically eligible to exhibit such roundness. 
On Binance perpetual futures, order step sizes are specified in the base asset and follow a power-of-ten format (e.g., 0.001 BTC for BTCUSDT). Although orders may be quoted in USDT, the exchange converts them to base-asset units internally, so trailing zeros are counted in the base-asset representation. Formally,
\begin{equation}
\mathrm{TZshare}(z)_t
\;=\;
\frac{\#\bigl\{k \in t \,:\, \mathrm{TZ}(V_k) \geq z\bigr\}}
     {\#\bigl\{k \in t \,:\, V_k \geq 10^z \cdot s_{\min}\bigr\}},
\label{eq:tzshare}
\end{equation}
where $V_k$ is the size of trade $k$, $s_{\min}$ is the minimum order
increment, and $\mathrm{TZ}(V_k)$ counts the trailing zeros in $V_k$.

This restriction to mechanically eligible trades adjusts for two confounding factors. First, the trade-size distribution shifts during burst windows. Without the eligibility restriction, the share of trailing-zero trades would mechanically rise with the share of large trades. Second, because the minimum increment is fixed in the base asset (e.g., $0.001$ BTC), the dollar value of any round quantity changes with the underlying price, so the mechanical-eligibility distribution varies over time. The restriction reduces behavioral variation in roundness from these compositional shifts.

We standardize $\mathrm{TZshare}(z)_t$ separately for each asset and each $z$ to have zero mean and unit variance, then estimate the following regression specification:
\begin{align*}
    \operatorname{TZshare}(z)_{t} &= \beta_0 +\beta_1 \mathbb{I}_{\{ b_t=1\}} + \beta_2 \mathbb{I}_{\{\operatorname{mod}(m_t,5)=0 \wedge b_t=1\}} + \beta_3 \mathbb{I}_{\{\operatorname{mod}(m_t,15)=0 \wedge b_t=1\}} \\&\,\quad + \beta_4 \mathbb{I}_{\{m_t=0 \wedge b_t=1\}}  + \epsilon_{t}
\end{align*}
The dummy indicators correspond to the periodic trading bursts identified earlier. This specification tests how the share of round-size orders systematically evolves during those specific bursts.

\begin{table}[!htbp]
\centering
\caption{Round-Number Bias at Periodic Boundary Openings}
\label{tab:TZshare2ExclusiveZ}
\footnotesize
\begin{tabularx}{\textwidth}{l *{5}{>{\centering\arraybackslash}X}}
\toprule
 & Constant & Top-of-hour & Quarter-hour & 5-minute & Every minute\\
 & & open $(b=1)$& open $(b=1)$ & open $(b=1)$ & open $(b=1)$ \\
 & \multicolumn{1}{c}{\scriptsize $(\beta_0)$}
 & \multicolumn{1}{c}{\scriptsize $(\beta_1+\beta_2+\beta_3+\beta_4)$}
 & \multicolumn{1}{c}{\scriptsize $(\beta_1+\beta_2+\beta_3)$}
 & \multicolumn{1}{c}{\scriptsize $(\beta_1+\beta_2)$}
 & \multicolumn{1}{c}{\scriptsize $(\beta_1)$} \\
\midrule
\multicolumn{6}{l}{\textbf{BTC}} \\
$\geq3$ trailing zeros & 0.0058*** & -0.1411*** & -0.0976*** & -0.0399*** & -0.0265*** \\
$\geq2$ trailing zeros & 0.0090*** & -0.2012*** & -0.1447*** & -0.0662*** & -0.0428*** \\
$\geq1$ trailing zero  & 0.0063*** & -0.0716*** & -0.0616*** & -0.0401*** & -0.0351*** \\
\midrule
\multicolumn{6}{l}{\textbf{ETH}} \\
$\geq3$ trailing zeros & 0.0045*** & -0.1112*** & -0.0800*** & -0.0379*** & -0.0198*** \\
$\geq2$ trailing zeros & 0.0045*** & -0.0771*** & -0.0634*** & -0.0355*** & -0.0223*** \\
$\geq1$ trailing zero  & 0.0032**\phantom{*}  & -0.0406*** & -0.0383*** & -0.0233*** & -0.0169*** \\
\midrule
\multicolumn{6}{l}{\textbf{XRP}} \\
$\geq3$ trailing zeros & 0.0075*** & -0.1769*** & -0.1136*** & -0.0644*** & -0.0348*** \\
$\geq2$ trailing zeros & 0.0073*** & -0.1946*** & -0.1122*** & -0.0637*** & -0.0329*** \\
$\geq1$ trailing zero  & 0.0043*** & -0.1173*** & -0.0612*** & -0.0350*** & -0.0203*** \\
\midrule
\multicolumn{6}{l}{\textbf{SOL}} \\
$\geq3$ trailing zeros & 0.0014\phantom{***} & -0.0657*** & -0.0281*** & -0.0145*** & -0.0032\phantom{***} \\
$\geq2$ trailing zeros & 0.0031*** & -0.0540*** & -0.0418*** & -0.0230*** & -0.0145*** \\
$\geq1$ trailing zero  & 0.0027*** & -0.0672*** & -0.0403*** & -0.0199*** & -0.0128*** \\
\midrule
\multicolumn{6}{l}{\textbf{DOGE}} \\
$\geq3$ trailing zeros & 0.0072*** & -0.1515*** & -0.0945*** & -0.0622*** & -0.0339*** \\
$\geq2$ trailing zeros & 0.0075*** & -0.1576*** & -0.0998*** & -0.0651*** & -0.0356*** \\
$\geq1$ trailing zero  & 0.0049*** & -0.1102*** & -0.0683*** & -0.0454*** & -0.0225*** \\
\midrule
\multicolumn{6}{l}{\textbf{ADA}} \\
$\geq3$ trailing zeros & 0.0069*** & -0.1469*** & -0.0968*** & -0.0517*** & -0.0325*** \\
$\geq2$ trailing zeros & 0.0089*** & -0.1763*** & -0.1245*** & -0.0732*** & -0.0428*** \\
$\geq1$ trailing zero  & 0.0015*** & -0.1252*** & -0.0248*** & -0.0494*** & \phantom{-}0.0010\phantom{***} \\
\bottomrule
\end{tabularx}
\medskip
\begin{minipage}[t]{\textwidth}
\setstretch{1}\footnotesize
\textit{Notes:} The table reports estimates from regressions using the full sample from January 1, 2021 to October 31, 2024. The dependent variable is $\operatorname{TZshare}(z)_{t}$, defined as the share of trades with at least $z$ trailing zeros among mechanically eligible trades. It is standardized to mean zero and unit variance in each row before estimation. Observations in bins where $z$ trailing zeros cannot mechanically arise because only small orders arrive were excluded from the sample. The reported columns correspond to the first 10 seconds of the top of the hour, quarter-hour marks excluding the top of the hour, 5-minute marks excluding quarter-hours, and ordinary minutes, respectively. The omitted category is all non-opening 10-second intervals. The coefficients shown are cumulative effects: $(\beta_1+\beta_2+\beta_3+\beta_4)$, $(\beta_1+\beta_2+\beta_3)$, $(\beta_1+\beta_2)$, and $(\beta_1)$. Standard errors are HAC with 30 lags. Statistical significance at the 1\%, 5\%, and 10\% levels is indicated by ***, **, and *, respectively.
\end{minipage}
\end{table}

Table \ref{tab:TZshare2ExclusiveZ} reports the regression estimates. The dummy coefficients are negative and highly significant across nearly all combinations of asset and threshold, with the hierarchical attenuation predicted under the algorithmic-participation hypothesis: smallest at the $b=1$ baseline of ordinary minutes, larger at 5-minute boundaries, larger still at quarter-hour boundaries, and largest at the top of the hour. For BTC at $z \geq 2$, the cumulative effect strengthens from $-0.04$ standard deviations at ordinary openings to $-0.20$ standard deviations at the top of the hour; the same hierarchy holds for the other assets and for $z \geq 3$. Taken together with the timing evidence in Section \ref{subsec:intrahour_periodicity}, the hierarchical decline in trade-size roundness provides diagnostic evidence of greater algorithmic participation at more salient clock-time boundaries.

A natural concern is whether this decline reflects algorithmic participation specifically or merely a shift in the composition of boundary order flow. Large institutional orders (which need not be round-sized), clustered liquidations, or funding-related arbitrage aligned with the eight-hour settlement schedule could each lower roundness without implying scheduled algorithms, and aggregate trade data cannot fully separate these channels. Two features are nonetheless difficult to attribute to such events alone: the decline strengthens monotonically with boundary salience, the prominence of a clock mark as a coordination point (ordinary $<$ five-minute $<$ quarter-hour $<$ top-of-hour), tracking the standardized calendar grid rather than any single liquidation or funding event; and calendar-aligned execution is itself the mechanism of interest, whether run by automated systems or by execution desks following bar-based conventions. We therefore treat the decline as consistent with periodic algorithmic participation rather than as proof of it.

\section{The Autocorrelation Map and Phase-Specific Dependence}\label{sec:ACM}

The conventional (Pearson) autocorrelation function (ACF) measures the correlation between returns at interval $t$ and $t + k$ by averaging across all $t$. When dependence varies with clock phase, this aggregation can obscure correlations concentrated at a small set of boundary locations. This concern is central in our setting because Section \ref{sec:rounding_bias} shows that trading activity changes sharply at quarter-hour openings. A phase-resolved measure is therefore needed to determine whether serial dependence is similarly concentrated at those boundaries. Appendix Figure~\ref{fig:acf_sacf_comparison} provides a preliminary illustration for 10-second BTC returns.

\subsection{The Autocorrelation Map (ACM)}\label{subsec:acm_definition}

Periodic dependence in time series has long been studied as cyclostationary (periodically correlated) processes (\citealp{Gladyshev1961}; see \citealp{GardnerNapolitanoPaura2006,Napolitano2016} for surveys), and the phase-resolved autocovariance we use is a standard object in that literature. The diagnostic tools there, however, operate in the frequency domain: \citet{Hurdgerr1991} detect periodicities from the discrete Fourier transform, and recent high-frequency work follows suit, with \citet{Wuetal2025} using Fourier analysis to identify dominant periodicities in equity volumes indexed by frequency rather than by within-cycle position.

Such summaries are well suited to identifying \emph{which} cyclic frequencies are present but not \emph{where in the cycle} the dependence concentrates, which is exactly the object of interest here: the quarter-hour boundary matters as a clock position, not merely as a 15-minute frequency. Our contribution is therefore methodological in a narrow sense. We use the term Autocorrelation Map (ACM) for the time-domain display of this object, a heatmap of the phase-resolved sign-based autocovariance indexed by lag $k$ and clock phase $m$. The sign-based formulation makes the display informative: it admits an exact decomposition of the conventional robust ACF into phase-specific components, which Pearson-based autocorrelations do not under the pronounced heteroskedasticity of these markets.

Using the notation introduced earlier, let $r_{(d,h,m)}$ and $\operatorname{OF}_{(d,h,m)}$ denote the one-minute log return and aggregate signed order flow for minute $(d,h,m)$. To measure dependence robustly, we use the sign-based correlation measure,
$$
\tau(k,m) \equiv \mathbb{E}\!\left[\operatorname{sign}\!\left(x_{(d,h,m)}x_{(d,h,m+k)}\right)\right],
$$
where $x$ represents either returns or order flow. Here, $m+k$ denotes advancing the clock by $k$ minutes, with the hour and date indices updated as necessary. A key advantage of $\tau$ is that it is scale-invariant, so it does not require estimation of volatility or other scale parameters, which are themselves highly time-varying and strongly periodic in these markets. This sign-based measure is closely related to the autocorrelation of order signs studied by \citet{LilloFarmer2004}, who analyze a $\pm 1$ symbolic order-flow series but do not explicitly cast it as a robust correlation measure.\footnote{Under suitable symmetry conditions, this sign-based correlation admits a one-to-one mapping to the Pearson correlation; see \citet{HansenLuo-RobustCorr:2023}.}

\paragraph{Application to 1-minute bars} We estimate the ACM for returns using its sample analog:
$$
\hat{\tau}_r(k,m) = \frac{1}{N_{k,m}}\sum_{d=1}^{D} \sum_{h=0}^{23} \mathrm{sign}\left(r_{(d,h,m)}r_{(d,h,m+k)}\right)
$$
where $k$ represents the lag in minutes and $N_{k,m}$ is the number of valid origins, i.e., date-hour pairs $(d,h)$ for which the observation at clock position $m+k$ lies within the sample; $N_{k,m}\approx HD$, with $D$ the number of days in the sample and $H=24$. The corresponding ACM estimator for signed order flow is:
$$
\hat{\tau}_v(k,m) = \frac{1}{N_{k,m}}\sum_{d=1}^{D} \sum_{h=0}^{23} \mathrm{sign}\left(\operatorname{OF}_{(d,h,m)}\operatorname{OF}_{(d,h,m+k)}\right).
$$

An additional advantage of the sign-based ACM is that it admits a clean aggregation property. The conventional robust ACF at lag $k$ is the weighted average of the phase-specific ACM rows, with weights given by the relative frequency of each clock phase. Thus, the ACM provides an exact decomposition of the aggregate robust ACF into its phase-specific components. This property does not generally hold for Pearson autocorrelations, because Pearson correlations require normalization by scale, and these scales vary systematically across clock phases. Under heteroskedasticity, the corresponding Pearson autocorrelation generally converges to a variance-weighted object rather than the simple average of phase-specific correlations; see \citet{HansenLuo-RobustCorr:2023}.

\subsection{One-Minute ACM Evidence}\label{subsec:acm_1min}

\begin{figure}[!htbp]
    \centering
    \includegraphics[width=1\textwidth]{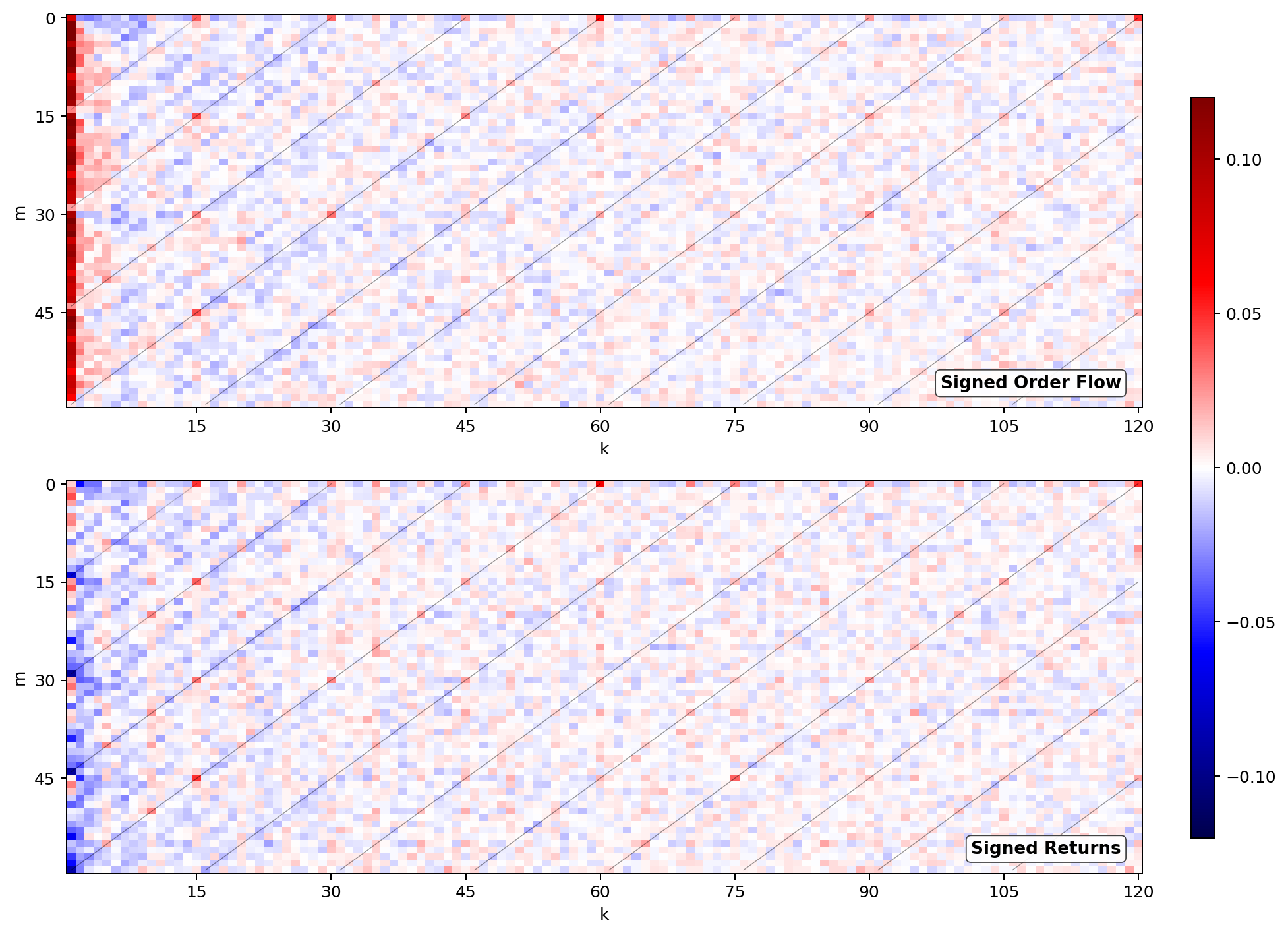}
    \caption{Autocorrelation Map (ACM) for 1-minute signed order flow and returns. The ACM plots the robust phase-conditioned autocorrelation $\hat{\tau}(k,m)$ as a function of lag $k$ and minute-of-the-hour $m$. The top panel reports signed order flow and the bottom panel reports returns. The black guide lines connect locations associated with quarter-hour phases. The figure reveals a pronounced lattice-like 15-minute structure that is obscured in the conventional autocorrelation function.}
    \label{fig:interminute_pattern_bymin}
\end{figure}

Figure \ref{fig:interminute_pattern_bymin} presents the 1-minute ACM for BTC order flow and returns. Each row $m$ corresponds to the minute-of-the-hour. To highlight the periodic structure, solid black lines connect all coordinates $(k, m)$ where the destination minute $m+k$ aligns with a quarter-hour mark (i.e., multiples of 15).

The 1-minute ACM reveals several salient patterns. First, the order-flow panel (top) displays a prominent red column at lag $k=1$, consistent with the well-documented long-memory property of order flow. However, this $k=1$ persistence weakens noticeably at $m=14,29,44,59$. These precise locations coincide with the bursts of algorithmic trading identified in Section~\ref{sec:PeriodicAlgo}, suggesting that periodic algorithms inject distinct temporal dynamics that disrupt the standard persistence observed in surrounding intervals.

Second, the return panel (bottom) reveals a starkly different pattern at lag $k=1$. While most minutes show little correlation, the quarter-hour marks exhibit pronounced negative autocorrelation (blue cells). For instance, the blue cell at $(k,m)=(1,14)$ indicates a negative correlation between the 1-minute returns at $m=14$ and $m=15$. This divergence suggests a temporary market friction. Under normal conditions, adaptive liquidity provision absorbs persistent order flow, preventing return autocorrelation \citep{LilloFarmer2004}. The sharp negative autocorrelation at quarter-hour marks is consistent with liquidity provision not fully absorbing the structural breaks in order flow at these boundaries, leaving price reversals that recur at the quarter-hour phase. Such reversals are consistent with the adverse selection that algorithmic order flow can impose on liquidity suppliers in equilibrium \citep{BiaisFoucaultMoinas2015}.

Lastly, both panels display a clear ``lattice'' structure of positive autocorrelation, visible as distinct red dots along the solid black lines. For example, at a lag of $k=15$, strong positive autocorrelation is evident at minutes $m=0, 15, 30, 45$. This pattern repeats at multiples of $k=15$, linking the 15-minute volume periodicity to autocorrelated algorithmic trading strategies deployed at these specific times.

\subsection{High-Resolution 10-Second ACM Evidence}\label{subsec:acm_10sec}

While the 1-minute aggregation provides a clear overview, it intermingles the sharp algorithmic bursts with the subsequent 50 seconds of standard trading. To isolate the exact footprint of these algorithms, we apply the ACM at a higher temporal resolution. 

We partition each minute into $B=6$ subperiods of 10 seconds each, indexed by $b$. We calculate the 10-second return $r_{(d,h,m;b)}$ and 10-second signed volume $v_{(d,h,m;b)}$. The high-resolution ACM estimator for returns is:
$$
\hat{\tau}_r(k,m;b) = \frac{1}{N_{k,m}}\sum_{d=1}^{D} \sum_{h=0}^{23} \mathrm{sign}\left(r_{(d,h,m;b)}r_{(d,h,m+k;b)}\right).
$$
A parallel estimator, $\hat{\tau}_v(k,m;b)$, is constructed for signed order flow. 

\begin{figure}[!htbp]
    \centering
    {\includegraphics[width=1.00\textwidth]{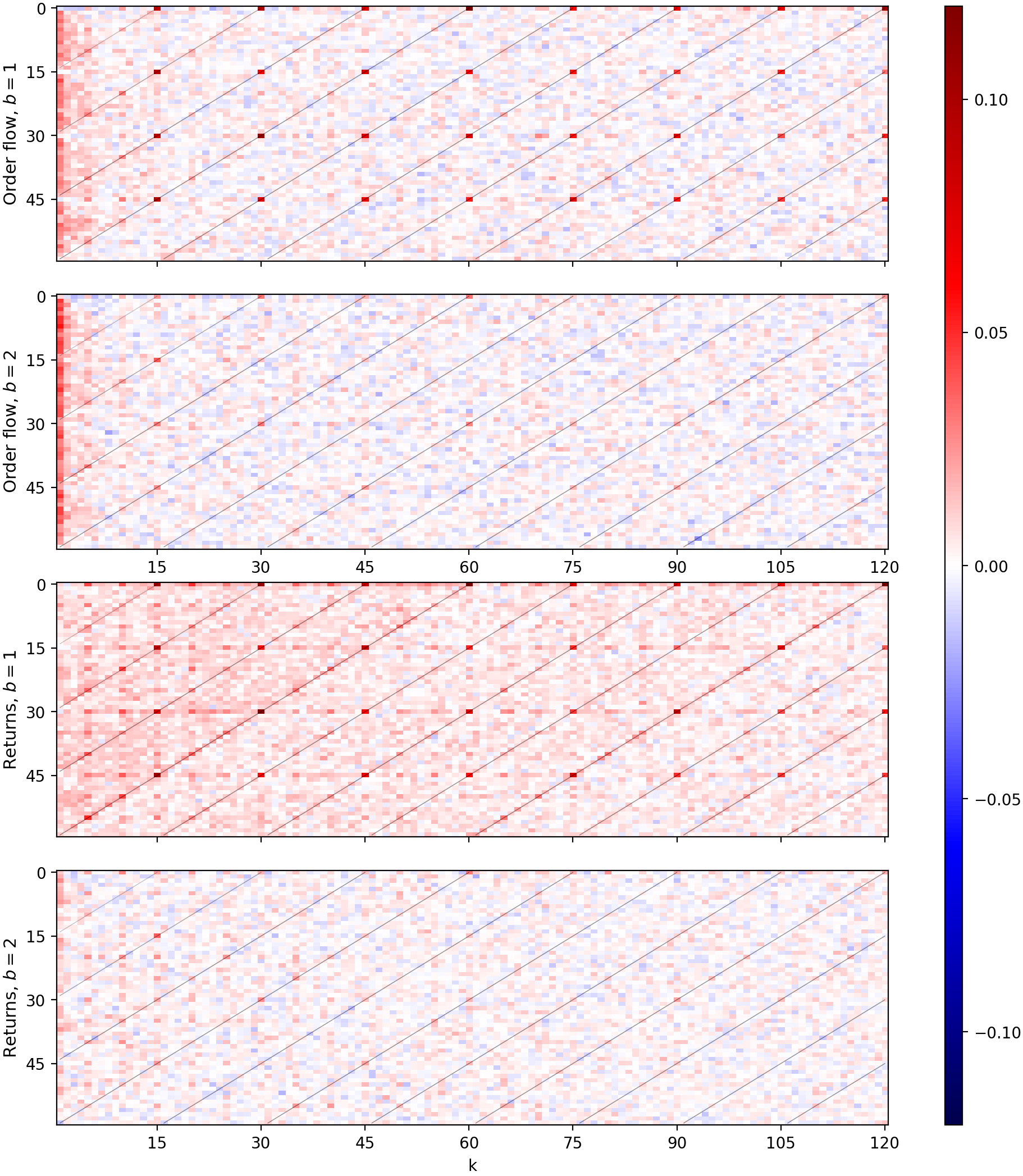}}\\
    \caption{Autocorrelation Maps (ACM) for 10-second signed order flow and returns. The upper two panels plot the robust phase-conditioned autocorrelation $\hat{\tau}_{v}(k,m;b)$ for BTC signed order flow at the 10-second frequency; the lower two panels plot the corresponding map for returns. Within each panel, the heatmap uses either the first 10-second interval of each minute ($b=1$) or the second interval ($b=2$). The strongest phase-specific dependence is concentrated in the opening 10 seconds of quarter-hour windows.}
    \label{fig:acm10s_combined}
\end{figure}

Figure \ref{fig:acm10s_combined} compares the 10-second ACM in the first and second 10-second intervals of each minute for signed order flow and returns. The quarter-hour lattice is overwhelmingly concentrated in the $b=1$ peak window and nearly disappears by $b=2$, showing that periodic dependence is sharply localized within the minute. The same within-minute decay appears across all six Binance contracts and in BTC and ETH data from Bybit (Appendix Figures~\ref{fig:Cross_asset_ACM_1m}--\ref{fig:bybit_ACM_returns_10s}).

\subsection{Interpreting the Lattice Structure in Order Flow and Returns}\label{subsec:lattice_interpretation}

Taken together, the heatmaps in Figures \ref{fig:interminute_pattern_bymin} and \ref{fig:acm10s_combined} show that the quarter-hour bursts documented in Section \ref{subsec:intrahour_periodicity} do not merely reflect elevated trading volume; they also carry a structured pattern of phase-specific serial dependence in order flow and returns. Synthesizing the evidence across the order-flow and return panels yields three observations.

First, the usual one-minute persistence in opening-window order flow weakens across quarter-hour boundaries. In the $b=1$ order-flow heatmap of Figure \ref{fig:acm10s_combined}, the lag-one autocorrelations at $m \in \{14, 29, 44, 59\}$ drop toward zero, indicating that order flow in the first 10 seconds of a quarter-hour is only weakly related to order flow in the corresponding 10-second window one minute earlier.

Second, the phase-specific dependence persists across longer horizons. The dark red dots at quarter-hour coordinates $k \in \{15, 30, \ldots\}$ and $m \in \{0, 15, 30, 45\}$ indicate positive autocorrelation in order flow across successive quarter-hour cycles. 

Finally, the return lattice raises the possibility that periodicity extends beyond synchronized taker flow to liquidity supply. In the $b=1$ panels of Figure \ref{fig:acm10s_combined}, return dependence forms a more extensive lattice than signed order-flow dependence, appearing not only at quarter-hour coordinates but also along the finer five-minute grid. Our order-flow measure captures only the direction of taker-initiated trades, whereas returns reflect the interaction of taker flow with standing limit orders, which provide liquidity in a limit order book \citep{Foucaultetal2005}. The additional structure may therefore reflect algorithmic liquidity provision that itself follows periodic schedules, through adjustments to limit-order placement, cancellation, or depth at salient clock-time boundaries. Because our data do not include order-book updates, we do not examine this mechanism directly.

A natural concern is whether the lattice reflects genuine quarter-hour structure or merely an artifact of imposing any regularly spaced four-point grid. We therefore compare the true grid (minutes 0, 15, 30, 45) against shifted grids that preserve the 15-minute spacing but are anchored at non-boundary phases (e.g., minutes 2, 17, 32, 47), computing the mean ACM value at the corresponding phase points across lags that are multiples of 15. Were the dependence a generic feature of evenly spaced sampling, the shifted grids would concentrate comparably. Instead, across all six assets and for both order flow and returns, the phase-point mean is largest under the true alignment and falls sharply under every placebo shift, tying the dependence to the true boundary phase rather than to an arbitrary grid (Appendix Figure~\ref{fig:Placebo}).

The ACM establishes that the 15-minute periodicity reflects phase-specific serial dependence rather than a mere surge in average volume. A natural next question is whether this structured, predictable order flow translates into predictable returns, and if so, what information drives it. Section~\ref{sec:forecasting} addresses both questions.

\section{Forecasting Quarter-Hour Opening Returns}\label{sec:forecasting}

The ACM in Section \ref{sec:ACM} shows that quarter-hour opening returns are autocorrelated in sample with their own same-phase lags; this section shows that they are also forecastable out of sample. We compare three predictor sets: lagged quarter-hour opening returns (motivated by the ACM), technical indicators (market-state predictors), and their combination. The combined model consistently outperforms either component alone. Thus, the predictability of the opening window is not exhausted by same-phase recurrence, but the technical indicators do not merely replace the lagged quarter-hour benchmark; rather, the two sources of information are complementary.

This result clarifies how to interpret phase-specific order-flow persistence. Prior work attributes persistent order flow either to order splitting by a single trader or to correlated decisions among traders who observe related signals \citep{Carteaetal2015}. Our evidence is more naturally aligned with the second channel, although it cannot rule out the first. Periodic rebalancing provides a related precedent, generating return autocorrelation and seasonality in equity markets \citep{Bogousslavsky2016}. The fact that technical and market-state variables improve out-of-sample forecasts of quarter-hour opening returns indicates that the periodic burst is not merely a mechanical continuation of same-phase order flow. Rather, part of the ACM pattern is linked to information available before the boundary. Without trader identifiers, we cannot separate same-trader persistence from cross-trader correlation, but the evidence shows that the pattern contains more than lagged same-phase dependence. More broadly, high-frequency return forecasting models should account for clock-time periodicity rather than treating intraday time as homogeneous.

\subsection{Response and Predictor Variables}

We construct a calendar-time version of the open-to-transaction return in
\citet{Sahaliaetal2025}. Let \(T\) denote the forecast origin, set
\(\Delta=10\) seconds, and let \(\mathcal{D}^{txn}\) be the set of transaction
timestamps. For each \(s\in\mathcal{D}^{txn}\), let \(P^{txn}_{s}\) and
\(V^{txn}_{s}\) denote the transaction price and traded volume. Relative to their construction, we make two adaptations. First, we evaluate the
return on a fixed 10-second calendar-time grid. Second, we replace their
equally weighted average of forward transaction prices with a volume-weighted
average. Because the Binance aggregate-trade data do not contain matched
contemporaneous quotes, we use the latest transaction price before \(T\) as
the opening reference price,\footnote{
\citet{Sahaliaetal2025} use the bid-ask mid-price. Binance provides
bid-ask data only from May 16, 2023 to March 30, 2024, which is
substantially shorter than our full sample. A parallel analysis using the
mid-price specification over this restricted window delivers similar conclusions (Appendix Table~\ref{tab:midprice}). }
$$
P^{open}_{T}=P^{txn}_{\tau(T)},
\qquad
\tau(T)=\max\{s\in\mathcal{D}^{txn}:s\leq T\}.
$$
The forward transaction set and its total traded volume are
$$
\mathcal{I}^{txn}_{T,\Delta}
=\{s\in\mathcal{D}^{txn}:T<s\leq T+\Delta\},
\qquad
V_{T,\Delta}=\sum_{s\in\mathcal{I}^{txn}_{T,\Delta}}V^{txn}_{s}.
$$
For \(V_{T,\Delta}>0\), the return response is
$$
Y^{ret}_{T}
=R(T,\Delta)
=\frac{P^{vwap}_{T,\Delta}}{P^{open}_{T}}-1,
\qquad
P^{vwap}_{T,\Delta}
=\frac{1}{V_{T,\Delta}}\sum_{s\in\mathcal{I}^{txn}_{T,\Delta}}V^{txn}_{s}P^{txn}_{s},
$$
the volume-weighted average transaction price over the forward window.
Windows with \(V_{T,\Delta}=0\) are dropped.

Two considerations motivate this forward-window construction. First,
individual transaction prices are noisy at very short horizons, whereas
aggregating all trades in \((T,T+\Delta]\) produces a more stable prediction
target. Second, this target better matches the execution problem. A trader
acting at \(T\) generally cannot control the exact instant of execution within
the next few seconds, so the relevant object is the price realized over the
window rather than the price at a single transaction time. Volume weighting
gives larger trades proportionally greater influence, so the target reflects
the distribution of traded quantity within the window rather than the
distribution of transaction timestamps alone.

For directional classification, we use the binary response
$$
Y^{dir}_{T}=\mathbf{1}\{R(T,\Delta)>0\}.
$$

We now turn to the predictor variables, which we group into two blocks. Let \(T\) denote a quarter-hour forecast origin and let \(R_T \equiv R(T,\Delta)\), with \(\Delta=10\) seconds, denote the response defined above. The predictor vector is
$$
    X_T = (L_T^\top, Z_T^\top)^\top,
$$
where \(L_T\) contains boundary-aligned lagged returns and \(Z_T\) contains technical indicators constructed from trailing price and volume information. 

The two blocks are designed to isolate different sources of predictive content. The lag block captures phase-specific recurrence at the same quarter-hour clock phase, as suggested by the Autocorrelation Map. The technical-indicator block captures broader pre-boundary market conditions summarized by recent price and volume histories. This decomposition lets us ask whether quarter-hour return predictability is mainly a same-phase persistence effect, or whether it also reflects information already embedded in the market state before the boundary.

The first block consists of the 12 first-10-second returns at preceding quarter-hour boundaries:
$$
    L_T =
    \left(
    R_{T-15\mathrm{m}},
    R_{T-30\mathrm{m}},
    \ldots,
    R_{T-180\mathrm{m}}
    \right).
$$
Equivalently, for \(k=1,\ldots,12\), the \(k\)-th lag variable is
$$
    L_{k,T}=R(T-15k\text{ minutes},\Delta).
$$
Because \(T\) is restricted to quarter-hour boundaries, these lags align the response with its own prior realizations at the same within-hour phase. The block therefore provides a direct forecasting counterpart to the phase-specific dependence documented in Section~\ref{sec:ACM}.

The second block consists of 28 technical indicators (TI28) following \citet{Fiebergetal2025}. This block is motivated by a growing empirical literature documenting predictive content in technical indicators and technical trading rules in cryptocurrency markets. For example, \citet{HudsonUrquhart2021} document significant predictability and profitability for broad classes of technical trading rules across major cryptocurrencies, and \citet{Fiebergetal2025} show that price-volume trend signals constructed from technical indicators predict the cross-section of cryptocurrency returns. 

We include technical indicators not to identify an optimal technical trading rule, but to proxy for publicly observable price-volume signals available to algorithmic traders before scheduled execution times. The incremental contribution of this block over the lag block therefore asks whether quarter-hour opening return predictability is exhausted by same-phase recurrence, or whether it also reflects pre-boundary market states that may trigger or shape periodic algorithmic trading.

At each quarter-hour forecast origin $T$, we compute the technical indicators from a rolling history of 15-minute OHLCV bars, with the most recent bar spanning $[T-30\text{ min},T-15\text{ min})$. The baseline specification therefore excludes all price and volume information from the final 15 minutes before the forecast boundary, ensuring that the predictors are fully observable well in advance. In an unreported robustness check, updating the predictors using information through $T-1$ minute generally improves out-of-sample forecasting performance, while leaving the overall conclusions unchanged. The stability of the results is unsurprising because technical indicators evolve relatively slowly and the two information sets share most of their rolling history.

The 28 indicators cover four dimensions of the pre-boundary market state. Momentum is summarized by RSI, stochastic oscillators, stochastic RSI, and CCI; trend by moving-average deviations and MACD-based variables; volume pressure by volume moving-average deviations, volume MACD variables, and a Chaikin-type money-flow measure; and volatility by Bollinger-band location and width measures. Where appropriate, the indicators are expressed as proportional deviations or scale-adjusted quantities to improve comparability across assets and over time. The complete list of the 28 indicators and their exact formulas is provided in Section~\ref{sec:ia_timenu} of the Appendix. 

The complete predictor vector contains \(12+28=40\) variables. The lag variables capture phase-specific persistence at quarter-hour boundaries, while TI28 captures broader market-state information over horizons ranging from one hour to one day. The nested specifications below use these blocks to evaluate the separate and joint contribution of same-phase return history and pre-boundary market-state variables.

\subsection{Out-of-Sample Forecasting Design}

We follow the experimental design of \citet{Sahaliaetal2025}. We describe in turn the prediction model, the out-of-sample evaluation metrics, and the rolling-window protocol used for tuning and testing. 

\subsubsection{Penalized Linear Regression Model}
We adopt a linear model for the response and estimate it by LASSO, an $\ell_1$-penalized least-squares regression that shrinks coefficients toward, and often exactly to, zero, yielding a sparse and interpretable forecasting specification. All centering and scaling constants are estimated using the training sample only and held fixed throughout the test month, so the procedure contains no look-ahead information from the test period; the objective function and standardization details are given in Section~\ref{sec:ia_forecast_details} of the Appendix. LASSO is a transparent benchmark for forecasting with many correlated predictors and is widely used in financial applications with high-dimensional data; see, among others, \citet{fanetal2011}, \citet{ChincoJosephYe2019}, \citet{Sahaliaetal2025}, and \citet{aletietal2024}. Our objective is not to identify the globally best forecasting model, but to document predictable content in a simple, reproducible specification.

\subsubsection{Evaluation Metrics}

We evaluate out-of-sample performance separately for the continuous-return regression and the binary direction classification. The return regression is assessed using the out-of-sample $R^2$ relative to the zero forecast, \(R^2_{\mathrm{OOS}}\), which is positive whenever the model improves on the zero benchmark. The direction classification, estimated by $L_1$-penalized logistic regression, is assessed using two complementary metrics: the area under the ROC curve (AUC), a threshold-free measure of how well the predicted probabilities rank upward moves, with 0.5 corresponding to random ranking, and classification accuracy at the fixed 0.5 probability threshold. This separation is useful because return magnitude and return direction are distinct forecasting targets: a model may rank directional probabilities well even when its return-magnitude forecasts explain only a small fraction of realized variation. Formal definitions of all three metrics are provided in Section~\ref{sec:ia_forecast_details} of the Appendix.

\subsubsection{Rolling-Window Estimation and Evaluation}

Because the forecasting specifications contain many predictors and require
regularization, we use a rolling-window procedure for both estimation and
evaluation. The out-of-sample evaluation period runs from 2021-07-01 to
2024-10-31, yielding roughly 117,000 quarter-hour boundary observations per
asset. Models are refit on a monthly calendar grid. At each refit, the
specification is estimated using only the trailing six months of observations
strictly preceding the test month, and the fitted model is then used to
forecast all boundary observations in that month. This design produces 40
monthly refits for each asset and specification, and ensures that every
forecast is based only on information available at the forecast origin.

Hyperparameters are re-tuned at each monthly refit, using only the
corresponding training window. This allows the strength of regularization to
adapt over time as the relationship between predictors and returns evolves. For
the LASSO specifications, the penalty parameter $\lambda$ is selected from the
grid $10^{-5}$ to $10^{-1}$. Within each training window, $\lambda$ is chosen by
a chronological split: the most recent 20\% is held out as a tuning set on which
each candidate is scored by mean squared error, and the earlier 80\% is used for
fitting. The model is then re-estimated on the full window at the selected
$\lambda$ before forecasting the test month.

\subsection{Out-of-Sample Predictability Results}

Table~\ref{tab:no_buffer_specs} summarizes the out-of-sample performance of the three predictor specifications. The results show economically and statistically meaningful predictability of quarter-hour opening returns across all six assets. Panel~A reports \(R^2_{\mathrm{OOS}}\) for the continuous-return forecasts. The lag-only model, which uses twelve quarter-hour-spaced lags of the opening 10-second return, delivers a positive \(R^2_{\mathrm{OOS}}\) for every asset and a cross-sectional average of 2.464\%. This confirms that the phase-specific persistence documented above has genuine out-of-sample forecasting content.

The TI28 specification is slightly weaker on average, with a cross-sectional mean \(R^2_{\mathrm{OOS}}\) of 2.090\%, but its relative performance varies across assets. Technical indicators are the stronger standalone block for BTC, ETH, and DOGE, whereas lagged boundary returns are stronger for XRP, SOL, and ADA. ADA exhibits the largest lag-only \(R^2_{\mathrm{OOS}}\) in the panel, at 5.2\%.

The strongest results come from combining the two predictor blocks. The Lag + TI28 specification attains the highest \(R^2_{\mathrm{OOS}}\) for all six assets and raises the cross-sectional average to 3.37\%, more than one-third above the lag-only benchmark. This pattern indicates that quarter-hour opening return predictability is not exhausted by same-phase lagged returns. Broader pre-boundary market-state information contributes incremental predictive content beyond boundary-aligned lagged returns, a claim formalized by the incremental tests in Section~\ref{subsec:dm_tests}.

Panels~B and C report direction-prediction performance. The Lag + TI28 specification also performs best for directional forecasts, delivering the highest average AUC, 0.6011, and the highest average 0.5-threshold accuracy, 57.12\%. It attains the best AUC for every asset. The incremental contribution of TI28 is smaller for direction prediction than for continuous-return prediction, which is natural because the binary outcome discards magnitude information. Even so, adding TI28 improves AUC relative to the lag-only model for every contract. Overall, the results support a complementary-information interpretation: lagged boundary returns capture phase-specific persistence, while TI28 contributes additional information about the lower-frequency market state before the boundary.

The out-of-sample \(R^2\) values may appear modest, but two observations help place them in context. First, the relevant comparison is not only the size of the predictable component, but also its implementability. The return predictability documented by \citet{Sahaliaetal2025} for U.S.\ equities is larger, but it is concentrated at millisecond-to-second horizons and is highly sensitive to data timeliness. Exploiting that signal therefore requires low-latency feeds and co-location. Our forecasts are different. The baseline predictors are fixed at least 15 minutes before the scheduled boundary, allowing the signal to be computed and trading decisions to be prepared well in advance. In an unreported robustness check, updating the predictors using information through $T-1$ minute generally improves out-of-sample performance. The reported results therefore provide a conservative assessment of predictability and do not rely on ultra-low-latency data processing.

Second, the forecast should be interpreted as an input to execution and liquidity provision rather than as a standalone trading strategy. The predictable component is small relative to trading costs. The sign-weighted realized forecast target, $\mathbb{E}[\operatorname{sign}(\widehat{Y}_t)\,Y_t]$, averages about $0.5$~bp per boundary, about one tenth of a single standard-tier taker fee and one twentieth of a round trip.\footnote{Over the sample period, the standard-tier fee on Binance USDT-margined perpetual futures is $5.0$~bp for takers ($0.05\%$) and $2.0$~bp for makers ($0.02\%$); higher VIP tiers pay less. A
round trip pays two fees.} This quantity measures the predictable component in return units rather than the return to a directly implementable strategy, since the opening reference price is not necessarily attainable. The forecast is close to well calibrated, with Mincer--Zarnowitz slopes between $0.77$ and $0.94$ across the six contracts (Table~\ref{tab:forecast_magnitude} of the Appendix). Periodic activity is concentrated in the quarter-hour opening window, so a
trader or quote-setter operating at the boundary can use the forecast to adjust aggressiveness, delay submission, cancel quotes, or supply liquidity. Following \citet{Sahaliaetal2025}, we therefore assess economic relevance through the magnitude, robustness, and implementability of the predictable component itself, rather than
through a net-of-cost trading strategy that would require additional assumptions about market impact, queue priority, and execution rules.

\begin{table*}[!htbp]
\centering
\caption{Out-of-Sample Performance by Predictor Specification}
\label{tab:no_buffer_specs}
\scriptsize\setlength{\tabcolsep}{4pt}
\begin{tabular*}{\textwidth}{@{\extracolsep{\fill}}l r r r r r r r r}
\toprule
 & \multicolumn{1}{c}{$p$} & \multicolumn{1}{c}{BTC} & \multicolumn{1}{c}{ETH} & \multicolumn{1}{c}{XRP} & \multicolumn{1}{c}{SOL} & \multicolumn{1}{c}{DOGE} & \multicolumn{1}{c}{ADA} & \multicolumn{1}{c}{mean} \\
\midrule
\multicolumn{9}{l}{\textit{Panel A: continuous-return OOS $R^2$ (\%) vs zero}} \\
\midrule
Lag & 12 & 2.448 & 3.102 & 1.837 & 1.209 & 1.004 & 5.187 & 2.464 \\
TI28 & 28 & 3.068 & 3.765 & 1.365 & 0.879 & 1.203 & 2.259 & 2.090 \\
Lag + TI28 & 40 & \textbf{3.822} & \textbf{4.772} & \textbf{2.244} & \textbf{1.595} & \textbf{1.926} & \textbf{5.846} & \textbf{3.367} \\
\midrule
\multicolumn{9}{l}{\textit{Panel B: direction prediction AUC}} \\
\midrule
Lag & 12 & 0.5908 & 0.6090 & 0.5862 & 0.5756 & 0.5679 & 0.6153 & 0.5908 \\
TI28 & 28 & 0.5900 & 0.6033 & 0.5573 & 0.5713 & 0.5647 & 0.5783 & 0.5775 \\
Lag + TI28 & 40 & \textbf{0.6035} & \textbf{0.6223} & \textbf{0.5904} & \textbf{0.5878} & \textbf{0.5793} & \textbf{0.6230} & \textbf{0.6011} \\
\midrule
\multicolumn{9}{l}{\textit{Panel C: direction prediction accuracy at 0.5 threshold (\%)}} \\
\midrule
Lag & 12 & 56.42 & 57.84 & 56.05 & 55.49 & 54.94 & 58.01 & 56.46 \\
TI28 & 28 & 56.37 & 57.45 & 53.88 & 55.08 & 54.56 & 55.62 & 55.49 \\
Lag + TI28 & 40 & \textbf{57.30} & \textbf{58.56} & \textbf{56.27} & \textbf{56.35} & \textbf{55.61} & \textbf{58.67} & \textbf{57.12} \\
\bottomrule
\end{tabular*}
\begin{minipage}[t]{\textwidth}\setstretch{1}\footnotesize\textit{Notes.} Each cell reports the OOS metric on the quarter-hour first-10s sample (2021-07-01 to 2024-10-31, $\approx 117{,}000$ observations per asset). Bold marks the best specification per asset (and in the mean column). The number of
features $p$ counts the predictors entering each specification before LASSO
selection (Lag: 12 quarter-hour-spaced lags (3 hours) of the first-10s return;
TI28: 28 technical indicators on the 15-min boundary OHLCV grid). The
estimation protocol is a rolling-window walk-forward LASSO (for the continuous
return) or $L_1$-penalized logistic regression (for direction): the model is
re-estimated each month on the trailing six months of data and used to forecast
the following month, with the penalty parameter $\lambda$ selected on the
chronological 20\% tail of each training window.
\end{minipage}
\end{table*}

\subsection{Tests of Incremental Predictive Content}
\label{subsec:dm_tests}

Table~\ref{tab:dm_tests} reports Diebold--Mariano tests for the pairwise forecast comparisons suggested by Table~\ref{tab:no_buffer_specs}. For the quarter-hour first-10-second return forecast, we compute the loss differential $L^{(a)}_{t}-L^{(b)}_{t} = (Y_t-\widehat{Y}^{(a)}_t)^2-(Y_t-\widehat{Y}^{(b)}_t)^2$ over the out-of-sample window. The DM statistic is the HAC $t$-statistic for the mean loss differential, using lag~6. A positive statistic indicates that model~$b$ has a lower mean squared forecast error than model~$a$. We report the corresponding one-sided $p$-value. Although these are nested comparisons, the forecasts are generated from a fixed-length rolling window, under which the Diebold--Mariano statistic remains asymptotically normal for nested models \citep{GiacominiWhiteCPA}.

Two patterns emerge. First, Panel~A shows that both standalone predictor blocks have significant out-of-sample content. The Lag specification significantly improves on the zero forecast for all six contracts, with $p<0.01$ throughout and $t$-statistics ranging from 4.78 for SOL to 22.95 for ADA. This confirms that the boundary-aligned return persistence documented in Section~\ref{sec:ACM} carries out-of-sample predictive content for every contract. The TI28 specification is also significant against the zero forecast in all six contracts, again with $p<0.01$ throughout. Its strength is more uneven across assets: the evidence is strongest for ETH, BTC, and ADA, and weaker for SOL, DOGE, and XRP.

Second, Panel~B shows that TI28 contributes incremental information beyond the Lag specification. Adding TI28 to Lag reduces mean squared forecast error for every contract, with positive DM statistics ranging from 1.28 to 11.97. The improvement is significant at the 10\% level in all six contracts and at the 1\% level in five of the six. The increment is largest for ETH and BTC, which are also the contracts with the strongest standalone TI28 performance, and smallest for SOL and XRP.

Taken together, the DM evidence supports a complementary-information interpretation. Boundary-aligned lagged returns provide a strong and robust baseline, consistent with phase-specific persistence. Technical indicators add a separate component of predictive content that is not subsumed by same-phase own-history. Thus, quarter-hour opening return predictability is not purely a lagged-boundary phenomenon; it also reflects information in the pre-boundary market state.

\begin{table}[!htbp]
\centering
\caption{Diebold--Mariano Tests for the Quarter-Hour First-10s Return Forecast}
\label{tab:dm_tests}
\small
\setlength{\tabcolsep}{6pt}
\begin{tabular*}{\textwidth}{@{\extracolsep{\fill}}l rr c r}
\toprule
 & \multicolumn{2}{c}{Panel A: each block vs.\ zero}
 & & Panel B: increment on Lag \\
\cmidrule(lr){2-3}\cmidrule(lr){5-5}
Asset
  & Lag & TI28
  &
  & \makecell{+TI28} \\
\midrule
BTC
  & $13.55^{***}$ & $12.33^{***}$ &
  & $\phantom{-}9.65^{***}$ \\
ETH
  & $15.63^{***}$ & $16.68^{***}$ &
  & $11.97^{***}$ \\
XRP
  & $\phantom{1}7.58^{***}$ & $\phantom{1}6.33^{***}$ &
  & $\phantom{-}2.46^{***}$ \\
SOL
  & $\phantom{1}4.78^{***}$ & $\phantom{1}2.81^{***}$ &
  & $\phantom{-}1.28^{*}\phantom{^{**}}$ \\
DOGE
  & $\phantom{1}7.13^{***}$ & $\phantom{1}5.05^{***}$ &
  & $\phantom{-}4.75^{***}$ \\
ADA
  & $22.95^{***}$ & $10.76^{***}$ &
  & $\phantom{-}4.81^{***}$ \\
\midrule
\#\,assets with $p<0.05$
  & 6/6 & 6/6 &
  & 5/6 \\
\bottomrule
\end{tabular*}
\medskip
\begin{minipage}[t]{\textwidth}
\setstretch{1}\footnotesize
\textit{Notes.} Each cell reports the one-sided DM $t$-statistic (HAC standard errors with lag~$=6$), under the alternative hypothesis that the column model achieves a lower out-of-sample mean squared error than the row benchmark. Panel~A tests each predictor block alone against the zero forecast. Panel~B tests the increment of the TI28 block added on top of Lag. All models are estimated by the same rolling-window walk-forward LASSO as in Table~\ref{tab:no_buffer_specs} (six-month training window, monthly refit), on the QH first-10s sample (2021-07-01 to 2024-10-31; $\approx 117{,}000$ observations per asset). Significance: $^{***}\,p<0.01$, $^{**}\,p<0.05$, $^{*}\,p<0.10$.
\end{minipage}
\end{table}

\section{Information Content of Quarter-Hour Order Flow}\label{sec:Information_Content}

Sections~\ref{sec:ACM} and~\ref{sec:forecasting} showed that periodic algorithmic order flow concentrates at quarter-hour boundaries and that the opening returns are forecastable. This raises a sharper question: is the boundary order flow \emph{informative}? That is, does order imbalance at a quarter-hour opening carry information about subsequent returns, and over what horizon? Distinguishing order flow that contributes to price discovery from flow that moves prices only transiently is a central question in market microstructure. We address it by asking whether order imbalance at salient clock-time boundaries predicts subsequent returns, and whether this predictive content differs across ordinary intervals, one-minute, five-minute, and quarter-hour openings.

Our empirical design follows the predictive-regression approach of \citet{Boehmeretal2022}, adapted to an intraday periodic setting. We estimate horizon-specific regressions that relate current order imbalance to future cumulative returns. By interacting order imbalance with nested clock-time indicators, the specification compares baseline, one-minute, five-minute, and quarter-hour opening order flow within a single regression framework. The objective is not to estimate a structural impulse response, but to characterize how the predictive association between order imbalance and future returns varies across horizons and clock-time frequencies.

We proceed in two steps. Section~\ref{subsec:HSregression} asks whether quarter-hour opening order imbalance predicts future returns, and compares its predictive content with order flow at non-periodic times and at one-minute and five-minute openings. This comparison shows whether the quarter-hour effect is distinct from more general periodic order-flow patterns, rather than simply a stronger manifestation of them. Section~\ref{sec:6source} then decomposes quarter-hour opening order imbalance into components spanned by same-phase lags, observable price-volume state variables, and residual variation, and examines which components account for the medium-horizon predictive association.

\subsection{Horizon-Specific Predictive Regressions (Baseline)}\label{subsec:HSregression}

We estimate each forecast horizon separately. This direct-regression design traces how the predictive association between current imbalance and future cumulative returns changes with the horizon, without imposing the joint-dynamics restrictions of a structural VAR. The nested clock-time interactions let the predictive slope on order imbalance differ across progressively more salient boundary regimes.

For each predictive horizon $\ell \in \{ 30\text{s},1\text{m},5\text{m},15\text{m},30\text{m},1\text{h},2\text{h},4\text{h},8\text{h},12\text{h},24\text{h}\}$, we estimate the following specification\footnote{The four quarter-hour openings (minutes 0, 15, 30, and 45) enter a single fifteen-minute indicator. Because the round-number diagnostic of Section~\ref{sec:rounding_bias} peaks at the top of the hour, it is natural to ask whether the effect is confined to the minute-0 openings; the Appendix separates them with an additional top-of-hour interaction and finds the fifteen-minute effect essentially unchanged, while the incremental top-of-hour term carries no reliable predictive content especially at the medium horizons of interest, so we retain the more parsimonious specification here.} at the 10-second frequency:
\begin{equation}
\begin{aligned}
r_{t,t+\ell} &= \alpha_\ell 
+ \beta_\ell \operatorname{OI}_{t} 
+ \beta_\ell^{1\text{min}} \operatorname{OI}_{t}\,\mathbb{I}_{\{b_t=1\}} 
+ \beta_{\ell}^{5\text{min}} \operatorname{OI}_{t}\,\mathbb{I}_{\{\operatorname{mod}(m_t,5)=0 \,\wedge\, b_t=1\}} \\
&\quad + \beta_{\ell}^{15\text{min}} \operatorname{OI}_{t}\,\mathbb{I}_{\{\operatorname{mod}(m_t,15)=0 \,\wedge\, b_t=1\}}  + \gamma_\ell^\top \mathrm{W}_{t} + \epsilon_{t,\ell},
\end{aligned}
\label{eq:baseline_reg}
\end{equation}
where $r_{t,t+\ell}=\log P_{t+\ell}-\log P_{t}$ is the log return over horizon $\ell$, measured from the last transaction price observed within 10-second interval $t$, and $\operatorname{OI}_t$ is order imbalance over interval $t$. The minute-of-the-hour $m_t$ and within-minute subperiod $b_t$ are read from the calendar tuple $(d_t,h_t,m_t;b_t)$ associated with interval $t$.

The microstructure control vector $W_t$ contains two short-term reversal dummies, $\eta_t^+$ and $\eta_t^-$. These are constructed from the aggressor side of the last trade in each 10-second interval: one captures buyer-to-seller switches, and the other captures seller-to-buyer switches. These controls mitigate bid-ask-bounce effects so that the estimated predictive slopes are not driven by mechanical microstructure noise.

Because the periodic dummy indicators are hierarchically nested, the predictive slope for each clock-time regime is the corresponding cumulative coefficient, which we call the \textit{Cumulative Forecasting Effect} (CFE):
$$
\begin{aligned}
\operatorname{CFE}_{\text{base}}(\ell) &= \beta_\ell \\
\operatorname{CFE}_{1\text{min}}(\ell) &= \beta_\ell + \beta^{1\text{min}}_\ell \\
\operatorname{CFE}_{5\text{min}}(\ell) &= \beta_\ell + \beta^{1\text{min}}_\ell + \beta^{5\text{min}}_\ell \\
\operatorname{CFE}_{15\text{min}}(\ell) &= \beta_\ell + \beta^{1\text{min}}_\ell + \beta^{5\text{min}}_\ell + \beta^{15\text{min}}_\ell \\
\end{aligned}
$$
The CFE should be interpreted as a horizon-specific predictive association between current order imbalance and future returns for the corresponding clock-time regime, not as a structural impulse response.

We estimate these predictive regressions by ordinary least squares, separately for each asset and horizon, using the full sample. Confidence intervals are constructed using heteroskedasticity- and autocorrelation-consistent standard errors with a Bartlett kernel. The bandwidth increases with the forecast horizon $\ell$ to account for the overlap in forward returns.

\subsubsection{Predictive Content Across Frequencies}

\begin{figure}[!htbp]
    \centering
    \includegraphics[width=1\textwidth]{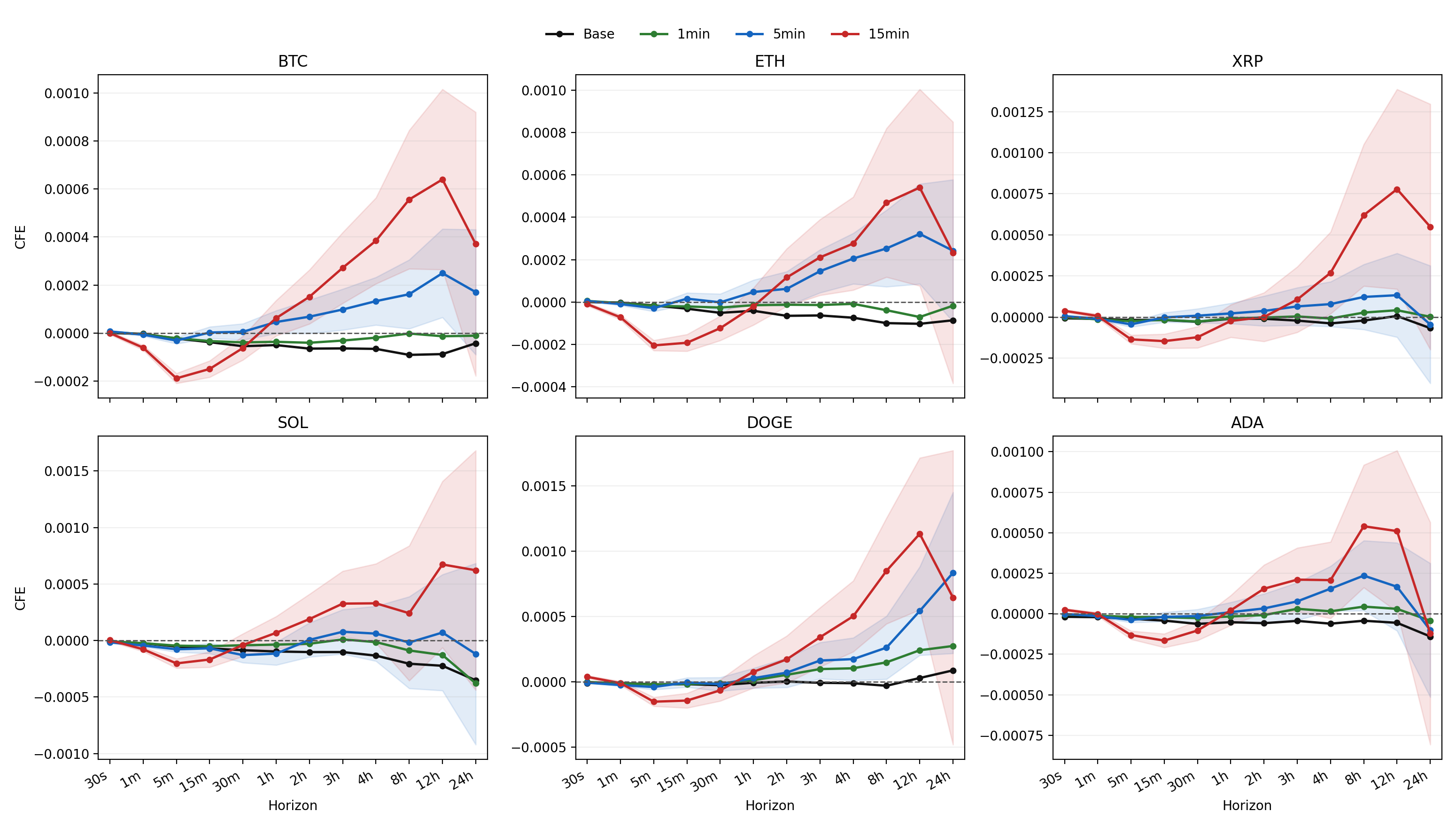}
    \caption{Cumulative Forecasting Effect (CFE) of order imbalance on future returns for six cryptocurrency markets. The plots show horizon-specific cumulative predictive slopes from direct forecasting regressions. Compared with baseline, one-minute, and five-minute openings, order imbalance measured during 15-minute boundary windows exhibits stronger medium-horizon predictive content.}
    \label{fig:irf_periodic}
\end{figure}

Figure \ref{fig:irf_periodic} presents the estimated CFEs for the six assets. Baseline order imbalance and order imbalance at one-minute openings exhibit little predictive association with subsequent returns. By contrast, $\operatorname{CFE}_{15\text{min}}(\ell)$ shows a clear horizon-dependent pattern in most markets: the coefficient is small or negative at short horizons but becomes positive at medium horizons in every market. Between four and twelve hours the estimates are significant at the $95\%$ confidence level for four of the six contracts at every horizon; SOL (at four and twelve hours) and ADA (at four hours) are significant at the $90\%$ level, and SOL at the eight-hour horizon is the only insignificant cell.

The horizon profile of the quarter-hour CFE is nonmonotonic. The CFE is negative over the first half hour, consistent with a short-run reversal following boundary imbalance. It subsequently turns positive and continues to rise, typically peaking between eight and twelve hours. Because $\operatorname{OI}_t$ is an observed imbalance rather than an identified order-flow shock, this pattern should not be interpreted as the causal price impact of trades executed during the opening 10 seconds. Rather, it shows that the direction of boundary order flow predicts cumulative returns over the subsequent several hours.

The main findings are robust to several alternative specifications. First, dropping the nested one-minute and five-minute interactions and keeping only the 15-minute interaction leaves the 15-minute coefficients positive and significant over medium horizons (Appendix Table~\ref{tab:cfe_15only}). Second, replacing the symmetric quarter-hour interaction with separate positive- and negative-imbalance terms yields coefficients with consistent signs (Appendix Figure~\ref{fig:hs_signsplit}). Third, the results are qualitatively unchanged when order imbalance is replaced with its sign (Appendix Table~\ref{tab:cfe_sign}). Fourth, inference is similar under \citet{Hodrick1992} standard errors, which are robust to the overlap in the forward returns (Appendix Table~\ref{tab:cfe_hodrick}). Fifth, the pattern is essentially unchanged when we exclude the three quarter-hours each day that coincide with funding settlement (00:00, 08:00, and 16:00 UTC), so the effect is not driven by the funding-settlement events (Appendix Table~\ref{tab:cfe_exsettle}). Finally, augmenting the specification with a top-of-hour interaction that absorbs the minute-0 openings leaves the quarter-hour effect, then identified from the minute-15, -30, and -45 openings alone, essentially unchanged, so the pooled quarter-hour coefficient is not driven by the top of the hour (Appendix Figure~\ref{fig:cfe_top_of_hour}).

Why does return-predictive content surface at the quarter-hour frequency but not at the one-minute and five-minute frequencies, despite comparable algorithmic signatures? A familiar noise-signal trade-off in the high-frequency literature (\citealp{BandiRussell:2008,HansenLunde:JBES2006}) suggests that signal extraction at very fine frequencies is dominated by microstructure noise. Recent intraday return forecasting models that adopt the 15-minute frequency explicitly cite higher-frequency microstructure violations as a reason to avoid finer sampling (\citealp{aletietal2024}). The quarter-hour boundary is also a focal coordination point in the standardized information architecture of electronic markets. The combination of noise attenuation favoring coarser frequencies and coordination conventions concentrating on the quarter-hour grid is consistent with the dissociation we document. Because the return-predictive content is concentrated at the quarter-hour frequency, the next section focuses on quarter-hour opening order imbalance and examines which components of that imbalance carry the medium-horizon predictive association.

\subsection{Decomposing Predictive Content of Quarter-Hour Order Flow}
\label{sec:6source}

Section \ref{subsec:HSregression} shows that quarter-hour opening order imbalance predicts returns over medium horizons. To examine which components of this imbalance carry the predictive association, we follow \citet{Boehmeretal2022} and project opening imbalance onto two blocks: its own twelve same-phase lags and the 28 technical indicators from Section~\ref{sec:forecasting}. The fitted values define the lagged-flow and public-signal components, and the remaining variation defines the residual component. Estimated over the full quarter-hour opening sample (roughly 131,700 observations per asset), the two blocks jointly span about 6\% of the variance of opening imbalance (4.6\% lagged flow, 0.6\% public signal), a non-negligible systematic component given the 10-second measurement window. The estimating equations, summary statistics of opening imbalance, and the full first-stage estimates are collected in Appendix Section~\ref{sec:ia_decomp}.

This decomposition is empirical rather than structural. The three components should not be read as mutually exclusive trading motives or information types. Lagged-flow variation may reflect repeated execution, responses to persistent public signals, or informed order splitting, while the residual may contain private information, omitted public signals, liquidity shocks, and noise.

In the second stage, we regress the forward cumulative return at four-, eight-, and twelve-hour horizons on the three estimated components, including the short-term reversal controls $\mathrm{W}_t$ of \eqref{eq:baseline_reg}. Because the components are generated regressors \citep{pagan1984econometric}, inference is based on a moving-block bootstrap that re-estimates the entire two-stage procedure on each resample, with four-day calendar blocks drawn jointly across the six contracts ($B=1000$) so that both serial and cross-asset dependence are preserved. Following \citet{Boehmeretal2022}, we summarize each component's economic magnitude by its interquartile effect, the change in the predicted horizon-$\ell$ return when the component moves from its 25th to its 75th percentile, and normalize these effects into contribution shares. In these terms, the lagged-flow component contributes a stable five to six basis points across horizons, whereas the public component grows from less than one basis point at four hours to 9.8 and 16.9 basis points at eight and twelve hours (Appendix Table~\ref{tab:stage2}).

Table~\ref{tab:decomp} summarizes the result. The salient feature is a \emph{rotation} in the dominant component as the horizon lengthens: at four hours the predictive content is carried mainly by the lagged-flow component (cross-asset mean share $58$ percent), whereas by twelve hours it is carried mainly by the public-signal component ($65$ percent). The rotation is not an artifact of one or two contracts driving the cross-asset average: between four and twelve hours, the public-signal share \emph{rises} and the lagged-flow share \emph{falls} in every one of the six contracts (Appendix Table~\ref{tab:decomp_full} reports the per-contract shares). Treating the cross-asset mean change in the public-signal share between the four- and twelve-hour horizons as a single statistic, the point estimate is $\widehat{\Delta}=45$ percentage points, with a $95$ percent percentile-bootstrap interval of $[12,52]$ that excludes zero; the rotation is positive in $996$ of the $1{,}000$ resamples (Appendix Table~\ref{tab:rotation_full}).

\begin{table}[!htbp]
\centering
\caption{Sources of quarter-hour order-imbalance predictability}
\label{tab:decomp}
\begin{threeparttable}
\small
\begin{tabular*}{\textwidth}{@{\extracolsep{\fill}}lccc}
\toprule
 & $\qquad\qquad$4h$\qquad\qquad$ & $\qquad\qquad$8h$\qquad\qquad$ & 12h \\
\midrule
\multicolumn{4}{l}{\textit{Panel A. Cross-asset mean contribution shares (\%)}} \\
 Lagged flow $\qquad\qquad\qquad$ & \textbf{58} [34,\,65] & 29 [17,\,47] & 22 [13,\,39] \\
 Public signal & 20 [12,\,49] & \textbf{55} [29,\,72] & \textbf{65} [41,\,77] \\
 Residual & 22 [12,\,30] & 16 [9,\,27] & 13 [8,\,24] \\
\addlinespace[0.6em]
\midrule
\multicolumn{4}{l}{\textit{Panel B. Rotation statistic: change in the public-signal share, 4h $\rightarrow$ 12h}} \\
\multicolumn{4}{l}{$\widehat{\Delta}=+45$~pp; \quad $95\%$ percentile interval $[12,\,52]$; \quad positive in $996/1{,}000$ resamples and in all six contracts} \\
\bottomrule
\end{tabular*}
\begin{tablenotes}[flushleft]
\setstretch{1}\footnotesize
\item\textit{Notes:} Panel~A reports cross-asset mean per-interquartile-range contribution shares, in percent, from the second-stage regression: for component $c$, $M^c_\ell=e^c_\ell\,\mathrm{IQR}(\widehat{\mathrm{OI}}^c_t)$ and shares are $s^c_\ell=|M^c_\ell|/\sum_j|M^j_\ell|$, so they summarize relative economic magnitudes rather than variance shares. Bold marks the largest component at each horizon. Panel~B reports the rotation statistic $\Delta$, the cross-asset mean of the within-contract change in the public-signal share between the four- and twelve-hour horizons. All intervals are from the joint moving-block bootstrap (common four-day calendar blocks across the six contracts, $B=1000$) that re-estimates both stages; because the four- and twelve-hour shares are correlated within each resample, the interval for $\Delta$ is computed from the bootstrap draws of the difference itself. Per-contract shares and the full rotation detail are reported in Appendix Tables~\ref{tab:decomp_full} and~\ref{tab:rotation_full}.
\end{tablenotes}
\end{threeparttable}
\end{table}

We deliberately do not lean on per-cell point estimates or single-series significance. Because the forward returns overlap and the public-signal regressor is a persistent function of slow-moving technical indicators, single-series predictive $t$-statistics are prone to size distortion, and the point shares are imprecisely estimated. Inference therefore rests on the cross-asset replication of the rotation under the joint moving-block bootstrap, which preserves both serial and cross-asset dependence.

These results indicate that the component associated with quarter-hour predictability changes with the return horizon. At the four-hour horizon, the contribution is concentrated in the lagged-flow component, linking short-horizon predictability primarily to the persistence of periodic order flow. At longer horizons, the public-signal component dominates in every contract. The decomposition, however, does not cleanly separate the underlying information sources. Because public signals can themselves be persistent and correlated with lagged order flow, some of their predictive content may be absorbed into the lagged-flow component. Moreover, public signals not captured by the 28 technical indicators may remain in either the lagged-flow or residual component. The estimated public-signal contribution should therefore be interpreted as a conservative measure of the predictive content associated with public signals.

\section{Conclusion}\label{sec:Conclusion}

We have documented a recurring quarter-hour microstructure pattern in cryptocurrency futures using trade data for six liquid contracts. Four results tie the pattern to periodic algorithmic execution. Trade-size roundness declines hierarchically within the burst windows, consistent with heightened algorithmic participation. The Autocorrelation Map reveals lattice-like dependence in signed order flow and returns, anchored at quarter-hour boundaries and concentrated in the first 10 seconds. Quarter-hour opening returns are forecastable out of sample, with the combined lagged-return and technical-indicator specification attaining an average $R^2$ of 3.4\% and an AUC of 0.60, and with the technical indicators adding significantly to the lagged-return benchmark. Finally, quarter-hour opening order imbalance predicts returns over four to twelve hours, with the dominant contribution rotating from persistent boundary flow to the component spanned by observable price-volume signals.

Two of our tools should be useful beyond this setting. The high-resolution roundness diagnostic adapts an established method for inferring trader type (\citealp{GarveyWu2014,Congetal2023}) to a sub-minute scale, robust to compositional shifts in trade size as prices change, and can localize algorithmic intensity to the brief windows in which it concentrates. The Autocorrelation Map complements spectral approaches to cyclostationary processes (\citealp{Hurdgerr1991,Wuetal2025}) by operating in the time-phase domain, making the location of phase-specific dependence visible and admitting an exact decomposition of the robust ACF into phase-specific components.

More broadly, our results show that trading synchronized around salient clock-time boundaries has consequences beyond localized bursts in activity. At quarter-hour openings, order-flow dependence is concentrated in the first few seconds, and boundary imbalance forecasts cumulative returns over four to twelve hours, while finer-frequency order flow carries little comparable predictive content. The predictive content of order flow—and potentially the adverse-selection exposure of liquidity suppliers—is therefore not uniform over clock time, giving liquidity providers an incentive to adjust quotes and inventory around quarter-hour openings. Because the forecast relies only on information available before the boundary, its construction does not require low-latency infrastructure. Boundary salience therefore shapes not only the concentration of trading activity but also the predictive content of aggregate order flow.

This interpretation has an important limitation. Our data are aggregate trades without trader identifiers, order-level messages, or account labels, so we link the bursts to algorithmic participation through timing and behavioral diagnostics rather than direct attribution. Order-level data would permit sharper tests and a finer decomposition across execution algorithms, market makers, and other trader types. A natural extension compares the same asset, for example a tokenized Treasury, equity, or commodity, across crypto-native and traditional venues to separate venue-specific from asset-specific effects.

\section*{Data Availability}

The analysis uses publicly available aggregate trade data for USDT-margined perpetual futures. Binance distributes its historical aggregate-trade files---each record containing a millisecond timestamp, price, quantity, and the buyer-maker indicator---through its public data portal, and the Bybit records used in the robustness analysis are likewise publicly available. The sample covers six contracts (BTC, ETH, XRP, SOL, DOGE, and ADA) from January~1, 2021 to October~31, 2024. No proprietary or access-restricted data were used. Replication code will be made available in accordance with the journal's code-sharing policy.

\begin{spacing}{1.00}
\footnotesize
\bibliographystyle{apalike}
\bibliography{prh,ChanKim}
\end{spacing}

\clearpage
\setcounter{section}{0}\setcounter{table}{0}\setcounter{figure}{0}\setcounter{equation}{0}
\renewcommand{\thesection}{A.\arabic{section}}
\renewcommand{\thetable}{A.\arabic{table}}
\renewcommand{\thefigure}{A.\arabic{figure}}
\renewcommand{\theequation}{A.\arabic{equation}}
\begin{center}
{\Large\textbf{Appendix}}
\end{center}
This Appendix collects supporting material for the main text: the extended descriptive statistics for the daily trading aggregates (Section~2), the magnitude of the
quarter-hour trading bursts (Section~3), additional robustness checks for the Autocorrelation Map
(Section~4), the full technical-indicator menu and the economic magnitude of the out-of-sample forecast (Section~5), and robustness checks for the horizon-specific regressions (Section~6).

\newpage

\section{Additional Descriptive Statistics}

This section accompanies Section~2.2 of the main text. Table~\ref{tab:daily_range}
reports the range of the daily trading aggregates summarized in Table 1 of the main
text.

\begin{table}[htbp]
\centering
\caption{Daily Trading Aggregates: Minimum, Median, and Maximum}
\label{tab:daily_range}
\footnotesize
\begin{tabularx}{\textwidth}{ll *{7}{>{\raggedleft\arraybackslash}X}}
\toprule
Contract & Statistic & \shortstack{Trades\\(M)} & \shortstack{DV\\(\$M)} & \shortstack{Trade\\Size (\$)} & \shortstack{Zero 10s\\Bins (\%)} & \shortstack{$|r|_{10s}$\\(bps)} & \shortstack{$|r|_{1m}$\\(bps)} & \shortstack{Buyer\\Share} \\
\midrule
BTC  & Min    & 0.19 & 1{,}320  & 4{,}997  & 0.00 & 0.08 & 0.37   & 0.460 \\
     & Median & 1.38 & 13{,}455 & 9{,}768  & 0.00 & 1.66 & 4.83   & 0.499 \\
     & Max    & 8.27 & 68{,}785 & 16{,}625 & 4.16 & 11.33 & 31.19 & 0.535 \\
\midrule
ETH  & Min    & 0.15 & 672      & 2{,}569  & 0.00 & 0.11 & 0.52   & 0.458 \\
     & Median & 1.05 & 6{,}365  & 6{,}022  & 0.00 & 2.04 & 5.87   & 0.499 \\
     & Max    & 9.63 & 63{,}463 & 10{,}372 & 4.16 & 16.48 & 47.83 & 0.565 \\
\midrule
XRP  & Min    & 0.06 & 99       & 885      & 0.00 & 0.84 & 1.72   & 0.306 \\
     & Median & 0.19 & 586      & 3{,}033  & 0.01 & 2.34 & 6.56   & 0.504 \\
     & Max    & 3.68 & 11{,}284 & 6{,}164  & 4.35 & 22.13 & 60.20 & 0.714 \\
\midrule
SOL  & Min    & 0.05 & 29       & 263      & 0.00 & 0.80 & 2.36   & 0.407 \\
     & Median & 0.35 & 963      & 2{,}502  & 0.01 & 3.36 & 9.43   & 0.500 \\
     & Max    & 5.57 & 17{,}287 & 10{,}753 & 11.08 & 30.51 & 85.12 & 0.637 \\
\midrule
DOGE & Min    & 0.04 & 11       & 205      & 0.00 & 0.77 & 1.66   & 0.340 \\
     & Median & 0.22 & 449      & 1{,}979  & 0.05 & 2.76 & 7.75   & 0.499 \\
     & Max    & 18.92 & 20{,}378 & 4{,}245 & 14.81 & 58.43 & 157.30 & 0.701 \\
\midrule
ADA  & Min    & 0.04 & 40       & 551      & 0.00 & 1.11 & 2.25   & 0.213 \\
     & Median & 0.15 & 344      & 2{,}010  & 0.09 & 2.59 & 7.24   & 0.500 \\
     & Max    & 2.72 & 4{,}879  & 3{,}865  & 6.34 & 20.48 & 58.63 & 0.708 \\
\bottomrule
\end{tabularx}
\smallskip

\begin{minipage}{\textwidth}
\setstretch{1}\small
\textit{Notes:} For each contract, the columns of Table~1 of the main text are computed
day by day over the 1{,}400 calendar days of the sample (January 1, 2021 to October 31,
2024), and the table reports the minimum, median, and maximum of the resulting daily
series. Trades is the daily number of trades in millions; DV is daily dollar volume in
millions of dollars; Trade Size is daily dollar volume divided by the daily trade count;
Zero 10s Bins is the percentage of 10-second intervals in the day with no trades;
$|r|_{10s}$ and $|r|_{1m}$ are the daily means of absolute 10-second and 1-minute log
returns in basis points; Buyer Share is the daily fraction of buyer-initiated trades.
\end{minipage}
\end{table}

\newpage
\section{Burst Intensity and Concentration}

This table accompanies Section~3 of the main text and quantifies the magnitude of the quarter-hour trading bursts documented there.

\begin{table}[htbp]
\centering
\caption{Quarter-Hour Burst Intensity and Concentration}
\label{tab:qh_burst}
\footnotesize
\begin{tabularx}{\textwidth}{l *{7}{>{\raggedleft\arraybackslash}X}}
\toprule
 & \multicolumn{4}{c}{\textit{Panel A: Burst Intensity \& Concentration}} & \multicolumn{3}{c}{\textit{Panel B: Economic Magnitude}} \\
\cmidrule(lr){2-5} \cmidrule(lr){6-8}
Contract & \shortstack{10s Trade\\Ratio} & \shortstack{Minute\\Trade Ratio} & \shortstack{QH Peak\\Share (\%)} & \shortstack{Non-QH Peak\\Share (\%)} & \shortstack{$|r|$\\Ratio} & \shortstack{Dollar Volume\\Ratio} & \shortstack{Avg Trade\\Size Ratio} \\
\midrule
BTC & 1.274 & 1.191 & 19.3 & 18.1 & 1.269 & 1.311 & 1.063 \\
ETH & 1.333 & 1.206 & 20.3 & 18.4 & 1.284 & 1.389 & 1.105 \\
XRP & 1.243 & 1.195 & 19.0 & 18.2 & 1.245 & 1.316 & 1.122 \\
SOL & 1.200 & 1.155 & 18.8 & 18.1 & 1.200 & 1.207 & 1.003 \\
DOGE & 1.200 & 1.194 & 18.4 & 18.3 & 1.232 & 1.233 & 1.013 \\
ADA & 1.319 & 1.230 & 20.1 & 18.8 & 1.298 & 1.482 & 1.211 \\
\midrule
Mean & 1.261 & 1.195 & 19.3 & 18.3 & 1.255 & 1.323 & 1.086 \\
\bottomrule
\end{tabularx}
\smallskip

\begin{minipage}{\textwidth}
\setstretch{1}\small
\textit{Notes:} 10s Trade Ratio compares mean trade count in the first 10-second bin ($b{=}1$, seconds 0--9) of quarter-hour minutes (minutes 0, 15, 30, 45) to the corresponding bin of non-quarter-hour minutes. Minute Trade Ratio compares total trades across the full minute. QH Peak Share is the fraction of all quarter-hour-minute trades concentrated in the first 10 seconds; Non-QH Peak Share is the analogous fraction for ordinary minutes. $|r|$ Ratio compares mean absolute 10-second log returns, $|r_{t}| = |\log(P_{t}/P_{t-1})| \times 10{,}000$ (in basis points), where $P_t$ is the last trade price of each 10-second bar. Dollar Volume Ratio compares mean dollar volume per 10-second bar. Avg Trade Size Ratio compares mean trade size (dollar volume divided by trade count). All ratios exceed unity at the 1\% level (Welch $t$-test).
\end{minipage}
\end{table}

\section{Technical-Indicator Menu}\label{sec:ia_timenu}

This section lists the 28 technical indicators (TI28) used in the forecasting design of Section~5 of the main text.

\emph{Momentum} (5 indicators): the 6-hour Wilder RSI computed from rolling 15-minute closing prices; the 6-hour Stochastic oscillator \(\%K\), together with its 1.5-hour smoothed \(\%D\); the 6-hour Stochastic RSI with double 1.5-hour smoothing; and the 6-hour Commodity Channel Index (CCI) computed from the typical price.

\emph{Trend} (9 indicators): seven price SMA-relative variables of the form
$$
\frac{\mathrm{close}_t}{\mathrm{SMA}_h(\mathrm{close})_t}-1,
$$
for horizons \(h \in \{1\mathrm{h},\,1.5\mathrm{h},\,3\mathrm{h},\,5\mathrm{h},\,8\mathrm{h},\,12\mathrm{h},\,1\mathrm{day}\}\); the price MACD line,
$$
\mathrm{MACD}^{p}_t
=
\frac{
\mathrm{EMA}_{2\mathrm{h}}(\mathrm{close})_t
-
\mathrm{EMA}_{8\mathrm{h}}(\mathrm{close})_t
}{
\mathrm{close}_t
};
$$
and the signal-line difference,
$$
\mathrm{MACDdiff}^{p}_t
=
\mathrm{MACD}^{p}_t
-
\mathrm{EMA}_{1.5\mathrm{h}}(\mathrm{MACD}^{p})_t.
$$

\emph{Volume} (10 indicators): seven volume SMA-relative variables of the form
$$
\frac{\mathrm{vol}_t}{\mathrm{SMA}_h(\mathrm{vol})_t}-1,
$$
for horizons \(h \in \{1\mathrm{h},\,1.5\mathrm{h},\,3\mathrm{h},\,4\mathrm{h},\,6\mathrm{h},\,8\mathrm{h},\,12\mathrm{h}\}\); a volume MACD line and its signal-line difference using the same 2-hour, 8-hour, and 1.5-hour EMA spans as the price MACD; and a normalized Chaikin money-flow measure that contrasts the 1-hour and 8-hour EMAs of the cumulative accumulation--distribution line, scaled by the 8-hour rolling-mean volume.

\emph{Volatility} (4 indicators): three proportional distances from the close to the lower, middle, and upper bands of a 6-hour, two-standard-deviation Bollinger envelope, together with the band width relative to the mid-band.

\section{Forecasting Model and Evaluation Details}\label{sec:ia_forecast_details}

This section accompanies Section~5.2 of the main text and collects the formal definitions of the forecasting model and the out-of-sample evaluation metrics.

\paragraph{LASSO objective and standardization.}
The prediction model is linear in the predictors and estimated by LASSO. For a regularization parameter $\lambda \geq 0$, the objective function is
$$
\hat{\beta} = \arg\min_{\beta} \left\{ \frac{1}{2n} \sum_{i=1}^{n} (\tilde{Y}_i - \tilde{X}_i^\top \beta)^2 + \lambda \sum_{j=1}^{p} |\beta_j| \right\},
$$
where $\tilde{Y}_i$ and $\tilde{X}_i$ denote the centered and standardized response and predictors. The $\ell_1$ penalty shrinks coefficients toward, and often exactly to, zero, providing automatic variable selection and a sparse forecasting specification. At each monthly refit, predictor means and standard deviations, together with the response mean and standard deviation $s_y$, are computed from the trailing six-month training window and then held fixed throughout the test month. Forecasts are formed in standardized units and converted back to return units using the training-window response moments. Standardizing the response does not affect the out-of-sample forecasts: scaling $Y$ by $s_y$ is exactly undone by the inverse transform, and is equivalent to leaving $Y$ unscaled under the correspondingly rescaled penalty $\lambda s_y$; the standardization is a numerical convenience, not a modeling choice. The realized return in the test month is used only to evaluate forecast accuracy, not to construct predictors, standardization constants, or forecasts.

\paragraph{Out-of-sample $R^2$.}
For the return regression, let \(Y_i^{\mathrm{ret}}\) denote the realized return and \(\widehat{Y}_i^{\mathrm{ret}}\) its forecast. The out-of-sample \(R^2\) compares the model's squared forecast error with that of the zero forecast,
$$
R^2_{\mathrm{OOS}}
=
1-
\frac{
\sum_i \left(Y_i^{\mathrm{ret}}-\widehat{Y}_i^{\mathrm{ret}}\right)^2
}{
\sum_i \left(Y_i^{\mathrm{ret}}\right)^2
}.
$$
It is bounded above by one, with \(R^2_{\mathrm{OOS}}=1\) for perfect prediction and \(R^2_{\mathrm{OOS}}>0\) whenever the model improves on the zero forecast.

\paragraph{AUC.}
For the direction classification, the \(L_1\)-penalized logistic model produces a predicted probability
$$
\widehat{p}_i
=
\Pr\!\left(Y_i^{\mathrm{dir}}=1 \mid X_i\right).
$$
AUC is computed from these predicted probabilities and measures the model's ability to rank upward-move observations above non-upward observations. Equivalently, it can be interpreted as
$$
\mathrm{AUC}
=
\Pr\!\left(
\widehat{p}_i > \widehat{p}_j
\mid
Y_i^{\mathrm{dir}}=1,\;
Y_j^{\mathrm{dir}}=0
\right),
$$
up to the usual treatment of ties. A value of 0.5 corresponds to random ranking, while values above 0.5 indicate positive directional ranking ability.

\paragraph{Accuracy.}
Classification accuracy applies the standard 0.5 probability threshold. The predicted class is
$$
\widehat{Y}_i^{\mathrm{dir}}
=
\mathbf{1}\{\widehat{p}_i \geq 0.5\},
$$
and accuracy is defined as
$$
\mathrm{Accuracy}
=
\frac{1}{n}
\sum_i
\mathbf{1}\left\{
\widehat{Y}_i^{\mathrm{dir}} = Y_i^{\mathrm{dir}}
\right\}.
$$
Accuracy is useful as a simple implementation metric, but it depends on the chosen threshold and on the unconditional class balance; AUC is the corresponding threshold-free measure.

\section{Economic Magnitude of the Out-of-Sample Forecast}

This section accompanies Section~5.3 of the main text and quantifies the economic size
of the out-of-sample quarter-hour forecast relative to trading fees.

\begin{table}[htbp]
\centering
\caption{Economic Magnitude of the Out-of-Sample Quarter-Hour Forecast}
\label{tab:forecast_magnitude}
\footnotesize
\begin{tabularx}{\textwidth}{l *{6}{>{\raggedleft\arraybackslash}X}}
\toprule
Contract & \shortstack{Gross per\\Trade (bps)} & \shortstack{Hit\\Rate (\%)} & \shortstack{MZ\\Slope} & \shortstack{Mean\\$|\widehat{Y}|$ (bps)} & \shortstack{SD\\$(\widehat{Y})$ (bps)} & N \\
\midrule
BTC  & 0.375 & 57.1 & 0.859 & 0.440 & 0.675 & 116{,}951 \\
ETH  & 0.532 & 58.1 & 0.908 & 0.587 & 0.889 & 116{,}951 \\
XRP  & 0.422 & 55.4 & 0.851 & 0.427 & 0.703 & 117{,}016 \\
SOL  & 0.543 & 55.6 & 0.774 & 0.615 & 0.941 & 117{,}016 \\
DOGE & 0.415 & 54.9 & 0.820 & 0.497 & 0.800 & 116{,}104 \\
ADA  & 0.775 & 58.2 & 0.940 & 0.785 & 1.132 & 116{,}655 \\
\midrule
Mean & 0.510 & 56.6 & 0.859 & 0.559 & 0.857 &  \\
\bottomrule
\end{tabularx}
\smallskip

\begin{minipage}{\textwidth}
\setstretch{1}\small
\textit{Notes:} The table reports the economic magnitude of the out-of-sample Lag+TI28
forecast $\widehat{Y}$ of the quarter-hour first-10-second return. These are the same
forecasts evaluated in Table~3 of the main text (rolling-window walk-forward LASSO,
monthly refit, out-of-sample window July 1, 2021 to October 31, 2024). Gross per Trade,
$\mathbb{E}[\operatorname{sign}(\widehat{Y}_t)\,Y_t]$, is the realized average pre-fee
return in basis points of trading the direction of the forecast at every quarter-hour
boundary. It measures the forecast's realized predictive content in return units, not
the dispersion of the forecast. Hit Rate is the directional accuracy
$\Pr[\operatorname{sign}(\widehat{Y}_t)=\operatorname{sign}(Y_t)]$. MZ Slope is the
Mincer--Zarnowitz calibration coefficient $b$ from the out-of-sample regression
$Y_t=a+b\,\widehat{Y}_t+\varepsilon_t$. A value of $b=1$ means the forecast is
perfectly calibrated, and $b<1$ means a forecast of a given size is followed on average
by a realized move of $b$ times that size. Since the slopes are near one, Gross per
Trade is close to the calibration-adjusted mean absolute forecast,
$b\times\text{Mean}\,|\widehat{Y}|$. Mean $|\widehat{Y}|$ and $\mathrm{SD}(\widehat{Y})$
give the size and dispersion of the forecast itself. For scale, the standard-tier fee
on Binance USDT-margined perpetual futures over the sample is $5.0$~bps for takers and
$2.0$~bps for makers, and a round trip pays two fees. The average gross per trade of
about $0.5$~bps is roughly one-tenth of a single taker fee. N is the number of
out-of-sample quarter-hour boundary observations.
\end{minipage}
\end{table}

\newpage

\section{Additional Robustness Checks}\label{sec:appendix_robustness}

This appendix collects robustness checks for the Autocorrelation Map (ACM) evidence in the main text: the cross-asset generalization of the main-text BTC maps, a cross-exchange replication on Bybit, a placebo phase-alignment check, and the within-minute block profile.

\begin{figure}[H]
    \centering
    \includegraphics[width=1\textwidth]{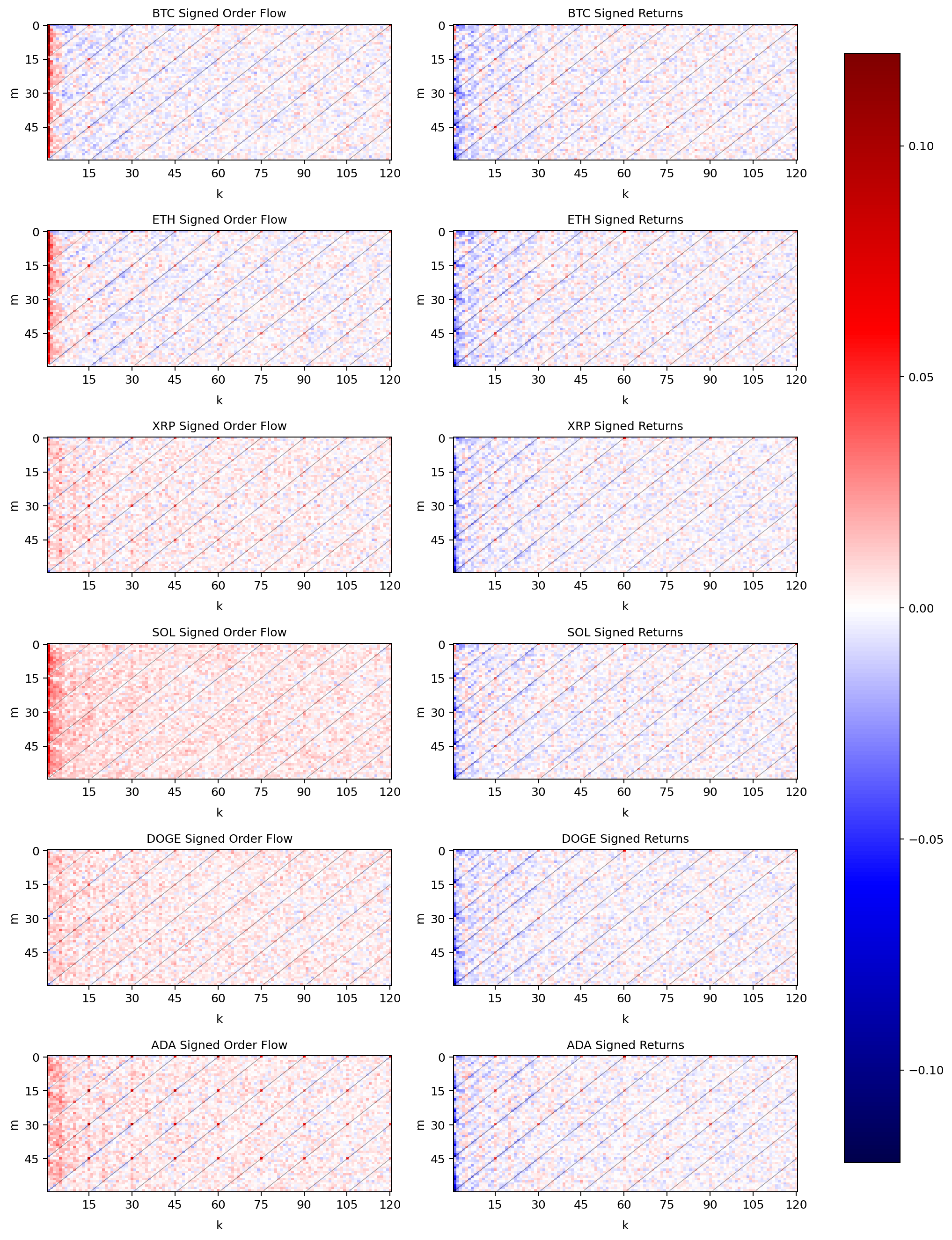}
    \caption{One-minute Autocorrelation Map (ACM) for the six contracts (BTC, ETH, XRP, SOL, DOGE, ADA).}
    \label{fig:Cross_asset_ACM_1m}
\end{figure}

\begin{figure}[H]
    \centering
    \includegraphics[width=1\textwidth]{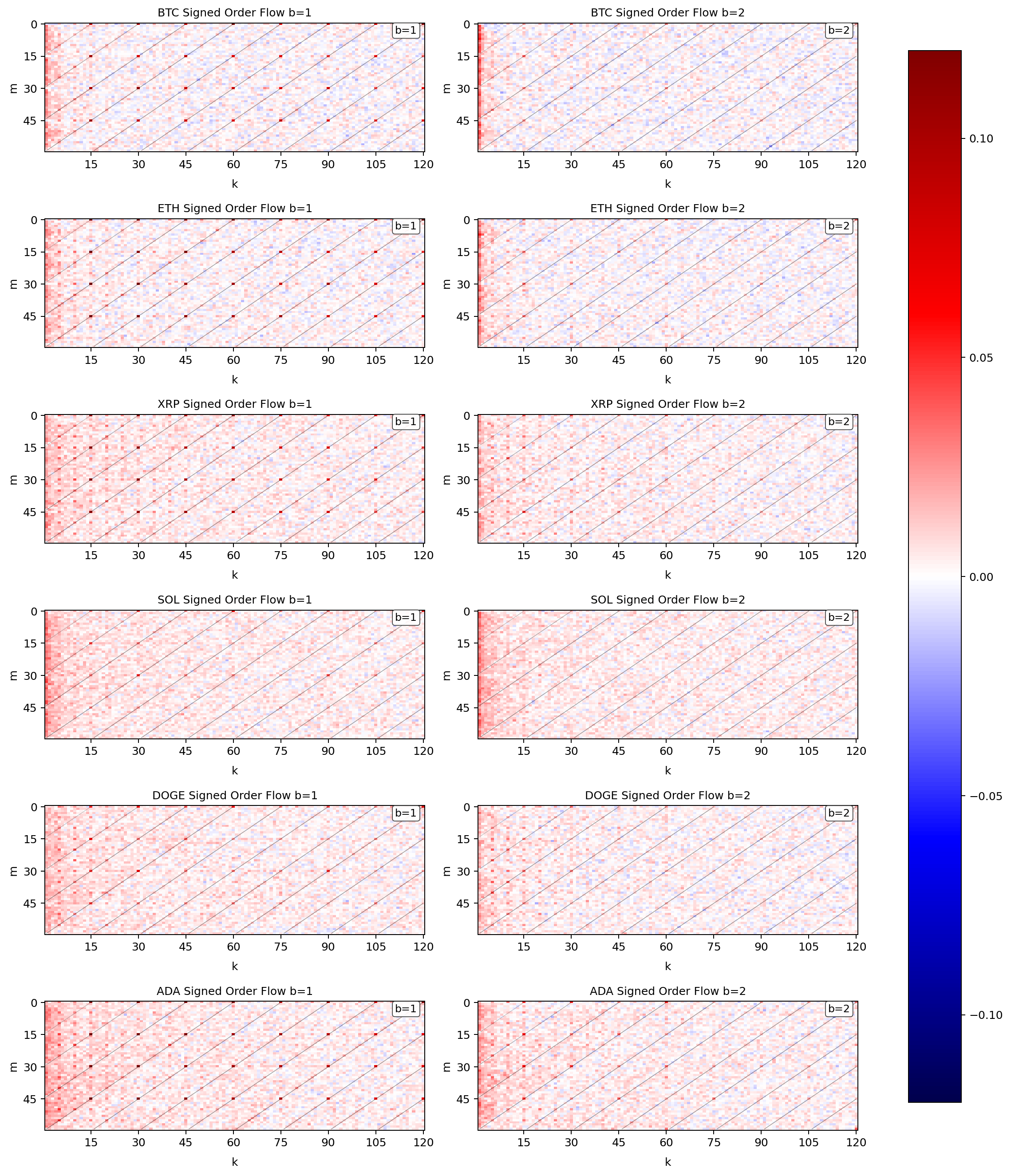}
    \caption{Ten-second signed order-flow ACM for the six contracts (BTC, ETH, XRP, SOL, DOGE, ADA).}
    \label{fig:Cross_asset_ACM_orderflow_10s}
\end{figure}

\begin{figure}[H]
    \centering
    \includegraphics[width=1\textwidth]{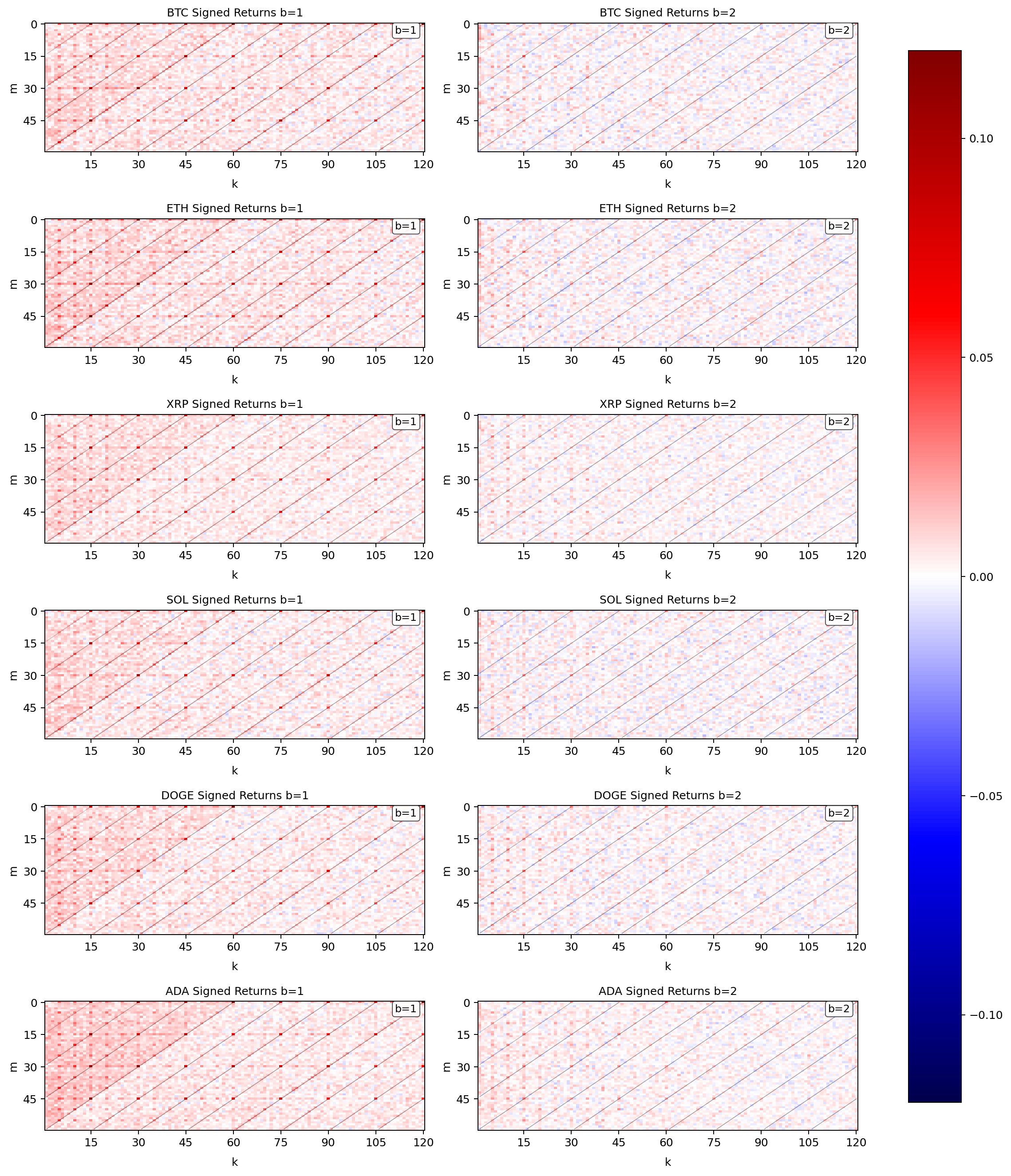}
    \caption{Ten-second return ACM for the six contracts (BTC, ETH, XRP, SOL, DOGE, ADA).}
    \label{fig:Cross_asset_ACM_returns_10s}
\end{figure}

Figures~\ref{fig:Cross_asset_ACM_1m}--\ref{fig:Cross_asset_ACM_returns_10s} reproduce the one-minute and 10-second ACMs for all six contracts. The lattice-like quarter-hour structure documented for BTC in the main text is present across assets, for both signed order flow and returns.

\begin{figure}[H]
    \centering
    \includegraphics[width=1\textwidth]{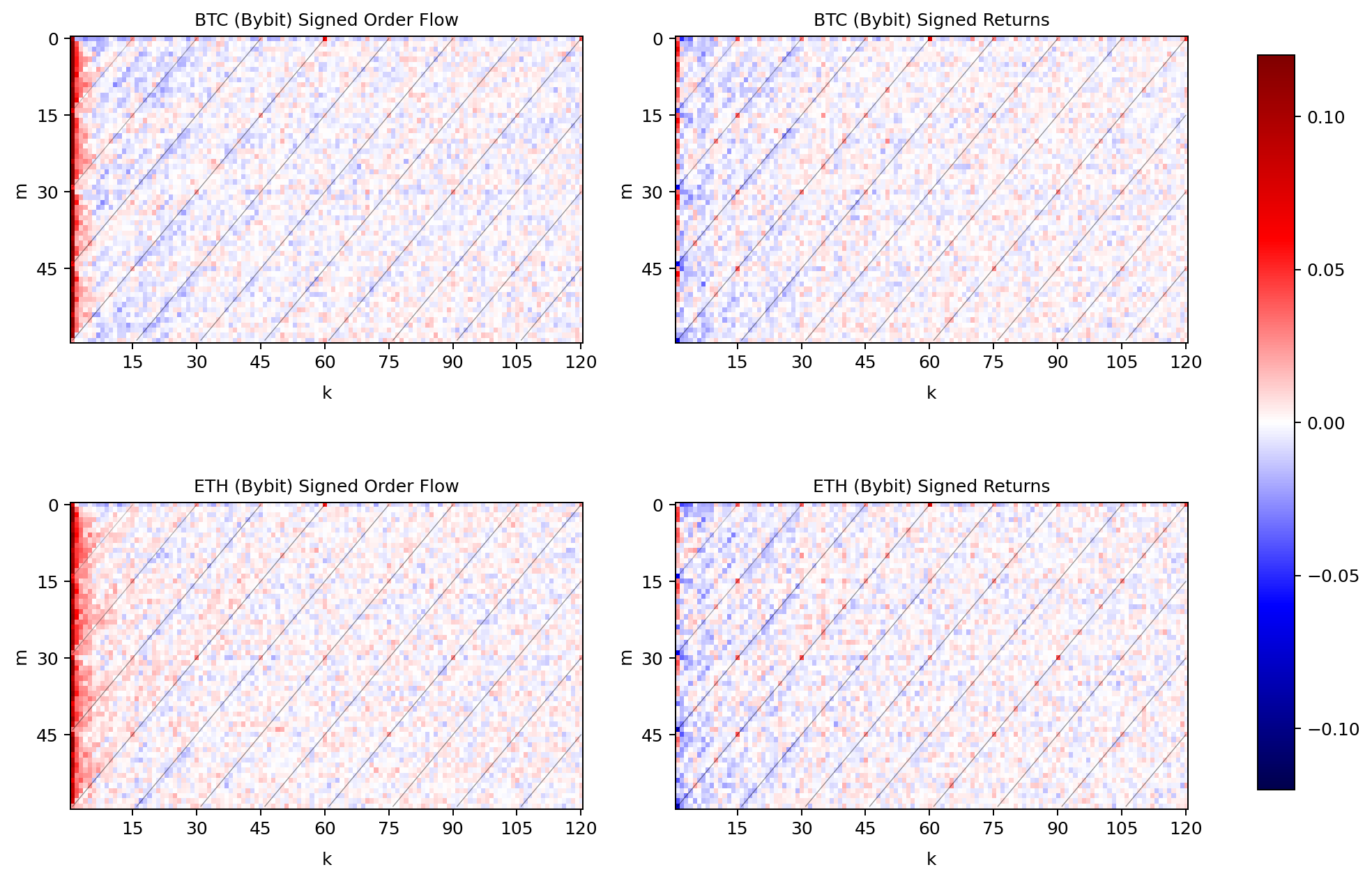}
    \caption{One-minute ACM for BTC and ETH on the Bybit exchange.}
    \label{fig:bybit_ACM_orderflow_1m}
\end{figure}

\begin{figure}[H]
    \centering
    \includegraphics[width=1\textwidth]{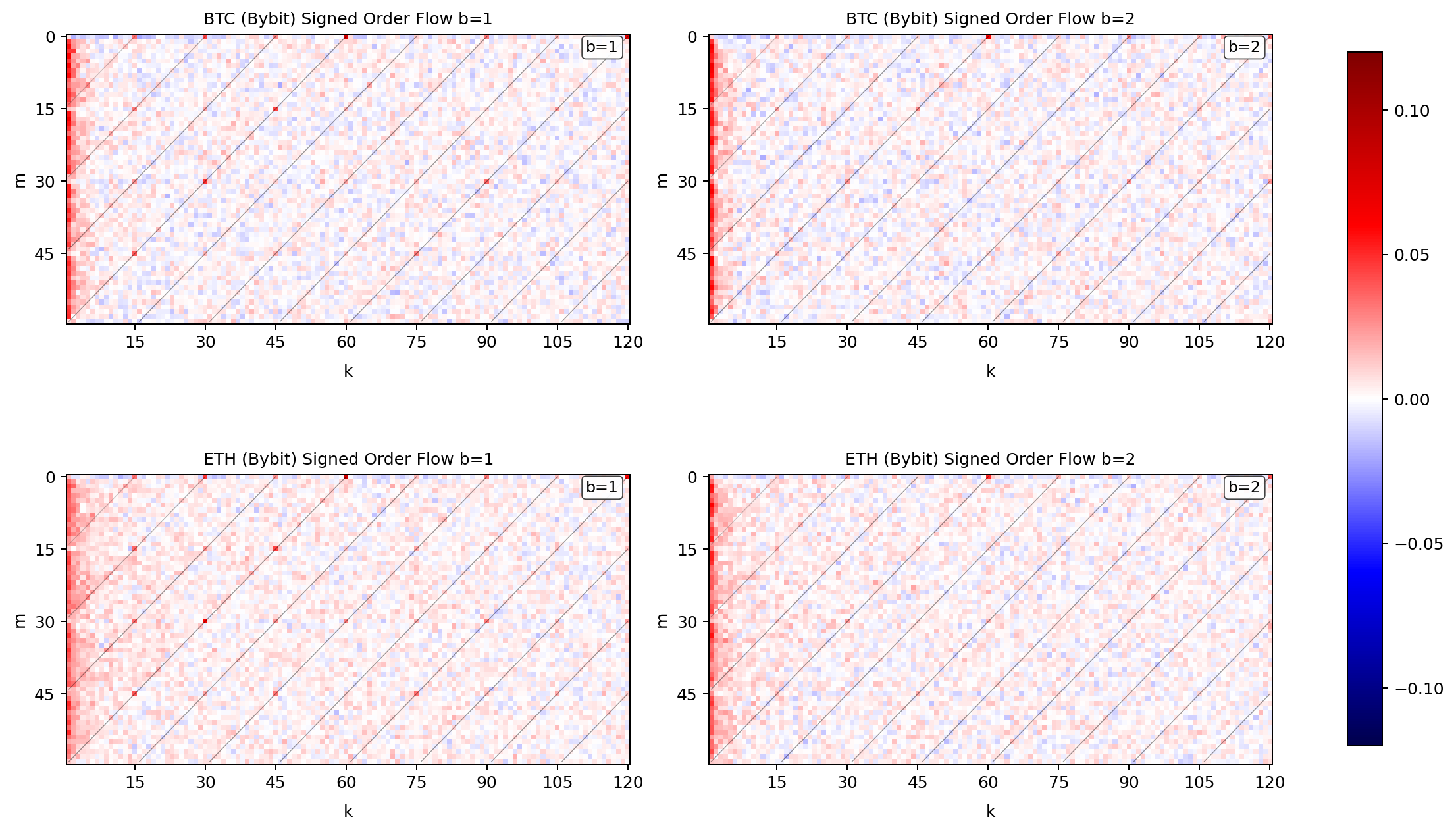}
    \caption{Ten-second signed order-flow ACM for BTC and ETH on the Bybit exchange.}
    \label{fig:bybit_ACM_orderflow_10s}
\end{figure}

\begin{figure}[H]
    \centering
    \includegraphics[width=1\textwidth]{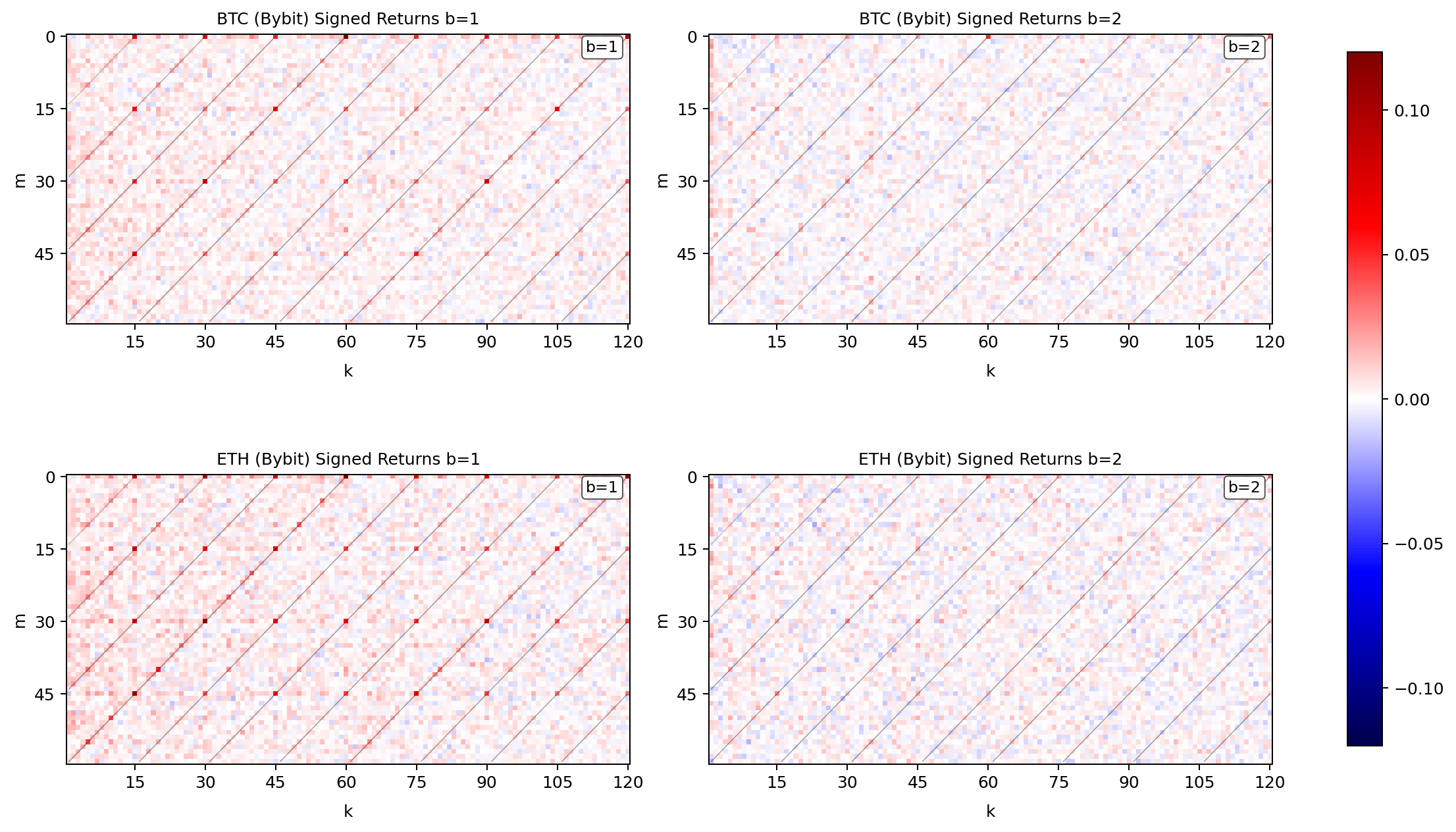}
    \caption{Ten-second return ACM for BTC and ETH on the Bybit exchange.}
    \label{fig:bybit_ACM_returns_10s}
\end{figure}

Figures~\ref{fig:bybit_ACM_orderflow_1m}--\ref{fig:bybit_ACM_returns_10s} replicate the maps for BTC and ETH on Bybit, a second major perpetual-futures venue. The same phase-specific dependence appears, indicating that the pattern is not specific to Binance.

\begin{figure}[H]
    \centering
    \includegraphics[width=1\textwidth]{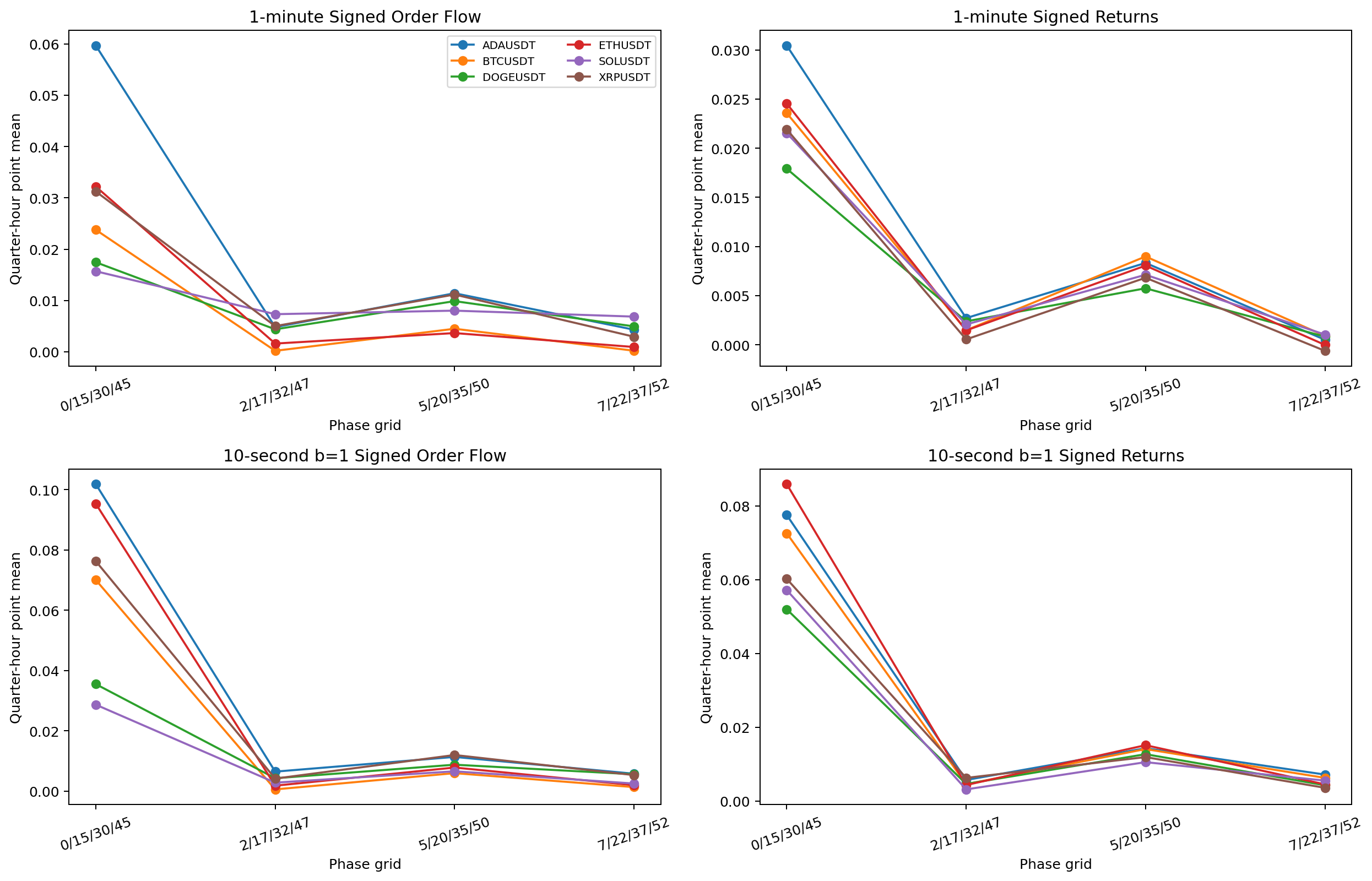}
    \caption{For each asset, the figure reports the mean ACM value at quarter-hour phase points, evaluated at lags $k=15, 30, 45, \ldots, 120$, where both the current phase and the implied future phase lie on the candidate phase grid. The top row uses one-minute ACMs for signed order flow and signed returns; the bottom row uses 10-second $b=1$ ACMs for signed order flow and signed returns. The true quarter-hour grid $\{0, 15, 30, 45\}$ is compared with placebo phase shifts $\{2, 17, 32, 47\}$, $\{5, 20, 35, 50\}$, and $\{7, 22, 37, 52\}$. Across assets, the point mean is generally largest under the true quarter-hour alignment and weakens under placebo shifts, indicating that the concentration of dependence is tied to the true phase rather than to an arbitrary four-point grid.}
    \label{fig:Placebo}
\end{figure}

Figure~\ref{fig:Placebo} reports the placebo phase-alignment check discussed in the main text. Across assets, the concentration of dependence at the true quarter-hour phase points exceeds that under shifted placebo grids.

\begin{figure}[H]
    \centering
    \includegraphics[width=1\textwidth]{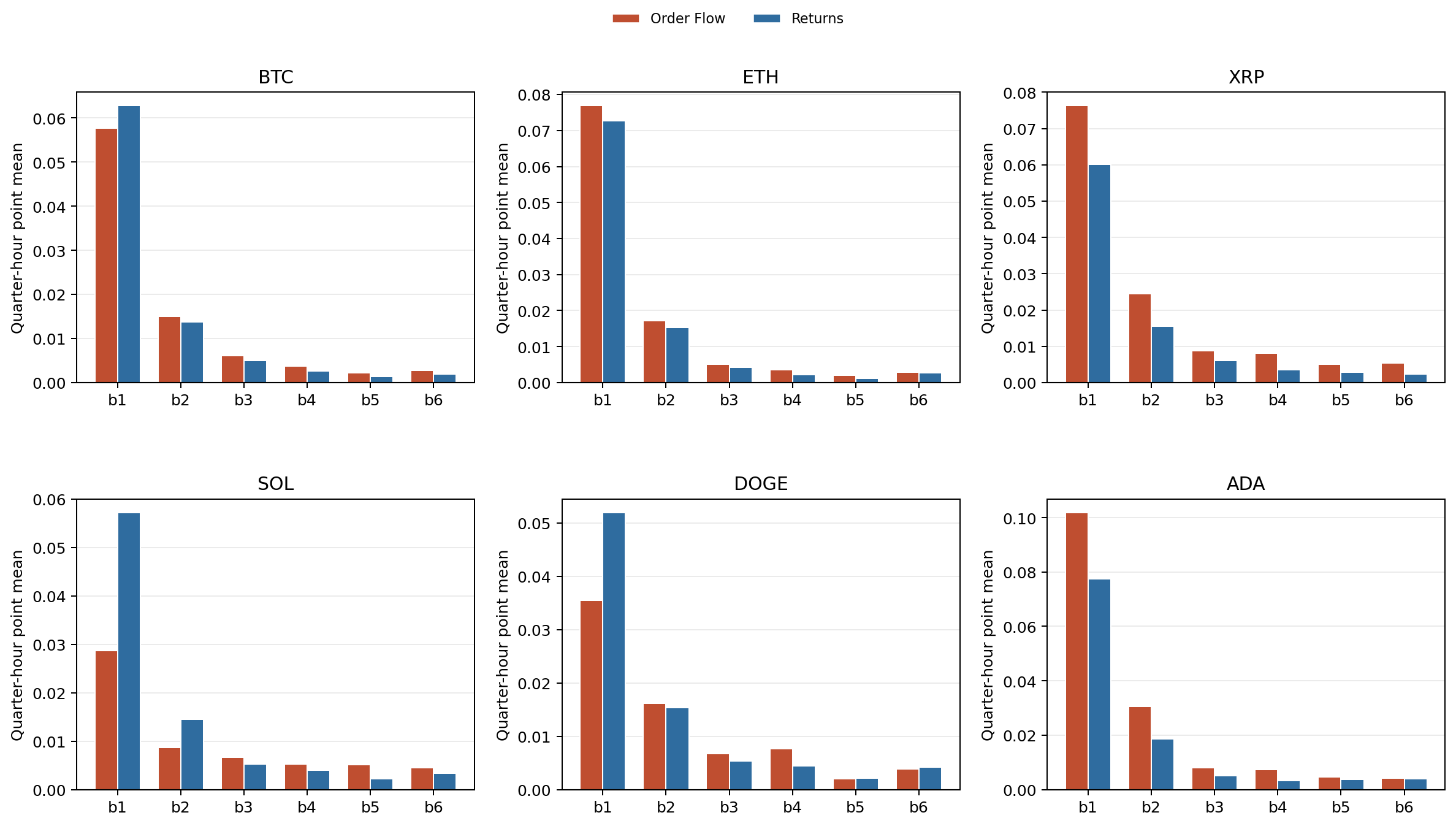}
    \caption{For each asset, the figure reports the mean ACM value at quarter-hour phase points for the 10-second blocks $b=1$ through $b=6$, shown separately for signed order flow and signed returns. For each ACM, the statistic averages the cells at lags $k=15, 30, 45, \ldots, 120$ for which both the current phase and the implied future phase lie on the true quarter-hour grid $\{0, 15, 30, 45\}$. Across assets, the point mean is largest in $b=1$, declines sharply in $b=2$, and remains much smaller in later blocks, especially for signed order flow. This shows that phase-specific dependence is concentrated in the first 10-second interval after the quarter-hour boundary rather than being spread uniformly across the minute.}
    \label{fig:b1b2_summary}
\end{figure}

Finally, Figure~\ref{fig:b1b2_summary} summarizes the within-minute profile of phase-specific dependence across the six 10-second blocks.

\newpage 

\section{Horizon-Specific Regression: Additional Robustness}

\begin{figure}[H]
    \centering
    \includegraphics[width=\textwidth]{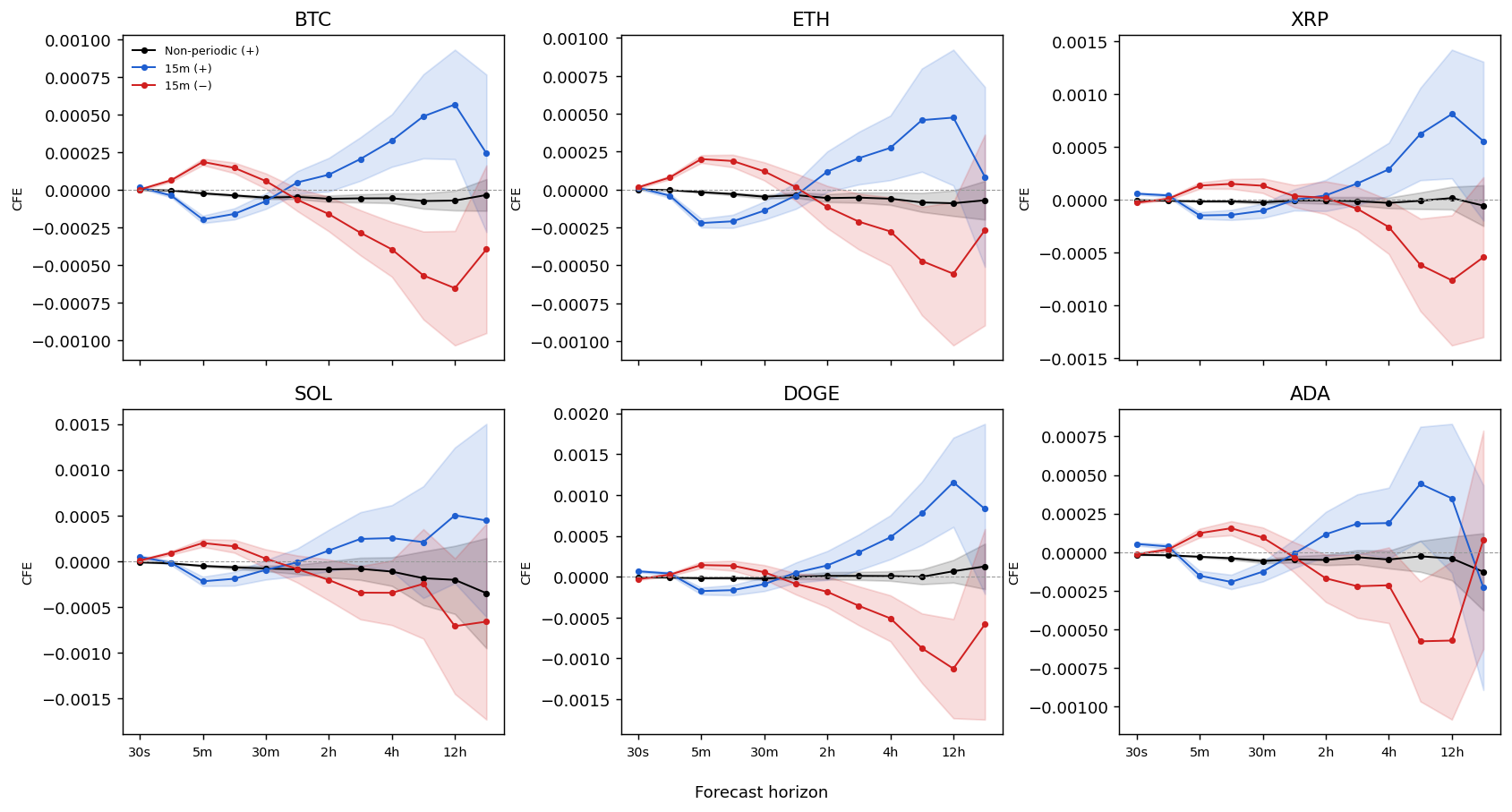}
    \caption{Sign-split horizon-specific regression of quarter-hour order imbalance. For each of the six contracts, the cumulative forecasting effect (CFE) is estimated separately for positive (blue) and negative (red) quarter-hour opening imbalance, with the non-periodic baseline in black; shaded bands are 95\% Newey--West HAC confidence intervals at the full overlap lag. At medium horizons, predictive content is present in many of the contracts for both positive and negative imbalance.}
    \label{fig:hs_signsplit}
\end{figure}

This section reports robustness checks for the horizon-specific regressions in
Section~6.1 of the main text. Figure~\ref{fig:hs_signsplit} estimates the quarter-hour
opening order-imbalance interaction separately for positive and negative imbalance
rather than as a single symmetric term. Tables~\ref{tab:cfe_15only}--\ref{tab:cfe_exsettle}
report the fifteen-minute cumulative forecasting effect, $\operatorname{CFE}_{15\min}(\ell)$,
at the four-, eight-, and twelve-hour horizons under four alternative specifications:
dropping the nested one- and five-minute interactions, replacing order imbalance with
its sign, computing \citet{Hodrick1992} standard errors, and excluding the quarter-hours
that coincide with funding settlement. Every specification includes the short-term
reversal controls $W_t=(\eta^+_t,\eta^-_t)$ of eq.~(2) of the main text; because returns
are computed from traded prices rather than mid-quotes, these controls absorb bid-ask-bounce
reversals. Across all four checks the medium-horizon effect is positive in every
contract-horizon cell. SOL at the eight-hour horizon is insignificant throughout, as in
the main text, and a few additional cells weaken to or slightly below the 10\% level in
the most demanding variants, most visibly at the four-hour horizon of the sign
specification, where the sign of imbalance overlaps with the sign-based reversal
controls.

\begin{table}[htbp]
\centering
\caption{Quarter-Hour CFE Robustness: Fifteen-Minute-Only Specification}
\label{tab:cfe_15only}
\footnotesize
\begin{tabularx}{\textwidth}{l *{3}{>{\raggedleft\arraybackslash}X}}
\toprule
Contract & 4h & 8h & 12h \\
\midrule
BTC  & $3.84^{***}$ [4.21] & $5.55^{***}$ [3.78] & $6.39^{***}$ [3.34] \\
ETH  & $2.76^{**}$ [2.46]  & $4.68^{***}$ [2.62] & $5.40^{**}$ [2.28] \\
XRP  & $2.68^{**}$ [2.11]  & $6.20^{***}$ [2.82] & $7.78^{**}$ [2.51] \\
SOL  & $3.29^{*}$ [1.84]   & $2.41$ [0.79]       & $6.73^{*}$ [1.79] \\
DOGE & $5.02^{***}$ [3.63] & $8.49^{***}$ [4.11] & $11.33^{***}$ [3.82] \\
ADA  & $2.07^{*}$ [1.72]   & $5.39^{***}$ [2.80] & $5.10^{**}$ [2.01] \\
\bottomrule
\end{tabularx}
\smallskip

\begin{minipage}{\textwidth}
\setstretch{1}\small
\textit{Notes:} The table reports $\operatorname{CFE}_{15\min}(\ell)$, the cumulative
predictive slope of quarter-hour opening order imbalance on future returns, at horizons
$\ell\in\{4\text{h},8\text{h},12\text{h}\}$, in basis points per unit of order
imbalance, with $t$-statistics in brackets. The specification drops the nested one- and
five-minute interactions of eq.~(2) of the main text and keeps only the fifteen-minute
interaction, so that $\operatorname{CFE}_{15\min}(\ell)=\beta_\ell+\beta^{15\min}_\ell$;
the reversal controls $W_t=(\eta^+_t,\eta^-_t)$ are retained. Standard errors are
Newey--West HAC with a Bartlett kernel and bandwidth equal to the forecast horizon in
ten-second steps. Sample: January 1, 2021 to October 31, 2024. Significance:
$^{***}\,p<0.01$, $^{**}\,p<0.05$, $^{*}\,p<0.10$.
\end{minipage}
\end{table}

\begin{table}[htbp]
\centering
\caption{Quarter-Hour CFE Robustness: Sign of Order Imbalance}
\label{tab:cfe_sign}
\footnotesize
\begin{tabularx}{\textwidth}{l *{3}{>{\raggedleft\arraybackslash}X}}
\toprule
Contract & 4h & 8h & 12h \\
\midrule
BTC  & $1.64^{***}$ [3.58] & $2.37^{***}$ [3.32] & $2.73^{***}$ [2.95] \\
ETH  & $0.78$ [1.27]       & $1.65^{*}$ [1.77]   & $2.38^{**}$ [1.98] \\
XRP  & $0.75$ [0.99]       & $2.61^{**}$ [2.12]  & $3.08^{*}$ [1.83] \\
SOL  & $1.44$ [1.60]       & $0.78$ [0.55]       & $3.64^{**}$ [2.01] \\
DOGE & $2.08^{**}$ [2.37]  & $4.16^{***}$ [3.01] & $4.80^{**}$ [2.51] \\
ADA  & $0.75$ [1.00]       & $3.07^{***}$ [2.65] & $3.44^{**}$ [2.25] \\
\bottomrule
\end{tabularx}
\smallskip

\begin{minipage}{\textwidth}
\setstretch{1}\small
\textit{Notes:} The table reports
$\operatorname{CFE}_{15\min}(\ell)=\beta_\ell+\beta^{1\min}_\ell+\beta^{5\min}_\ell+\beta^{15\min}_\ell$
from the nested specification of eq.~(2) of the main text with order imbalance
$\mathrm{OI}_t$ replaced by its sign $\operatorname{sign}(\mathrm{OI}_t)$ throughout.
Coefficients are in basis points per unit of $\operatorname{sign}(\mathrm{OI})$, so
magnitudes are not directly comparable to the raw-imbalance tables; $t$-statistics are
in brackets. Standard errors are Newey--West HAC with a Bartlett kernel and bandwidth
equal to the forecast horizon in ten-second steps. Sample: January 1, 2021 to October
31, 2024. Significance: $^{***}\,p<0.01$, $^{**}\,p<0.05$, $^{*}\,p<0.10$.
\end{minipage}
\end{table}

\begin{table}[htbp]
\centering
\caption{Quarter-Hour CFE Robustness: Hodrick (1992) Standard Errors}
\label{tab:cfe_hodrick}
\footnotesize
\begin{tabularx}{\textwidth}{l *{3}{>{\raggedleft\arraybackslash}X}}
\toprule
Contract & 4h & 8h & 12h \\
\midrule
BTC  & $3.85^{***}$ [3.62] & $5.56^{***}$ [3.34] & $6.40^{***}$ [3.02] \\
ETH  & $2.76^{**}$ [2.18]  & $4.69^{**}$ [2.36]  & $5.40^{**}$ [2.14] \\
XRP  & $2.69^{*}$ [1.92]   & $6.20^{***}$ [2.88] & $7.78^{***}$ [2.79] \\
SOL  & $3.29^{*}$ [1.68]   & $2.42$ [0.75]       & $6.73$ [1.58] \\
DOGE & $5.03^{***}$ [3.22] & $8.50^{***}$ [3.54] & $11.34^{***}$ [3.62] \\
ADA  & $2.08$ [1.40]       & $5.40^{**}$ [2.33]  & $5.10^{*}$ [1.77] \\
\bottomrule
\end{tabularx}
\smallskip

\begin{minipage}{\textwidth}
\setstretch{1}\small
\textit{Notes:} The table reports
$\operatorname{CFE}_{15\min}(\ell)=\beta_\ell+\beta^{1\min}_\ell+\beta^{5\min}_\ell+\beta^{15\min}_\ell$
from the nested specification of eq.~(2) of the main text, in basis points per unit of
order imbalance. Point estimates are identical to the baseline; the $t$-statistics in
brackets instead use \citet{Hodrick1992} 1B standard errors, which are robust to the
overlap in the forward returns: the covariance sums the regressors backward over the
ten-second steps spanned by the horizon against the one-step-ahead return innovation.
Sample: January 1, 2021 to October 31, 2024. Significance: $^{***}\,p<0.01$,
$^{**}\,p<0.05$, $^{*}\,p<0.10$.
\end{minipage}
\end{table}
\begin{table}[htbp]
\centering
\caption{Quarter-Hour CFE Robustness: Excluding Funding-Settlement Openings}
\label{tab:cfe_exsettle}
\footnotesize
\begin{tabularx}{\textwidth}{l *{3}{>{\raggedleft\arraybackslash}X}}
\toprule
Contract & 4h & 8h & 12h \\
\midrule
BTC  & $3.97^{***}$ [4.33] & $5.54^{***}$ [3.76] & $6.61^{***}$ [3.45] \\
ETH  & $2.79^{**}$ [2.47]  & $4.35^{**}$ [2.42]  & $5.01^{**}$ [2.12] \\
XRP  & $2.87^{**}$ [2.22]  & $6.18^{***}$ [2.78] & $7.79^{**}$ [2.52] \\
SOL  & $3.35^{*}$ [1.84]   & $2.36$ [0.76]       & $6.49^{*}$ [1.69] \\
DOGE & $5.10^{***}$ [3.65] & $8.24^{***}$ [3.92] & $10.93^{***}$ [3.60] \\
ADA  & $2.03^{*}$ [1.68]   & $5.31^{***}$ [2.76] & $4.93^{*}$ [1.94] \\
\bottomrule
\end{tabularx}
\smallskip

\begin{minipage}{\textwidth}
\setstretch{1}\small
\textit{Notes:} The table reports
$\operatorname{CFE}_{15\min}(\ell)=\beta_\ell+\beta^{1\min}_\ell+\beta^{5\min}_\ell+\beta^{15\min}_\ell$
from the nested specification of eq.~(2) of the main text, including the reversal
controls $W_t=(\eta^+_t,\eta^-_t)$, in basis points per unit of order imbalance, with
Newey--West $t$-statistics in brackets. The estimation sample excludes the three
quarter-hour openings each day that coincide with perpetual funding settlement, which
Binance applies at fixed UTC times (00:00, 08:00, and 16:00; Section~2 of the main
text). These are 3 of the 96 daily quarter-hour events, about 4{,}200 openings per
contract over the sample. Forward returns are computed on the full ten-second panel
before the exclusion, so horizons remain aligned, and the quarter-hour coefficients are
identified from the 93 non-settlement openings per day. The HAC bandwidth equals the
forecast horizon in ten-second steps. Sample: January 1, 2021 to October 31, 2024.
Significance: $^{***}\,p<0.01$, $^{**}\,p<0.05$, $^{*}\,p<0.10$.
\end{minipage}
\end{table}

\newpage 

\section{Decomposition of Quarter-Hour Opening Order Imbalance: Details and Supporting Tables}\label{sec:ia_decomp}

This section accompanies Section~6.2 of the main text. It states the two-stage decomposition, reports summary statistics of quarter-hour opening order imbalance (Table~\ref{tab:summ}), the first-stage estimates (Table~\ref{tab:stage1}), the second-stage estimates (Table~\ref{tab:stage2}), the per-contract contribution shares (Table~\ref{tab:decomp_full}), and the full block-bootstrap rotation test (Table~\ref{tab:rotation_full}).

\paragraph{Stage 1.} Let $\mathrm{OI}_t$ denote quarter-hour opening order imbalance. In the first stage, we estimate
\begin{equation}
\mathrm{OI}_t
=
a
+
\sum_{k=1}^{K}\phi_k\,\mathrm{OI}_{t-k}
+
\psi^\top X_t
+
u_t,
\qquad K=12,
\label{eq:oidecomp}
\end{equation}
by ordinary least squares over the full quarter-hour opening sample (roughly 131,700 observations per asset; this exceeds the count in Section~5 of the main text, where the out-of-sample evaluation window begins on July 1, 2021). The first block consists of 12 lags of the same first-window quarter-hour imbalance, and the second block consists of the 28 technical indicators $X_t$ constructed from past prices and volumes, using the same lag length and technical-indicator specification as in Section~5 of the main text. The three estimated components are
\begin{equation}
\widehat{\mathrm{OI}}^{\mathrm{lag}}_t
=
\sum_{k=1}^{K}\hat\phi_k\,\mathrm{OI}_{t-k},
\qquad
\widehat{\mathrm{OI}}^{\mathrm{pub}}_t
=
\hat\psi^\top X_t,
\qquad
\widehat{\mathrm{OI}}^{\mathrm{res}}_t
=
\hat{a}+\hat{u}_t
=
\mathrm{OI}_t
-
\widehat{\mathrm{OI}}^{\mathrm{lag}}_t
-
\widehat{\mathrm{OI}}^{\mathrm{pub}}_t ,
\label{eq:oicomp}
\end{equation}
where the residual component absorbs the estimated intercept $\hat{a}$, which is numerically negligible, so that $\mathrm{OI}_t = \widehat{\mathrm{OI}}^{\mathrm{lag}}_t + \widehat{\mathrm{OI}}^{\mathrm{pub}}_t + \widehat{\mathrm{OI}}^{\mathrm{res}}_t$ holds exactly.

\paragraph{Stage 2.} For each horizon $\ell$, we estimate
\begin{equation}
r_{t+1,\,t+1+\ell}
=
\alpha_\ell
+
e^{\mathrm{lag}}_\ell\,\widehat{\mathrm{OI}}^{\mathrm{lag}}_t
+
e^{\mathrm{pub}}_\ell\,\widehat{\mathrm{OI}}^{\mathrm{pub}}_t
+
e^{\mathrm{res}}_\ell\,\widehat{\mathrm{OI}}^{\mathrm{res}}_t
+
\gamma_\ell' \mathrm{W}_{t}
+
\varepsilon_{t,\ell},
\label{eq:stage2}
\end{equation}
by ordinary least squares, where $\mathrm{W}_t$ is the same vector of short-term reversal controls as in eq.~(2) of the main text. Because the three imbalance components are generated from the first-stage projection \eqref{eq:oidecomp}, conventional standard errors understate sampling uncertainty by ignoring the first-stage estimation. Inference is therefore based on a moving-block bootstrap that re-estimates the entire two-stage procedure (both \eqref{eq:oidecomp} and \eqref{eq:stage2}) on each resample. Blocks are four trading days of quarter-hour observations, long enough to span the overlapping forward-return horizons, and are drawn jointly across the six contracts (the same calendar blocks for every asset) so that cross-asset dependence of the estimates is preserved; we use $B=1000$ replications. The economic magnitude of component $c \in \{\mathrm{lag},\mathrm{pub},\mathrm{res}\}$ at horizon $\ell$ is summarized by the interquartile effect
\begin{equation}
M^c_\ell
=
e^c_\ell\,\mathrm{IQR}\!\left(\widehat{\mathrm{OI}}^c_t\right),
\label{eq:iqreffect}
\end{equation}
and the contribution shares are $s^c_\ell \equiv |M^c_\ell| / \sum_j |M^j_\ell|$. The rotation statistic of the main text is
\begin{equation}
\Delta
\;\equiv\;
\frac{1}{6}\sum_{i=1}^{6}\Bigl[\,s^{\mathrm{pub}}_{i}(12\mathrm{h}) - s^{\mathrm{pub}}_{i}(4\mathrm{h})\,\Bigr].
\label{eq:rotation}
\end{equation}

\begin{table}[htbp]
\centering
\caption{Summary Statistics of Quarter-Hour Opening Order Imbalance}
\label{tab:summ}
\begin{tabular*}{\textwidth}{@{\extracolsep{\fill}}l ccc ccc c}
\toprule
Asset &  Mean & SD & IQR & Min & Median & Max & AR(1)  \\
\midrule
BTC  &  $0.016$ & $0.48$ & $0.78$ & $-0.999$ & $0.022$ & $1.000$ & $0.135$  \\
ETH  &  $0.013$ & $0.51$ & $0.82$ & $-1.000$ & $0.019$ & $1.000$ & $0.200$  \\
XRP  & $0.015$ & $0.58$ & $0.98$ & $-1.000$ & $0.018$ & $1.000$ & $0.163$  \\
SOL  &  $0.006$ & $0.50$ & $0.78$ & $-1.000$ & $0.009$ & $1.000$ & $0.054$  \\
DOGE &  $0.007$ & $0.56$ & $0.91$ & $-1.000$ & $0.007$ & $1.000$ & $0.076$  \\
ADA  &  $0.009$ & $0.59$ & $1.00$ & $-1.000$ & $0.016$ & $1.000$ & $0.211$  \\
\midrule
Mean & $0.011$ & $0.54$ & $0.88$ & $-1.000$ & $0.015$ & $1.000$ & $0.140$  \\
\bottomrule
\end{tabular*}
\medskip
\begin{minipage}[t]{\textwidth}
\setstretch{1}\small
\textit{Notes:} $\mathrm{OI}$ denotes quarter-hour opening order imbalance, measured
over the first ten seconds of each quarter hour. $\mathrm{OI}$ is bounded in $[-1,1]$;
the extreme values correspond to openings in which all first-ten-second volume is on
one side of the market. AR(1) denotes the first-order autocorrelation at the
quarter-hour frequency. The last row is the cross-asset mean.
\end{minipage}
\end{table}

\begin{table}[htbp]
\centering
\caption{First-Stage Decomposition of Quarter-Hour Opening Order Imbalance}
\label{tab:stage1}
\begin{tabular*}{\textwidth}{@{\extracolsep{\fill}}l ccc c}
\toprule
 & \multicolumn{3}{c}{Variance share (\%)} & \\
\cmidrule(lr){2-4}
Asset & Lag & Public & Residual & $R^2$ (\%) \\
\midrule
BTC  & $4.3^{\ast\ast\ast}$ & $0.6^{\ast\ast\ast}$ & $94.2$ & $5.76$ \\
ETH  & $7.7^{\ast\ast\ast}$ & $0.6^{\ast\ast\ast}$ & $90.6$ & $9.35$ \\
XRP  & $5.5^{\ast\ast\ast}$ & $0.4^{\ast\ast\ast}$ & $93.6$ & $6.42$ \\
SOL  & $0.8^{\ast\ast\ast}$ & $0.5^{\ast\ast\ast}$ & $98.4$ & $1.64$ \\
DOGE & $1.3^{\ast\ast\ast}$ & $0.7^{\ast\ast\ast}$ & $97.6$ & $2.41$ \\
ADA  & $8.1^{\ast\ast\ast}$ & $0.7^{\ast\ast\ast}$ & $90.2$ & $9.84$ \\
\midrule
Mean & $4.6$ & $0.6$ & $94.1$ & $5.90$ \\
\bottomrule
\end{tabular*}
\medskip
\begin{minipage}[t]{\textwidth}
\setstretch{1}\small
\textit{Notes:} The decomposition estimates equation~\eqref{eq:oidecomp} in sample for each asset. The first three columns give the share of the variance of $\mathrm{OI}_t$ carried by the lagged-flow component (its twelve own quarter-hour lags), the public-signal component (the 28 technical indicators), and the residual. A block's variance share is zero only if all of its coefficients are zero, so $^{\ast\ast\ast}$ marks joint significance of that block at the $1\%$ level from a Newey--West HAC Wald test; the lagged-flow and public blocks are significant at the $1\%$ level for every asset. Variance shares need not sum to exactly $100$ because the lagged-flow and public blocks are not orthogonal, and $R^2$ equals $100$ minus the residual share. The last row is the cross-asset mean.
\end{minipage}
\end{table}

\begin{table}[htbp]
\centering
\caption{Second-Stage Predictive Regression on the Components of Quarter-Hour Opening Order Imbalance}
\label{tab:stage2}
\begin{tabular*}{\textwidth}{@{\extracolsep{\fill}}l ccc}
\toprule
 & \multicolumn{3}{c}{Horizon} \\
\cmidrule(lr){2-4}
 & 4h & 8h & 12h \\
\midrule
\multicolumn{4}{l}{\emph{Coefficient $e^c_\ell$ (bp per unit), block-bootstrap $t$ in parentheses}}\\
\quad Lagged-flow $\mathrm{OI}^{\mathrm{lag}}$ & $46.1\ (2.6)$ & $52.2\ (1.9)$ & $61.5\ (1.6)$ \\
\quad Public $\mathrm{OI}^{\mathrm{pub}}$ & $17.6\ (0.4)$ & $184.8\ (2.0)$ & $310.5\ (2.5)$ \\
\quad Residual $\mathrm{OI}^{\mathrm{res}}$ & $2.3\ (2.0)$ & $3.6\ (2.2)$ & $4.3\ (2.0)$ \\
\midrule
\multicolumn{4}{l}{\emph{Interquartile effect $e^c_\ell\,\mathrm{IQR}(\mathrm{OI}^c)$ (bp)}}\\
\quad Lagged-flow & $5.0$ & $5.9$ & $6.4$ \\
\quad Public & $0.9$ & $9.8$ & $16.9$ \\
\quad Residual & $1.9$ & $3.1$ & $3.5$ \\
\midrule
$R^2$ (\%) & $0.046$ & $0.126$ & $0.176$ \\
$N$ & $131{,}728$ & $131{,}712$ & $131{,}696$ \\
\bottomrule
\end{tabular*}
\medskip
\begin{minipage}[t]{\textwidth}
\setstretch{1}\small
\textit{Notes:} The regression relates the forward cumulative return to the three components of quarter-hour opening order imbalance, equation~\eqref{eq:stage2}, estimated in sample for each asset; entries are cross-asset means. The regression includes the reversal controls $\mathrm{W}_t$ of eq.~(2) of the main text. The top panel reports the regression coefficients $e^c_\ell$ in basis points per unit of the component, with block-bootstrap $t$-statistics in parentheses; because the components differ in scale, the bottom panel reports the comparable interquartile effect $e^c_\ell\,\mathrm{IQR}(\mathrm{OI}^c)$ in basis points. The $t$-statistics are from a joint moving-block bootstrap (common four-day blocks across the six contracts, $B=1000$) that re-estimates both stages. $R^2$ is the in-sample fit and $N$ the per-asset number of quarter-hour observations on the common grid. Horizons of four, eight, and twelve hours.
\end{minipage}
\end{table}

\begin{table}[htbp]
\centering
\caption{Per-Contract Contribution Shares of Quarter-Hour Order-Imbalance Predictability}
\label{tab:decomp_full}
\begin{threeparttable}
\setlength{\tabcolsep}{7.0pt}
\small
\begin{tabular*}{\textwidth}{@{\extracolsep{\fill}}lccc}
\toprule
Asset & Lagged flow & Public signal & Residual \\
\midrule
\multicolumn{4}{l}{\textit{Panel A. 4-hour horizon}} \\
BTC  & \textbf{62} & 14 & 24 \\
ETH  & \textbf{57} & 29 & 14 \\
XRP  & \textbf{72} & 4  & 24 \\
SOL  & \textbf{55} & 18 & 27 \\
DOGE & \textbf{63} & 12 & 25 \\
ADA  & \textbf{42} & 40 & 18 \\
\midrule
Mean & \textbf{58} [34,\,65] & 20 [12,\,49] & 22 [12,\,30] \\
\addlinespace[0.6em]
\midrule
\multicolumn{4}{l}{\textit{Panel B. 8-hour horizon}} \\
BTC  & 28 & \textbf{57} & 15 \\
ETH  & 21 & \textbf{67} & 12 \\
XRP  & \textbf{43} & 33 & 24 \\
SOL  & 15 & \textbf{70} & 15 \\
DOGE & \textbf{45} & 37 & 18 \\
ADA  & 20 & \textbf{69} & 11 \\
\midrule
Mean & 29 [17,\,47] & \textbf{55} [29,\,72] & 16 [9,\,27] \\
\addlinespace[0.6em]
\midrule
\multicolumn{4}{l}{\textit{Panel C. 12-hour horizon}} \\
BTC  & 19 & \textbf{64} & 17 \\
ETH  & 15 & \textbf{75} & 9  \\
XRP  & 39 & \textbf{40} & 21 \\
SOL  & 23 & \textbf{63} & 14 \\
DOGE & 26 & \textbf{62} & 12 \\
ADA  & 12 & \textbf{84} & 4  \\
\midrule
Mean & 22 [13,\,39] & \textbf{65} [41,\,77] & 13 [8,\,24] \\
\bottomrule
\end{tabular*}
\begin{tablenotes}[flushleft]
\setstretch{1}\small
\item\textit{Notes:} This table reports per-interquartile-range contribution shares, in percent, from
the second-stage regression \eqref{eq:stage2}. For component $c$, $M^c_\ell=e^c_\ell\,\mathrm{IQR}(\widehat{\mathrm{OI}}^c_t)$, equation~\eqref{eq:iqreffect},
and shares are $s^c_\ell=|M^c_\ell|/\sum_j|M^j_\ell|$, so they summarize relative economic magnitudes rather
than variance shares. The last row in each panel reports the cross-asset mean, with a 95 percent
interval in brackets from a joint moving-block bootstrap (common four-day blocks across the six
contracts, $B=1000$) that re-estimates both stages. Per-asset entries are point estimates; their
bootstrap intervals are wide and omitted for brevity. Bold entries indicate the largest component
within each asset-horizon row.
\end{tablenotes}
\end{threeparttable}
\end{table}

\newpage 
\begin{table}[htbp]
\centering
\caption{Block-Bootstrap Test of the Rotation Statistic}
\label{tab:rotation_full}
\begin{threeparttable}
\small
\begin{tabular*}{\textwidth}{@{\extracolsep{\fill}}cccccc}
\toprule
\multicolumn{6}{l}{\textit{Panel A. Cross-asset rotation} $\Delta=\tfrac{1}{6}\sum_{i}\bigl[s^{\mathrm{pub}}_{i}(12\mathrm{h})-s^{\mathrm{pub}}_{i}(4\mathrm{h})\bigr]$}\\
\midrule
$\widehat{\Delta}$ (pp) & \multicolumn{2}{c}{$95\%$ percentile interval} & \multicolumn{3}{c}{Resamples with $\Delta>0$} \\
\midrule
$+45$ & \multicolumn{2}{c}{$[12,\,52]$} & \multicolumn{3}{c}{$996/1{,}000$} \\
\addlinespace[0.6em]
\multicolumn{6}{l}{\textit{Panel B. Within-contract change} $\Delta_i$ \textit{(pp); positive in all six}}\\
\midrule
BTC & ETH & XRP & SOL & DOGE & ADA \\
\midrule
$+50$ & $+46$ & $+35$ & $+44$ & $+50$ & $+43$ \\
\bottomrule
\end{tabular*}
\begin{tablenotes}[flushleft]
\setstretch{1}\small
\item\textit{Notes:} The rotation statistic $\Delta$, equation~\eqref{eq:rotation}, is the cross-asset mean of the within-contract change in the public-signal contribution share between the four- and twelve-hour horizons; the public-signal share rises from a cross-asset mean of $20$ to $65$ percent (Table~\ref{tab:decomp_full}). All inference is from the joint moving-block bootstrap: common four-day calendar blocks across the six contracts, $B=1000$ replications, re-estimating both stages \eqref{eq:oidecomp}--\eqref{eq:stage2} on each resample. Because the four- and twelve-hour shares are correlated within each resample, the percentile interval is computed from the bootstrap draws of the \emph{difference} $\Delta$, not from the separate level intervals in Table~\ref{tab:decomp_full}; the level intervals (twelve-hour public share $65$ with interval $[41,77]$, four-hour $20$ with $[12,49]$) overlap, whereas the interval for their difference, $[12,52]$, excludes zero. Panel~B reports the point change $\Delta_i$ for each contract.
\end{tablenotes}
\end{threeparttable}
\end{table}

\clearpage

\section{Variable Definitions}

Table~\ref{tab:vardefs} collects the notation used throughout the paper.
\begin{table}[H]
\centering
\caption{Variable definitions.}
\label{tab:vardefs}
\footnotesize
\renewcommand{\arraystretch}{1.05}
\begin{tabularx}{\textwidth}{@{}lX@{}}
\toprule
Symbol & Definition \\
\midrule
\multicolumn{2}{@{}l}{\textit{Panel A: prices, returns, and order flow (main text, Section~2)}}\\
$P_t,\ r_t$ & Last traded price in calendar-time interval $t$ and the log return $r_t=\ln P_t-\ln P_{t-1}$ (10-second or 1-minute interval). \\
$V_k,\ D_k$ & Size of trade $k$ in interval $t$ and its direction: $D_k=+1$ if buyer-initiated (\texttt{isBuyerMaker}=False), $-1$ if seller-initiated. \\
$\mathrm{OF}_t$ & Net signed order flow, $\mathrm{OF}_t=\sum_{k\in t}V_kD_k$. \\
$\mathrm{OI}_t$ & Order imbalance, $\mathrm{OI}_t=\mathrm{OF}_t/\sum_{k\in t}V_k\in[-1,1]$. \\
$d_t,h_t,m_t,b_t$ & Calendar coordinates of $t$: day, hour, minute-of-hour $m\in\{0,\dots,59\}$, within-minute 10-second block $b\in\{1,\dots,6\}$. \\
\addlinespace
\multicolumn{2}{@{}l}{\textit{Panel B: trade-size roundness (main text, Section~3)}}\\
$\mathrm{TZ}(V_k)$ & Number of trailing zeros in the base-asset size of trade $k$, measured relative to the minimum order increment $s_{\min}$ (e.g., $0.001$ BTC). \\
$\mathrm{TZshare}(z)_t$ & Share of mechanically-eligible trades in $t$ with $\ge z$ trailing zeros, $z\in\{1,2,3\}$; standardized per asset and per $z$. \\
\addlinespace
\multicolumn{2}{@{}l}{\textit{Panel C: Autocorrelation Map (main text, Section~4)}}\\
$\tau(k,m)$ & Sign-based phase-conditioned autocorrelation, $\mathbb{E}[\operatorname{sign}(x_{(d,h,m)}\,x_{(d,h,m+k)})]$, $x\in\{r,\mathrm{OF}\}$; $\hat\tau_r,\ \hat\tau_v$ denote the sample estimators for returns and signed order flow (lag $k$, phase $m$, block $b$). \\
\addlinespace
\multicolumn{2}{@{}l}{\textit{Panel D: forecasting design (main text, Section~5)}}\\
$\Delta$ & Forecast horizon, $\Delta=10$ seconds. \\
$Y^{ret}_T,\ Y^{dir}_T$ & Return response $Y^{ret}_T=R(T,\Delta)$, the forward return from the pre-$T$ price to the volume-weighted average price over $(T,T+\Delta]$, and its direction indicator $Y^{dir}_T=\mathbf{1}\{R(T,\Delta)>0\}$. \\
$L_T$ & Lag block: 12 quarter-hour-spaced first-10s returns, $R(T-15k\text{m},\Delta)$, $k=1,\dots,12$. \\
$Z_T$ (TI28) & 28 technical indicators (momentum, trend, volume, volatility) on the trailing 15-minute OHLCV grid, following \citet{Fiebergetal2025}. \\
$R^2_{\mathrm{OOS}}$, ROC AUC, Accuracy & Out-of-sample $R^2$ vs.\ the zero-return benchmark; direction-forecast ranking and 0.5-threshold classification metrics. \\
\addlinespace
\multicolumn{2}{@{}l}{\textit{Panel E: informational content (main text, Section~6)}}\\
$\operatorname{CFE}(\ell)$ & Cumulative forecasting effect: cumulative predictive slope on $\mathrm{OI}_t$ for a clock-time regime, on the forward cumulative log return $r_{t+1,\,t+1+\ell}$ over horizon $\ell$. \\
$W_t=(\eta^+_t,\eta^-_t)$ & Microstructure controls: short-term reversal dummies from the last-trade aggressor side. \\
$\widehat{\mathrm{OI}}^{\mathrm{lag}}_t,\ \widehat{\mathrm{OI}}^{\mathrm{pub}}_t,\ \widehat{\mathrm{OI}}^{\mathrm{res}}_t$ & Lagged-flow, public-signal, and residual components of quarter-hour opening imbalance (first-stage projection, main text). \\
$M^c_\ell$ & Interquartile economic effect of component $c$ at horizon $\ell$, $e^c_\ell\,\mathrm{IQR}(\widehat{\mathrm{OI}}^c_t)$. \\
\bottomrule
\end{tabularx}
\end{table}

\newpage

\section{Mid-Price versus Trade-Price Anchoring}

This section accompanies the discussion of the opening reference price in Section~5.1 of
the main text, where we anchor the quarter-hour return on the last transaction price
because the aggregate-trade data lack matched quotes over the full sample.
\citet{Sahaliaetal2025} instead use the bid-ask mid-price. On the window for which Binance
bid-ask data are available, Table~\ref{tab:midprice} re-estimates the out-of-sample design
under both anchors. The choice does not affect the conclusions of Section~5.

\begin{table}[htbp]
\centering
\caption{Out-of-Sample Performance under Mid-Price versus Trade-Price Anchoring}
\label{tab:midprice}
\footnotesize
\begin{tabularx}{\textwidth}{ll *{7}{>{\raggedleft\arraybackslash}X}}
\toprule
Anchor & Specification & BTC & ETH & XRP & SOL & DOGE & ADA & Mean \\
\midrule
\multicolumn{9}{l}{\textit{Panel A: continuous-return out-of-sample $R^2$ (\%) vs.\ zero}} \\
Mid-price   & Lag       & $5.17$ & $6.80$ & $7.66$ & $0.39$ & $3.04$ & $6.78$ & $4.97$ \\
Mid-price   & TI28      & $0.10$ & $0.12$ & $-0.19$ & $0.11$ & $-0.52$ & $0.77$ & $0.07$ \\
Mid-price   & Lag+TI28  & $5.34$ & $6.91$ & $7.88$ & $0.64$ & $2.72$ & $6.93$ & $5.07$ \\
Trade-price & Lag       & $3.65$ & $5.17$ & $0.56$ & $0.34$ & $-0.11$ & $4.58$ & $2.36$ \\
Trade-price & TI28      & $1.37$ & $1.90$ & $-0.36$ & $-0.05$ & $0.93$ & $1.51$ & $0.88$ \\
Trade-price & Lag+TI28  & $3.77$ & $5.54$ & $0.41$ & $0.21$ & $1.67$ & $4.37$ & $2.66$ \\
\addlinespace
\multicolumn{9}{l}{\textit{Panel B: direction-prediction AUC}} \\
Mid-price   & Lag       & $0.6087$ & $0.6406$ & $0.6163$ & $0.5624$ & $0.5518$ & $0.6376$ & $0.6029$ \\
Mid-price   & TI28      & $0.5741$ & $0.5825$ & $0.5380$ & $0.5413$ & $0.5464$ & $0.5830$ & $0.5609$ \\
Mid-price   & Lag+TI28  & $0.6133$ & $0.6479$ & $0.6189$ & $0.5649$ & $0.5648$ & $0.6449$ & $0.6091$ \\
Trade-price & Lag       & $0.6087$ & $0.6399$ & $0.5914$ & $0.5645$ & $0.5506$ & $0.6311$ & $0.5977$ \\
Trade-price & TI28      & $0.5703$ & $0.5829$ & $0.5233$ & $0.5397$ & $0.5421$ & $0.5753$ & $0.5556$ \\
Trade-price & Lag+TI28  & $0.6113$ & $0.6478$ & $0.5895$ & $0.5676$ & $0.5599$ & $0.6371$ & $0.6022$ \\
\addlinespace
\multicolumn{9}{l}{\textit{Panel C: direction-prediction accuracy at the 0.5 threshold (\%)}} \\
Mid-price   & Lag       & $58.24$ & $60.43$ & $59.24$ & $54.67$ & $54.15$ & $60.27$ & $57.83$ \\
Mid-price   & TI28      & $55.52$ & $56.22$ & $52.89$ & $53.29$ & $53.45$ & $56.10$ & $54.58$ \\
Mid-price   & Lag+TI28  & $58.65$ & $60.95$ & $59.64$ & $55.01$ & $54.65$ & $60.53$ & $58.24$ \\
Trade-price & Lag       & $58.26$ & $60.42$ & $56.27$ & $54.74$ & $54.21$ & $59.52$ & $57.24$ \\
Trade-price & TI28      & $55.34$ & $56.19$ & $51.68$ & $53.19$ & $52.87$ & $55.79$ & $54.18$ \\
Trade-price & Lag+TI28  & $58.39$ & $60.99$ & $56.20$ & $55.13$ & $54.37$ & $60.06$ & $57.52$ \\
\bottomrule
\end{tabularx}
\smallskip

\begin{minipage}{\textwidth}
\setstretch{1}\small
\textit{Notes:} The table repeats the out-of-sample forecasting design of Section~5 of the
main text on the window for which Binance bid-ask data are available, May~16, 2023 to
March~30, 2024 (roughly $14{,}500$ quarter-hour boundary observations per asset).
\emph{Mid-price} anchors the quarter-hour return on the bid-ask mid-price immediately
before the boundary; \emph{Trade-price} anchors it on the last transaction price, as in the
main text. Both are estimated on the identical window with the same rolling-window
walk-forward protocol (six-month training, monthly refit, LASSO for the continuous return
and $\ell_1$-penalized logistic regression for direction). Lag uses the 12 quarter-hour
lags, TI28 the 28 technical indicators, and Lag+TI28 the 40-feature union. Panel~A reports
the out-of-sample $R^2$ against the zero forecast; the shorter window is not comparable to
the full-sample $R^2$ of Table~3 and is used only to compare the two anchors. Under both
anchors the combined specification is strongest for every asset, and the direction metrics
are nearly identical (mean AUC $0.609$ versus $0.602$); if anything the mid-price anchor is
slightly stronger, consistent with the trade-price anchor carrying bid-ask-bounce noise
that the mid-price removes.
\end{minipage}
\end{table}

\section{Is the Quarter-Hour Effect a Top-of-Hour Effect?}
The round-number diagnostic of Section~3 is strongest at the top of the hour, yet the
regressions of Sections~5 and~6 pool the four quarter-hour openings (minutes 0, 15, 30, and
45). One might therefore ask whether the pooled quarter-hour coefficient is driven by the
minute-0 phase alone rather than by a genuine fifteen-minute periodicity. The
funding-settlement exclusion (Appendix Table~\ref{tab:cfe_exsettle}) does not answer
this, because it removes only 3 of the 24 daily top-of-hour openings.

\begin{figure}[H]
    \centering
    \includegraphics[width=\textwidth]{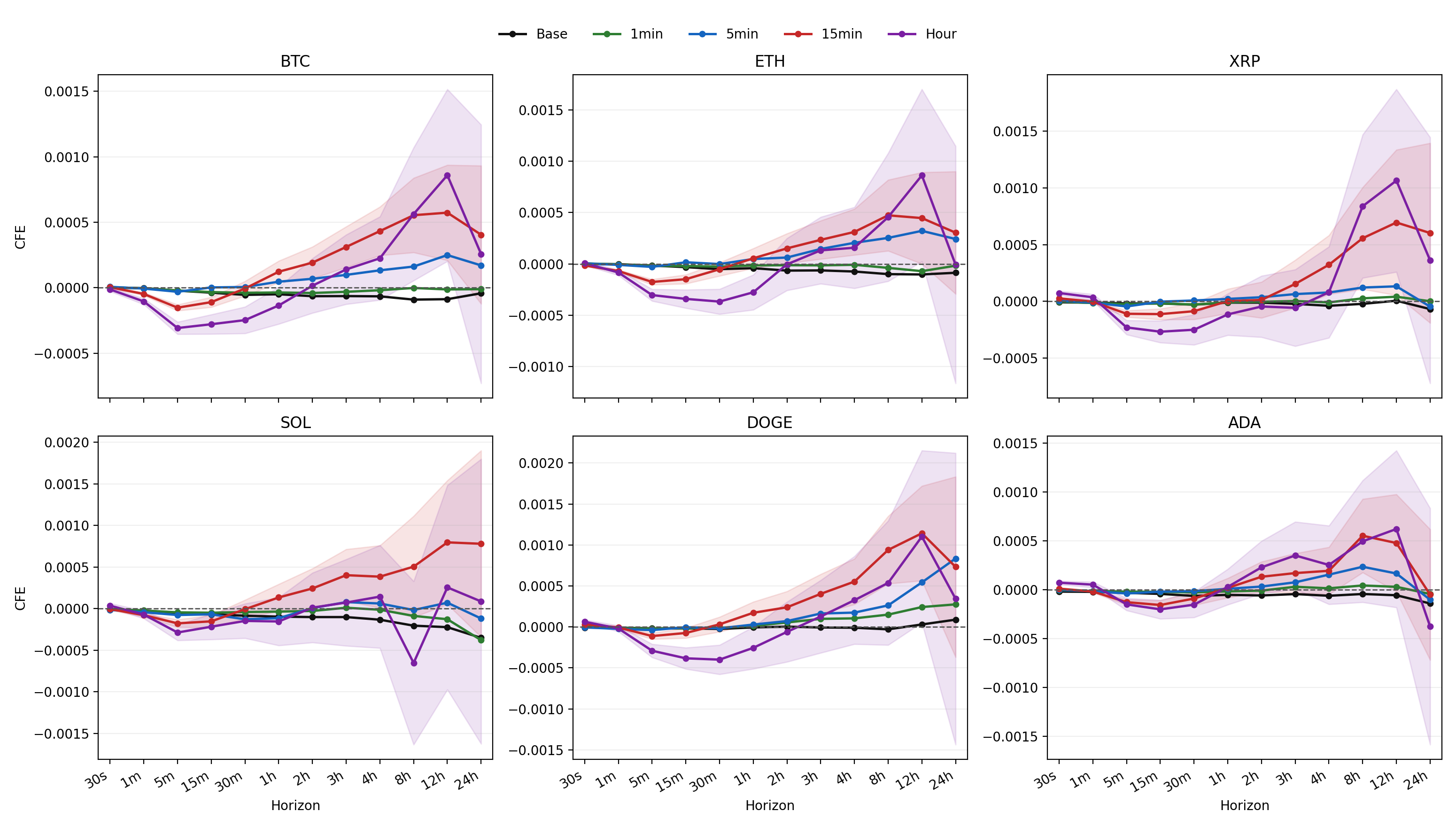}
    \caption{Cumulative forecasting effect (CFE) of quarter-hour opening order imbalance, with
    the top of the hour separated out, at horizons from 30 seconds to 24 hours. The five nested
    curves are Baseline, 1min, 5min, 15min, and Hour. Shaded bands are 95\% Newey--West HAC intervals (FFT-Bartlett kernel, bandwidth equal to the
    forecast horizon) on the 15min and Hour curves. Sample: January 1, 2021 to October 31, 2024.}
    \label{fig:cfe_top_of_hour}
\end{figure}

We augment the nested specification of eq.~(2) of the main text with a top-of-hour
interaction, $\mathrm{OI}_t\,\mathbb{I}\{m_t=0 \wedge b_t=1\}$, which absorbs the minute-0
openings. The fifteen-minute cumulative effect is then identified from the minute-15, -30,
and -45 openings alone, and a fifth curve adds the top-of-hour increment:
\begin{align*}
\operatorname{CFE}_{15\min}(\ell) &= \beta_\ell+\beta^{1\min}_\ell+\beta^{5\min}_\ell+\beta^{15\min}_\ell, \\
\operatorname{CFE}_{\mathrm{hour}}(\ell) &= \operatorname{CFE}_{15\min}(\ell)+\beta^{60\min}_\ell.
\end{align*}
The reversal controls $W_t=(\eta^+_t,\eta^-_t)$ are included as in the main text.
Figure~\ref{fig:cfe_top_of_hour} shows that the quarter-hour curve, now purged of the hour, is
essentially unchanged: positive and peaking at medium horizons in every market. The
top-of-hour increment is statistically insignificant at the four-hour horizon for every
contract and mixed in sign at longer horizons, with a sharp negative excursion for SOL at
eight hours. The quarter-hour predictability is therefore not a top-of-hour effect; the
strongest medium-horizon content sits at the fifteen-minute phase, and the minute-0 openings
add noise rather than signal.

\newpage

\section{Conventional versus Phase-Specific Autocorrelations}

\begin{figure}[H]
    \centering
    \subfloat[First 10s of each hour]{\includegraphics[width=0.33\textwidth]{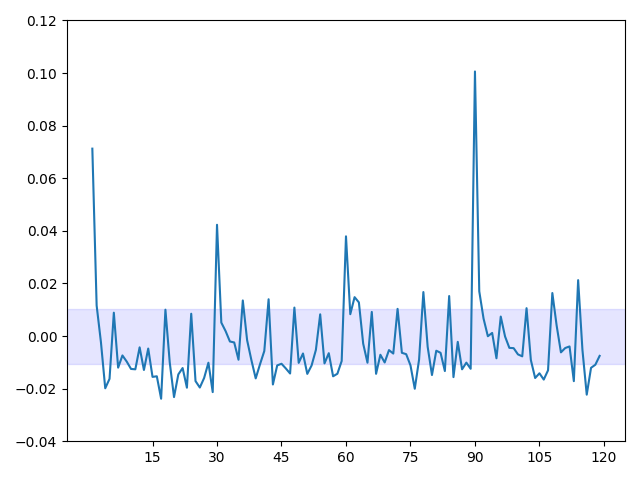}}
    \hfill
    \subfloat[Second 10s of each hour]{\includegraphics[width=0.33\textwidth]{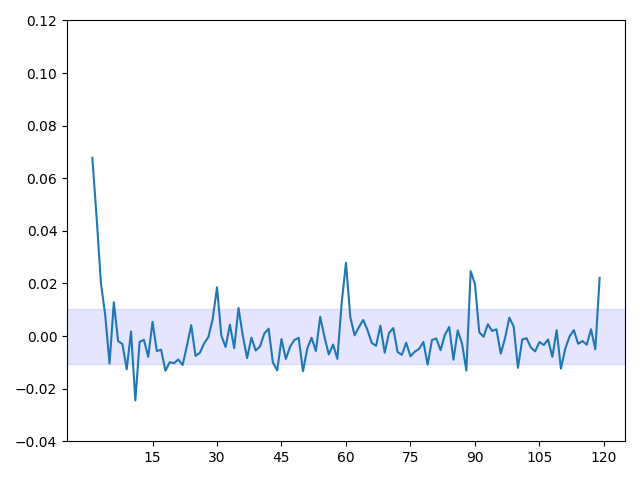}}
    \hfill
    \subfloat[Third 10s of hour]{\includegraphics[width=0.33\textwidth]{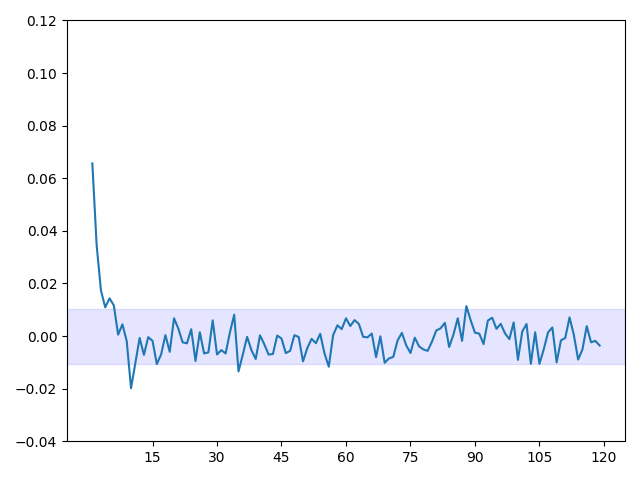}}
    \vspace{1ex}

    \begin{minipage}[c]{0.66\textwidth}
        \centering
        \subfloat[All 10-second returns]{\includegraphics[width=\textwidth]{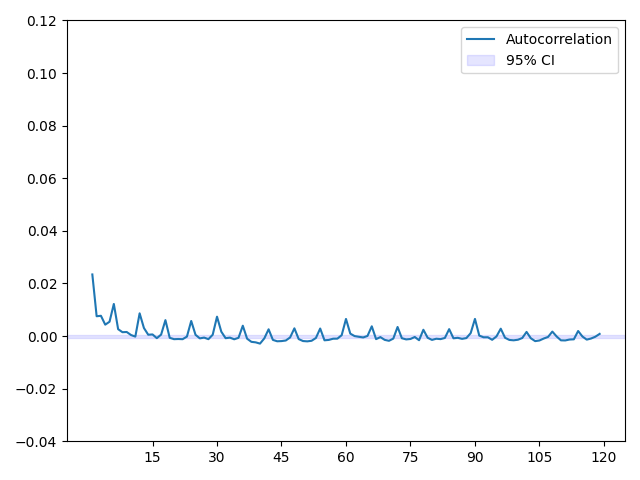}}
    \end{minipage}
    \hfill
    \begin{minipage}[c]{0.33\textwidth}
        \centering
        \subfloat[Minute 22, seconds 30--39]{\includegraphics[width=\textwidth]{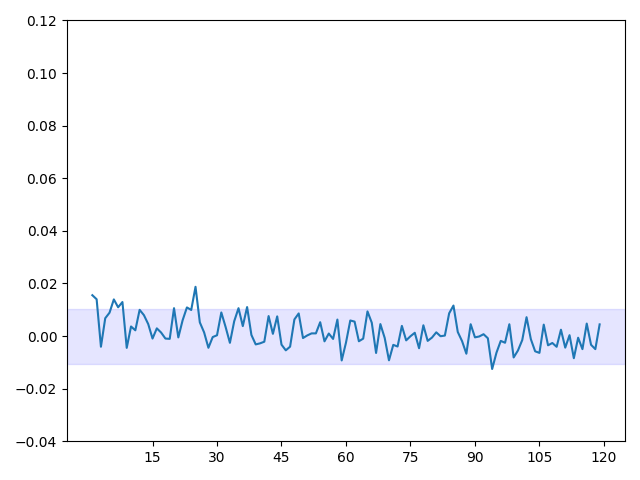}}\\
        \subfloat[Minute 52, seconds 30--39]{\includegraphics[width=\textwidth]{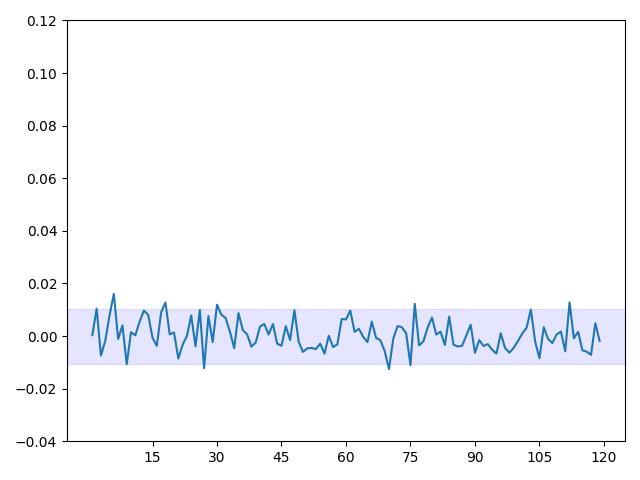}}
    \end{minipage}

    \caption{Conventional versus phase-specific autocorrelation in 10-second BTC returns. Panels (a), (b), and (c) report the phase-specific autocorrelations for the first, second, and third 10-second intervals of each hour, respectively. The spike at $k=90$ corresponds to 15 minutes.
    Panel (d) reports the conventional robust autocorrelation function (ACF) for the sign of 10-second returns.
    Panels (e) and (f) present autocorrelations for two other 10-second windows (minute 22, seconds 30--39, and minute 52, seconds 30--39) to illustrate the persistence of the phase-specific dependence across different times of the hour. Conditioning on the recurrent clock phase reveals strong, persistent dependence concentrated at the start of the 15-minute intervals that is almost entirely obscured in the aggregate autocorrelation.}
    \label{fig:acf_sacf_comparison}
\end{figure}

This section accompanies Section~4.1 of the main text. Panel (d) of Figure~\ref{fig:acf_sacf_comparison} plots the conventional robust autocorrelation function (ACF) for the sign of 10-second BTC returns. The aggregate plot displays a positive autocorrelation at $k=1$, a standard finding in high-frequency data, followed by a series of small but regular spikes at lags corresponding to one-minute intervals ($k = 6, 12, 18, \ldots$). Conditioning on clock time reveals the underlying structure. Panel (a) isolates returns observed exclusively during the first 10 seconds of each hour (seconds 0--9); despite the much smaller effective sample, the correlations exhibit a pronounced and regular periodic structure, with the spike at $k=90$ corresponding to 15 minutes. Panels (b) and (c) apply the same logic to the second and third 10-second intervals of the hour, showing that this dependence weakens substantially just seconds later. Panels (e) and (f) show that the same phase-specific pattern arises in windows away from the top of the hour.

\newpage

\section{Trade Counts by Second of the Hour}

\begin{figure}[H]
    \centering
    \includegraphics[width=0.32\textwidth]{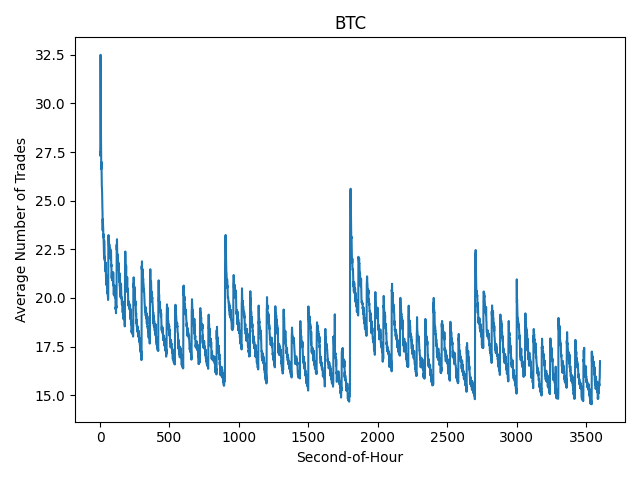}
    \includegraphics[width=0.32\textwidth]{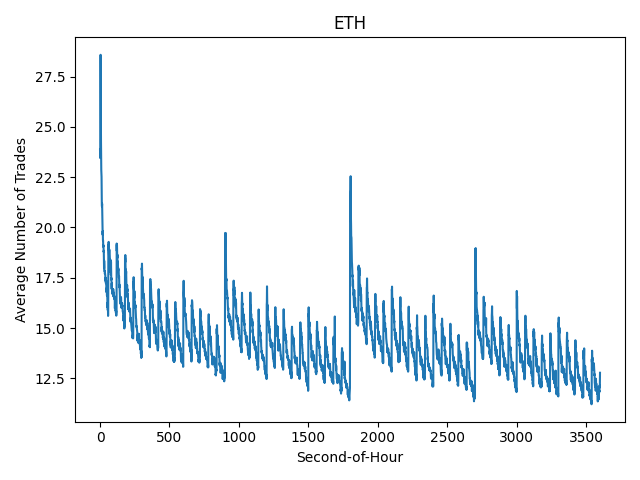}
    \includegraphics[width=0.32\textwidth]{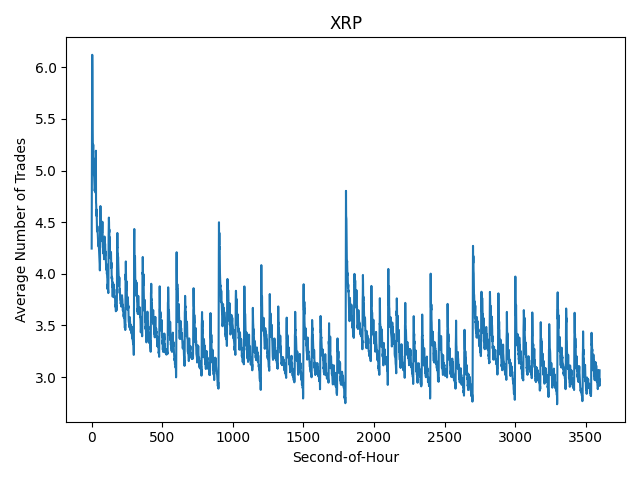}
    \includegraphics[width=0.32\textwidth]{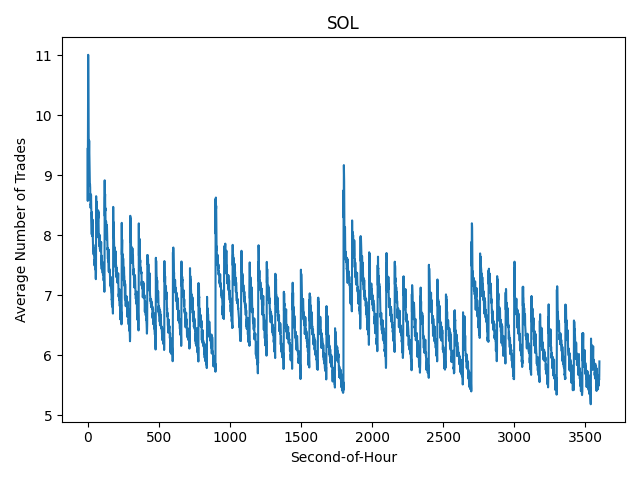}
    \includegraphics[width=0.32\textwidth]{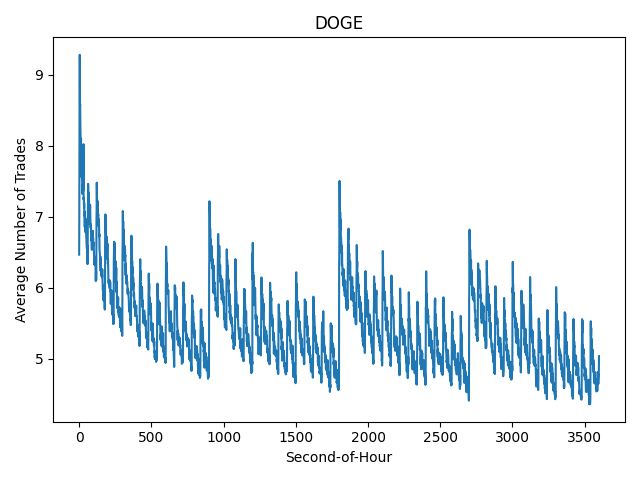}
    \includegraphics[width=0.32\textwidth]{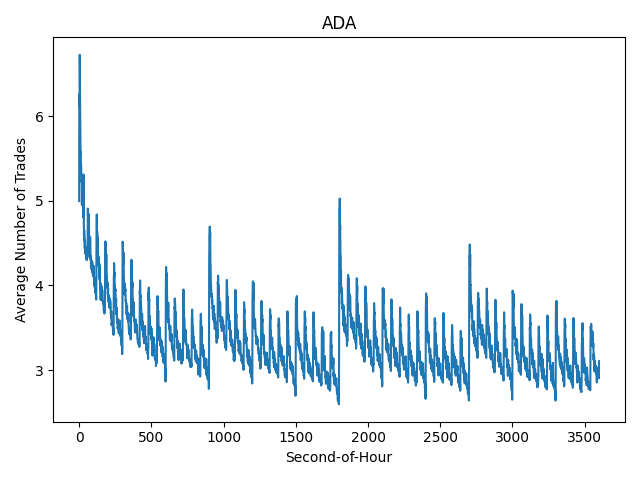}
    \caption{Each panel plots the average number of trades by second of the hour for BTC, ETH, XRP, SOL, DOGE, and ADA. Trading activity peaks sharply at the start of each hour, with additional recurring surges at the 15-, 30-, and 45-minute marks. Smaller micro-spikes also appear at one-minute intervals throughout the hour, revealing a layered intrahour periodic structure in trade arrival intensity. Sample period: January 1, 2021 to October 31, 2024.}
    \label{fig:TradingSecondOfHour}
\end{figure}

This section accompanies Section~3.1 of the main text. Figure~\ref{fig:TradingSecondOfHour} displays the average number of trades by second of the hour for the six contracts, corroborating the layered intrahour periodicity described there.

\end{document}